\documentclass[aps,prd,reprint,preprintnumbers,superscriptaddress,nofootinbib,longbibliography,floatfix]{revtex4-2}

\pdfoutput=1

\usepackage{graphicx, subcaption}
\usepackage{dcolumn}
\usepackage{bm}
\usepackage{xcolor}
\usepackage{amsmath}
\usepackage{amssymb}
\usepackage{multirow}
\usepackage{caption}
\usepackage{comment}
\usepackage{float}
\usepackage{tabularx}
\usepackage{booktabs}
\usepackage[normalem]{ulem}
\useunder{\uline}{\ul}{}

\usepackage{tikz}
\usetikzlibrary{arrows,shapes}
\usetikzlibrary{trees}
\usetikzlibrary{matrix,arrows} 				
\usetikzlibrary{positioning}				
\usetikzlibrary{calc,through}				
\usetikzlibrary{decorations.pathreplacing}  
\usepackage{pgffor}							

\usetikzlibrary{decorations.pathmorphing}	
\usetikzlibrary{decorations.markings}
\usetikzlibrary{snakes}
\tikzset{
    vector/.style={decorate, decoration={snake}, draw},
	provector/.style={decorate, decoration={snake,amplitude=2.5pt}, draw},
	antivector/.style={decorate, decoration={snake,amplitude=-2.5pt}, draw},
    fermion/.style={draw=black, postaction={decorate},
        decoration={markings,mark=at position .55 with {\arrow[draw=black]{>}}}},
    fermionbar/.style={draw=black, postaction={decorate},
        decoration={markings,mark=at position .55 with {\arrow[draw=black]{<}}}},
    fermionnoarrow/.style={draw=black},
    gluon/.style={decorate, draw=black,
        decoration={coil,amplitude=3pt, segment length=5pt}},
    scalar/.style={dashed,draw=black, postaction={decorate},
        decoration={markings,mark=at position .55 with {\arrow[draw=black]{>}}}},
    scalarbar/.style={dashed,draw=black, postaction={decorate},
        decoration={markings,mark=at position .55 with {\arrow[draw=black]{<}}}},
    scalarnoarrow/.style={dashed,draw=black},
    electron/.style={draw=black, postaction={decorate},
        decoration={markings,mark=at position .55 with {\arrow[draw=black]{>}}}},
	bigvector/.style={decorate, decoration={snake,amplitude=4pt}, draw},
 vertex/.style = {shape=circle, fill=black,minimum size=130pt,inner sep=0pt}
}

\tikzstyle{block} = [draw, rectangle, 
    minimum height=3em, minimum width=6em]

\usepackage{hyperref}
\hypersetup{
  colorlinks=true,
  citecolor=blue,
  linkcolor=blue,
  urlcolor=blue
}

\DeclareRobustCommand{\Sec}[1]{Sec.~\ref{#1}}

\DeclareRobustCommand{\Tab}[1]{Table~\ref{#1}}

\DeclareRobustCommand{\Fig}[1]{Fig.~\ref{#1}}

\renewcommand{\cp}{c_{\phi}}
\newcommand{\cdp}{c_{ \phi d}}
\newcommand{\ctp}{c_{t\phi}}
\newcommand{\op}{\mathcal{O}_{\phi}}
\newcommand{\odp}{\mathcal{O}_{ \phi d}}
\newcommand{\otp}{\mathcal{O}_{t\phi}}

\begin{document}

\title{Constraining the Higgs Potential with \\Neural Simulation-based Inference for Di-Higgs Production}
\author{Radha Mastandrea}
\email{rmastand@berkeley.edu}
\affiliation{Department of Physics, University of California, Berkeley, CA 94720, USA}
\affiliation{Physics Division, Lawrence Berkeley National Laboratory, Berkeley, CA 94720, USA}

\author{Benjamin Nachman}
\email{bpnachman@lbl.gov}
\affiliation{Physics Division, Lawrence Berkeley National Laboratory, Berkeley, CA 94720, USA}
\affiliation{Berkeley Institute for Data Science, University of California, Berkeley, CA 94720, USA}

\author{Tilman Plehn}
\email{plehn@uni-heidelberg.de}
\affiliation{Institut für Theoretische Physik, Universität Heidelberg, Germany}
\affiliation{Interdisciplinary Center for Scientific Computing (IWR), Universit\"at Heidelberg, Germany}

\date{\today}

\begin{abstract}

Determining the form of the Higgs potential is one of the most exciting challenges of modern particle physics. Higgs pair production directly probes the Higgs self-coupling and should be observed in the near future at the High-Luminosity LHC. We explore how to improve the sensitivity to physics beyond the Standard Model through per-event kinematics for di-Higgs events.  In particular, we employ machine learning through simulation-based inference to estimate per-event likelihood ratios and gauge potential sensitivity gains from including this kinematic information. In terms of the Standard Model Effective Field Theory, we find that adding a limited number of observables can help to remove degeneracies in Wilson coefficient likelihoods and significantly improve the experimental sensitivity. 

\end{abstract}

\maketitle

\tableofcontents

\section{Introduction}

The discovery of the Higgs boson at the Large Hadron Collider (LHC) in 
2012~\cite{Aad:2012tfa,Chatrchyan:2012ufa} completed the Standard Model (SM) and marked the beginning of an era of new measurements to characterize the Higgs boson properties.  Measurements of Higgs boson production and decays are a powerful probe for 
beyond-the-Standard Model (BSM) physics, and such searches are increasingly necessary given that the SM is unable to explain all known phenomena. 

The most pressing fundamental question about the Higgs sector is the form of its potential.  The structure of the potential is linked to crucial questions like the stability of the universe~\cite{Isidori:2001bm, Degrassi:2012ry, Bednyakov_2015}, the observed matter anti-matter asymmetry~\cite{Grojean:2004xa,Reichert:2017puo,Anisha:2022hgv}, and dark matter~\cite{Biekotter:2023mpd}. The SM assumes the simplest, renormalizable potential
\begin{align}
  V(\phi)  = \mu^2(\phi^\dagger\phi) + \lambda (\phi^\dagger\phi)^2 \,,
\end{align}
for an SU(2) doublet $\phi$. This potential describes trilinear and quartic self-couplings of the physical Higgs scalar $h$, with the coupling strength related to the Higgs vacuum expectation value ($v = 246$~GeV) and mass ($m_h=125$~GeV)~\cite{Workman:2022ynf},
\begin{align}
   \lambda_{3h} = \frac{3m_h^2}{v}
    \qquad \text{and} \qquad 
    \lambda_{4h} = \frac{3m_h^2}{v^2}\,.
\label{eq:sm_self}
\end{align}
Many BSM hypotheses shift these self-couplings and thus can be tested by measuring these shifts, either through quantum effects~\cite{DiVita:2017eyz} or through direct multi-Higgs production.

ATLAS and CMS are capable of probing $hh$ production and have performed many searches for both resonant and non-resonant di-Higgs boson production. In this way, they have placed limits on many BSM models affecting or extending the Higgs sector. 

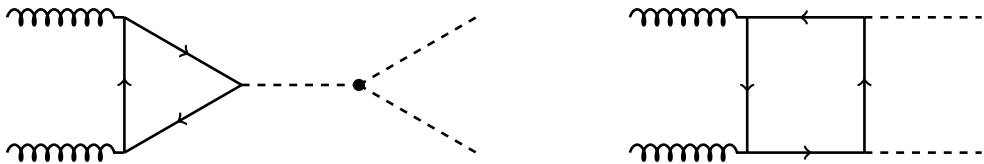
\begin{figure*}[t]
\centering
\begin{tikzpicture}[line width=1. pt,scale=1.8]
	\draw[gluon] (-1.732,0.5) -- (-0.866,0.5);
	\draw[gluon] (-1.732,-0.5) -- (-0.866,-0.5);
	\draw[fermion] (0,0) -- (-0.866,-0.5);
	\draw[fermion] (-0.866,0.5)-- (0,0);
    \draw[fermion] (-0.866,-0.5)-- (-0.866,0.5);
    \draw[scalarnoarrow] (0,0) -- (0.866,0);
    \draw[scalarnoarrow] (0.866,0) -- (1.732,0.5);
    \draw[scalarnoarrow] (0.866,0) -- (1.732,-0.5);
     \filldraw (0.866,0) circle (1pt) ;
 \end{tikzpicture}
\hspace*{0.1\textwidth}
\begin{tikzpicture}[line width=1. pt,scale=1.8]
	\draw[gluon] (-1.732,0.5) -- (-0.866,0.5);
	\draw[gluon] (-1.732,-0.5) -- (-0.866,-0.5);
	\draw[fermion] (-0.866,0.5) -- (-0.866,-0.5);
	\draw[fermion]  (0,0.5) -- (-0.866,0.5);
    \draw[fermion] (-0.866,-0.5)-- (0,-0.5);
    \draw[fermion]  (0,-0.5) -- (0,0.5);
    \draw[scalarnoarrow] (0,0.5) -- (0.866,0.5);
    \draw[scalarnoarrow] (0,-0.5) -- (0.866,-0.5);
 \end{tikzpicture}
\caption{Interfering Feynman diagrams associated with $hh$ production in the Standard Model.}
\label{fig:sm_diagrams}
\end{figure*}

We focus on non-resonant Higgs-pair production~\cite{Baur:2002rb,Baur:2003gp,Chang:2019ncg}, one of the main motivations for the High-Luminosity (HL) LHC~\cite{ZurbanoFernandez:2020cco,Faus-Golfe:2022cgm} and for a future precision hadron collider~\cite{FCC:2018byv}. 
Effective field theory in terms of the Higgs doublet (SMEFT)~\cite{Buchmuller:1985jz, Grzadkowski:2010es, Brivio:2017vri} or the physical Higgs field (HEFT)~\cite{Feruglio:1992wf, Alonso:2012px, Buchalla:2013rka} provides the appropriate framework for these analyses. Such effective theories 
rest on the fundamental assumption that any UV-completion of the SM will not 
just modify the scalar Higgs-self coupling, but induce all Wilson coefficients 
allowed by its underlying symmetry. A number of theoretical studies have explored this possibility~\cite{Goncalves:2018qas, Chang:2018uwu, Chang:2019vez, Agrawal:2019bpm, Chang:2019ncg, Mangano:2020sao,Chai:2022zeq, DaRold:2023hst, Papaefstathiou:2023uum} and recent experimental limits can be found in Refs.~\cite{CMS:2022dwd, ATLAS:2022jtk}.

Given the small $hh$ production rate, it will be essential to make the most of every candidate event. Many studies have relied on differential information to inform cuts to balance the number of $hh$-signal versus background events for a counting analysis.  This is effective, but we know that there is much more information available in the full event kinematics~\cite{Baur:2002rb,Kling:2016lay,Goncalves:2018qas}.  

For a fixed BSM hypothesis, the most powerful test statistic is the likelihood ratio with the SM. Collision events are independent, so the log-likelihood ratio 
factorizes into a rate term and a sum of per-event shape terms.
%
%
Modern machine learning gives us access to the per-event likelihood ratios using neural simulation-based inference (nSBI)~\cite{sbi_pnas}. These likelihood ratios can be estimated with neural networks using simulated data at a variety of (B)SM parameter points. Our goal is to demonstrate that a multi-dimensional per-event analysis is a promising avenue to increase  the sensitivity of current and future data to the $hh$ signal.

Several studies have explored the effects of certain HEFT Wilson coefficients on the shapes of relevant kinematic distributions, particularly $m_{hh}$ and $p_{T_{hh}}$~\cite{Carvalho_2016, Buchalla:2018yce, Capozi:2019xsi, Alasfar:2023xpc}. Such studies reflect a growing trend in particle physics towards using nSBI for exploring EFT signatures, as also evidenced by the creation of several repositories that publicize useful code for carrying out these analyses~\cite{Brehmer:2019xox, Kong:2022rnd, GomezAmbrosio:2022mpm}.

In this paper, we explore how SMEFT Wilson coefficients associated with $hh$ production can be better constrained by integrating per-event shape information, similar to earlier studies for associated Higgs production~\cite{Brehmer:2019gmn}. We consider the HL-LHC and a future 100~TeV hadron collider and attempt to constrain a set of three dimension-6 Wilson coefficients.

The paper is organized as follows. In \Sec{sec:theory}, we review relevant SM and SMEFT aspects for $hh$ production at the two colliders. In \Sec{sec:dataset}, we explain the event generation and selection procedure, as well as provide an estimation of the expected background yields. In \Sec{sec:analysis_procedure}, we describe the rate and shape analyses used to constrain the Wilson coefficients.  Numerical results are presented in \Sec{sec:results}.  The paper ends with an outlook in \Sec{sec:conclusions}.

\section{SMEFT setup}
\label{sec:theory}

\begin{figure*}[t]
\centering
\begin{subfigure}[b]{.35\linewidth}
\begin{tikzpicture}[line width=1. pt,scale=1.8]
	\draw[gluon] (-1.732,0.5) -- (-0.866,0.5);
	\draw[gluon] (-1.732,-0.5) -- (-0.866,-0.5);
	\draw[fermion] (0,0) -- (-0.866,-0.5);
	\draw[fermion] (-0.866,0.5)-- (0,0);
    \draw[fermion] (-0.866,-0.5)-- (-0.866,0.5);
    \draw[scalarnoarrow] (0,0) -- (0.6,0);
    \draw[scalarnoarrow] (0.6,0) -- (1.4,0.5);
    \draw[scalarnoarrow] (0.6,0) -- (1.4,-0.5);
   \filldraw[fill=white, ultra thick] (0.6,0) circle (3pt) ;
 \end{tikzpicture}
\caption{Contribution from $\op$ and $\odp$.}
\label{fig:BSM_1}
\end{subfigure}
\begin{subfigure}[b]{.35\linewidth}
\begin{tikzpicture}[line width=1. pt,scale=1.8]
	\draw[gluon] (-1.732,0.5) -- (-0.866,0.5);
	\draw[gluon] (-1.732,-0.5) -- (-0.866,-0.5);
	\draw[fermion] (0,0) -- (-0.866,-0.5);
	\draw[fermion] (-0.866,0.5)-- (0,0);
    \draw[fermion] (-0.866,-0.5)-- (-0.866,0.5);
    \draw[scalarnoarrow] (0,0) -- (0.6,0);
    \draw[scalarnoarrow] (0.6,0) -- (1.4,0.5);
    \draw[scalarnoarrow] (0.6,0) -- (1.4,-0.5);
   \filldraw[fill=white, ultra thick] (0,0) circle (3pt) ;
 \end{tikzpicture}
\caption{Contribution from $\odp$ and $\otp$}
\label{fig:BSM_2}
\end{subfigure}
\begin{subfigure}[b]{.35\linewidth}
\begin{tikzpicture}[line width=1. pt,scale=1.8]
	\draw[gluon] (-1.732,0.5) -- (-0.866,0.5);
	\draw[gluon] (-1.732,-0.5) -- (-0.866,-0.5);
	\draw[fermion] (-0.866,0.5) -- (-0.866,-0.5);
	\draw[fermion]  (0,0.5) -- (-0.866,0.5);
    \draw[fermion] (-0.866,-0.5)-- (0,-0.5);
    \draw[fermion]  (0,-0.5) -- (0,0.5);
    \draw[scalarnoarrow] (0,0.5) -- (0.866,0.5);
    \draw[scalarnoarrow] (0,-0.5) -- (0.866,-0.5);
    \filldraw[fill=white, ultra thick] (0,-0.5) circle (3pt) ;
 \end{tikzpicture}
\caption{Contribution from $\odp$ and $\otp$}
\label{fig:BSM_32}
\end{subfigure}
\begin{subfigure}[b]{.35\linewidth}
\begin{tikzpicture}[line width=1. pt,scale=1.8]
	\draw[gluon] (-1.732,0.5) -- (-0.866,0.5);
	\draw[gluon] (-1.732,-0.5) -- (-0.866,-0.5);
	\draw[fermion] (0.3,0) -- (-0.866,-0.5);
	\draw[fermion] (-0.866,0.5)-- (0.3,0);
        \draw[fermion] (-0.866,-0.5)-- (-0.866,0.5);
    \draw[scalarnoarrow] (0.3,0) -- (1.4,0.5);
    \draw[scalarnoarrow] (0.3,0) -- (1.4,-0.5);
   \filldraw[fill=white, ultra thick] (0.35,0) circle (3pt) ;
 \end{tikzpicture}
\caption{Contribution from $\otp$}
\label{fig:BSM_4}
\end{subfigure}
\caption{Effect of the three dimension-6 operators considered in our analysis on 
the $hh$ Feynman diagrams. The BSM vertices are denoted by open circles. For
diagram (c), the BSM vertex can be on either of the $tth$ junctions.}
\label{fig:np_diagrams}
\end{figure*}
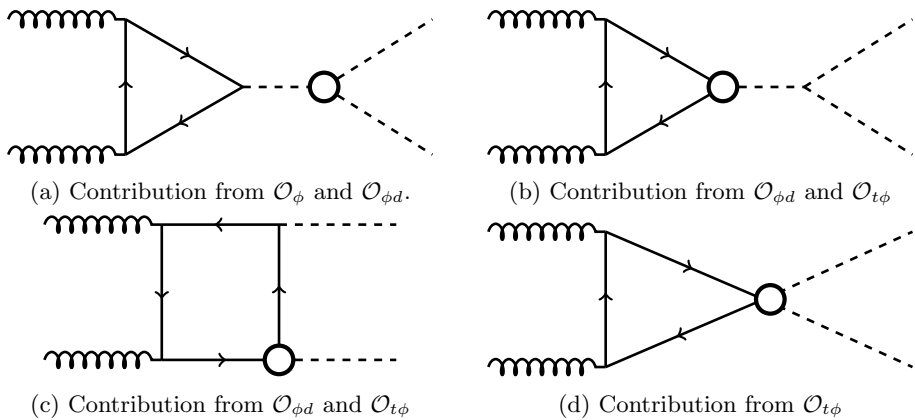

In this section, we briefly review some relevant theoretical background to $hh$ production within the SMEFT framework and define the operators we choose to vary.

SMEFT is a consistent quantum field theory framework for parameterizing the effects of new physics on the known Standard Model fields. The SMEFT Lagrangian is written as 
\begin{align}
    \mathcal{L}_\text{SMEFT} 
    = \mathcal{L}_\text{SM} + \sum_{i} \frac{c_i}{\Lambda^{d_i-4}} \mathcal{O}_i,
\label{eq:def_smeft}
\end{align}
where $\Lambda$ is the matching scale to the UV-theory, $\mathcal{O}_i$ is an operator of dimension $d_i$ composed of Standard Model fields, and $c_i$ is the Wilson coefficient governing the operator's coupling strength.

In Eq. \eqref{eq:def_smeft} we see that for instance at 
dimension 6 there is a degeneracy in the way we assign 
a given effect on the Wilson coefficient $c_i$ and the scale 
$\Lambda$. This degeneracy is only broken by matching 
to a full theory with given masses and couplings. Throughout
this paper we report SMEFT limits on $c_i$ for the fixed scale
\begin{align}
  \frac{c_i}{\Lambda^2} 
  \; \stackrel{\text{limits}}{\longrightarrow}
  \frac{c_i}{(1~\text{TeV})^2} \; .
\end{align}

At dimension 6, there exist 59 independent SMEFT operators, ignoring the flavor structure~\cite{Grzadkowski:2010es}. Global analyses of these operators is a realistic task for the LHC~\cite{Ellis:2020unq,Ethier:2021bye,Elmer:2023wtr}, while the 44,807 dimension-8 operators~\cite{Borowka:2018pxx} are unlikely to be a realistic framework for global LHC analyses~\cite{Murphy:2020rsh}. To eventually combine $hh$ results with such a global analysis~\cite{Biekotter:2018jzu}, we limit ourselves to dimension-6 operators.

For Higgs physics, there exist two complementary EFT descriptions. While the SMEFT treats the Higgs as part of a SU(2) doublet, the HEFT framework uses the physical Higgs boson and the Goldstones as the degrees of freedom. Different BSM models may be matched more easily to the SMEFT or the HEFT treatment, depending on the degree that  the SM-Higgs is responsible for the gauge boson masses; a small comparison of models is given in Chapter~2 of Ref.~\cite{Alasfar:2023xpc} and Section 5.4 of Ref.~\cite{Brivio:2017vri}. We use the SMEFT framework, assuming that the SM-Higgs really forms an SU(2) doublet with the electroweak Goldstones. The event generation framework is implemented in \textsc{MadGraph}~\cite{Alwall_2014} with the \textsc{SMEFT@NLO}~\cite{Degrande:2020evl} model.


At the LHC, the main SM contribution to $hh$ production is gluon-gluon fusion, with the Feynman diagrams presented in \Fig{fig:sm_diagrams}. The triangle diagram is sensitive to the trilinear Higgs vertex, and the box diagram can enhance this sensitivity through a cancellation at threshold~\cite{Plehn:1996wb}. When including SMEFT operators up to dimension 6, the SM-coupling of Eq. \eqref{eq:sm_self} is modified to~\cite{Plehn:2015dqa}
\begin{align}
\lambda_{3h} = 
\frac{3m_h^2}{v}  
\left[ 1-\frac{\cdp v^2}{\Lambda^2} 
- \frac{2 \cp v^4}{ \Lambda^2 m_h^2} 
+ \frac{4\cdp v^2}{3\Lambda^2m_h^2}\sum_{j<k}^3(p_j p_k)\right ] \; ,
\label{eq:trilinear_full}
\end{align}
where $\cp$ and $\cdp$ are the Wilson coefficients associated with the operators
\begin{align}
\op = (\phi^\dagger\phi)^3 
\quad \text{and} \quad 
\odp = \partial_\mu(\phi^\dagger\phi)\partial^\mu(\phi^\dagger\phi)\; .
\end{align}
We see that these two operators directly affect the trilinear Higgs 
coupling, but in different ways. In addition, $\cdp$ changes the 
field normalization of the physical Higgs, and with that, all physical
Higgs couplings.

The complete set of operators contributing to Higgs pair production is 
given in Tab.~\ref{tab:smeft_hh_operators}~\cite{Degrande:2020evl}.
The operator $\mathcal{O}_{\phi D}$ contributes the same way 
as $\mathcal{O}_{\phi d}$, but it violates custodial symmetry and is 
therefore strongly constrained by electroweak precision data.

Because the two Feynman diagrams depend on the top Yukawa coupling 
differently, we include the modified top Yukawa coupling through 
\begin{align}
\otp = (\phi^\dagger \phi) \overline{Q}t\tilde{\phi} + \text{h.c.}
\end{align}
While a BSM-induced $\mathcal{O}_{\phi G}$ contributes non-negligibly to single 
Higgs and double Higgs production, the induced $ggh$ and $gghh$ couplings 
are related through low-energy theorems, which means the BSM contribution to  
$hh$ production will be strongly constrained by single Higgs production 
and structurally similar to the modified top Yukawa~\cite{Buschmann:2014sia}.
Similarly, while $\mathcal{O}_{tG}$ will produce a number of new Higgs vertices, the coefficient is most strongly bounded by single-Higgs and $tt$ processes.
%
The effect of our three operators on the 
$hh$ Feynman diagrams is summarized in \Fig{fig:np_diagrams}.

\begin{table}[b!]
    \centering
    \begin{small} \begin{tabular}{lcc}
    \toprule
    Operator & Explicit form & \textsc{SMEFT@NLO} ID \\ 
     \midrule
        $\op$  & $(\phi^\dagger\phi)^3$ & \texttt{cp}\\
       $\odp$  & $\partial_\mu(\phi^\dagger\phi)\partial^\mu(\phi^\dagger\phi)$& \texttt{cdp} \\
         $\mathcal{O}_{\phi D}$  & $(\phi^\dagger D^\mu \phi)^\dagger (\phi^\dagger D_\mu \phi)$& \texttt{cpDC} \\
        $\otp$& $\phi^\dagger\phi \overline{Q}t\tilde{\phi}$ + h.c.  & \texttt{ctp} \\
      $\mathcal{O}_{\phi G}$& $(\phi^\dagger\phi)G^{\mu\nu}_A G^A_{\mu\nu}$  & \texttt{cpG} \\
      $\mathcal{O}_{tG}$ & $ig_s(\overline{Q}\tau^{\mu\nu}T_At)\tilde{\phi}G^A_{\mu\nu}$ + h.c. & \texttt{ctG}\\
       \bottomrule
    \end{tabular} \end{small}
    \caption{Dimension-6 SMEFT operators contributing to the $gg\rightarrow hh$. For each operator, we provide the name of the Wilson coefficient in the \textsc{SMEFT@NLO}~\cite{Degrande:2020evl} model for \textsc{MadGraph}.}
    \label{tab:smeft_hh_operators}
\end{table}




Current experimental constraints on the Wilson coefficients $\cdp$ and $\ctp$ 
have been found through global fits to LHC data~\cite{Ellis_2021,Brivio:2022hrb}. 
For instance, Ref.~\cite{Brivio:2022hrb} gives the current (profiled) 95\% confidence bounds
\begin{align}
 \cdp \in [-2.23, 3.28] 
 \quad \text{and} \quad 
 \ctp \in [-3.56, 5.75] \; ,
 \end{align}
where we note that the definition of 
$\mathcal{O}_{\phi d}$ used in Ref.~\cite{Brivio:2022hrb}  differs by the one 
used in the generating \textsc{SMEFT@NLO} \cite{Degrande:2020evl} \textsc{MadGraph} model by a factor of $1/2$.
In the future, tight bounds on $\ctp$ will come from the measurement of 
$tth$ production. 

Current bounds on $\cp$ are much looser. Since $hh$ production has not yet 
been observed at the LHC, the most recent limits come from the 2022 CMS 
and ATLAS summaries~\cite{CMS:2022dwd, ATLAS:2022jtk}, which constrain the 
Higgs self-coupling modification 
$\kappa_\lambda \in [-1.24, 6.49]$ and $\in [-1.4, 6.1]$, respectively. 
Assuming a cutoff scale $\Lambda = 1$~TeV, the first interval can be converted 
to $\cp \in [-12.95, 5.39]$, but this limit is not based on a 
global EFT analysis, and is neither model-independent nor conservative.
 
\section{Event generation}
\label{sec:dataset}

\begin{table*}[t]
\begin{small} \begin{tabular}{lrlllrlll}
\toprule
& \multicolumn{4}{l}{HL-LHC, 14TeV, 3ab$^{-1}$}                                  & \multicolumn{4}{l}{Future Collider, 100 TeV, 30ab$^{-1}$}                         \\ \midrule
             & \multicolumn{2}{c}{Signal} & \multicolumn{2}{c}{Background} & \multicolumn{2}{c}{Signal} & \multicolumn{2}{c}{Background} \\
             & Events     & Retention     & Events       & Retention       & Events     & Retention     & Events       & Retention       \\
Start        & 257        & 100\%         & --           & --              & 89,604     & 100\%         & --           & --              \\
+ tagging \& efficiencies    &   95     &  37.1\% &     4.65$\times 10^{4}$         &  100\%            &       29,600        &          33.0\%       &   5.16$\times 10^{6}$   &              100\%         \\
+ kinematic cuts    &    49        &   18.9\%            &         1.43$\times 10^{4}$      &    30.8\%         &        11,100    &      12.3\%             &      1.58$\times 10^{6}$          &      30.6\%           \\
+ $m_h$ windows   &    15        &  5.89\%             &         4.09$\times 10^{2}$      &    0.88\%         &      3,950    &      4.40\%             &      4.02$\times 10^{4}$          &        0.78\%         \\
+ angular cuts &    13        &     4.92\%          &      4.37$\times 10^{1}$           &        0.094\%         &       3,600     &          4.02\%    &      4.34$\times 10^{3}$          &      0.084\%           \\ \bottomrule
\end{tabular} \end{small}
\caption{Signal and background cut flows for the HL-LHC ($S/B = 0.30$) and the Future Collider ($S/B = 0.83$). Both the signal and background event yields reflect the NLO prediction,}
\label{tab:event_yields}
\end{table*}

In this section, we provide details on the event generation procedure 
for the $hh$ and background samples used in this analysis. 

For the decay channel, we choose the process
\begin{align}
 pp \to hh \to 
  (b\bar{b}) \, (\gamma\gamma) \; .
\end{align}
The $h\rightarrow b\bar{b}$ decay is ideal because it has the largest branching 
fraction at 58\%; the $h\rightarrow\gamma\gamma$ channel has a much smaller 
branching rate of 0.227\%, but it benefits from the excellent photon identification and 
resolution~\cite{Baur:2003gp}. At the LHC, it is expected that a measurement of 
$hh$ decay will be made through combining five channels~\cite{CMS:2022dwd, ATLAS:2022jtk}:
$\bar{b}b\bar{b}b$,  $b\bar{b}\tau\tau$~\cite{Baur:2003gpa,Dolan:2012rv,Baglio:2012np}, 
$b\bar{b}\gamma\gamma$, $4l$, and $b\bar{b}ZZ$, with $b\bar{b}\gamma\gamma$ 
providing the best sensitivity along with $b\bar{b}\tau\tau$. Because 
of the large QCD backgrounds, 
the naively most promising $\bar{b}b\bar{b}b$ channel will likely be most useful for 
on-shell modifications of Higgs pair production.  

We consider two colliders to probe the Higgs self-coupling directly:
\begin{itemize}
    \item The HL-LHC, $\sqrt{s}$ = 14 TeV to 3~ab$^{-1}$~\cite{ZurbanoFernandez:2020cco}, and
    \item a future hadron collider, $\sqrt{s}$ = 100 TeV to 30~ab$^{-1}$~\cite{FCC:2019}.
\end{itemize}
%

\subsection{SMEFT signal}
\label{sec:evt_sig}

The $gg \to hh$ production cross sections in the SM are 
known to NLO in QCD~\cite{Dawson:1998py,Grigo:2013rya,Maltoni:2014eza,Borowka:2016ypz,Borowka:2016ehy,Baglio:2018lrj,Baglio:2020ini}, approximate NNLO~\cite{deFlorian:2015moa,Grigo:2015dia,deFlorian:2016uhr}, N$^3$LO~\cite{Chen:2019lzz,Chen:2019fhs}, and NLO including 
parton showers~\cite{Heinrich:2019bkc}.
We use the NLO rate prediction 
32.81(7) fb (+13.5\%, -12.5\%) for the HL-LHC and 1140(2) fb (+10.7\%, -10.0\%) at 
100~TeV \cite{Mangano:2020sao,Baglio:2020wgt}. The combined branching ratio for $hh\rightarrow b\bar{b}\gamma\gamma$ is 0.262\%~\cite{LHCHiggsCrossSectionWorkingGroup:2016ypw}.
The expected SM event yields are given in \Tab{tab:event_yields}. 

To generate $hh$ events, we use \textsc{MadGraph} 3.5.1 \cite{Alwall_2014} with 
the \textsc{SMEFT@NLO}~\cite{Degrande:2020evl} model, which assumes a finite top quark mass and is LO with respect to the $ggh$ vertex. Note that \textsc{SMEFT@NLO} canonically normalizes the dynamical Higgs field, so the derivative correction to the trilinear coupling visible in the last term of Eq. \eqref{eq:trilinear_full} is not modeled. We use \textsc{MadSpin}~\cite{Artoisenet_2013} for the Higgs decays
and \textsc{Pythia}~8.306 \cite{Bierlich:2022pfr} for the underlying event, parton 
shower, and hadronization. To simulate detector effects, we use 
\textsc{Delphes}~3.5.0 \cite{de_Favereau_2014} with the HL-LHC card. The detector simulation uses 
a particle flow-like reconstruction and clusters jets with the 
anti-$k_t$~\cite{Cacciari:2008gp} algorithm with $R = 0.4$ using 
\textsc{FastJet}~\cite{Cacciari:2011ma}.

For SMEFT extensions, precision predictions for $hh$ production exist 
to NLO with subleading operators~\cite{Heinrich:2023rsd},
combined with parton shower~\cite{Heinrich:2020ckp},
including truncation uncertainties~\cite{Heinrich:2022idm}, 
and approximate NNLO precision~\cite{deFlorian:2021azd}.
As to be discussed in \Sec{sec:morphing}, we generate signal events at a set of 
10 points in the 3D Wilson coefficient space, such that event weights at an 
arbitrary parameter point can be calculated using the morphing basis. At 
the SM point, we generate 10$^6$ (1.5$\times 10^6$) events for the HL-LHC (100~TeV), 
and at each of the other 9 points, we generate 5$\times 10^5$ (7.5$\times 10^5$) 
events. 


After event generation, we carry out a number of analysis selections to 
guarantee the event acceptance and enhance the 
$hh$-signal rate relative to the background ratio. The 
choice of cuts leaves us a lot of freedom. Looser cuts will lower $S/B$, which is 
less ideal for a cut-and-count analysis, but not a problem for ML-based analyses. We 
leave further explorations of this trade-off to future studies. For the selections, 
we follow a similar strategy as Ref. \cite{Goncalves:2018qas}. We define three 
categories of selections:
\begin{enumerate}
    \item \textit{$b$-quality:} at least 2 $b$-tagged jets.
    
    \item \textit{kinematics:} at least two jets 
    and two photons with 
    \begin{align}
    p_T> 30~\text{GeV} 
    \quad \text{and} \quad 
    |\eta| < 2.4 \; .
    \end{align}
    The 
    four main analysis objects must have an angular separation of $\Delta R > 0.4$. These cuts select events with high triggering efficiencies.

    \item \textit{Higgs mass windows:} two mass windows, 
    \begin{align}
    |m_{b\bar{b}} - m_h| &< 25~\text{GeV} \notag \\
    |m_{\gamma\gamma} - m_h| &< 3~\text{GeV} \; ,
    \end{align}
    which strongly reduce the background events while minimally reducing the signal.

    \item \textit{angular cuts:} in addition to the acceptance cuts, 
    \begin{align}
        \Delta R_{b\gamma} > 1 
        \quad \text{and} \quad \Delta R_{\gamma\gamma} < 2 \; .
    \end{align}
    This reduces the background by a factor of $\sim$ 10 and the 
    signal by only a factor $\sim$ 1.2~\cite{Baur:2003gp}.
\end{enumerate} 
After event selection, we are left with 129k (176k) signal events for the HL-LHC 
(100~TeV), spread across the 10 morphing generation points.

\begin{table*}[t]
\begin{small} \begin{tabular}{llllllll}
\toprule
   &     & $pp\rightarrow b\bar{b}\gamma\gamma$ & $pp\rightarrow b\bar{b}j\gamma$ & $pp\rightarrow b\bar{b}jj$ & $pp\rightarrow jj\gamma\gamma$ & $pp\rightarrow Zh$ & Total  \\
\midrule
\multirow{4}{*}{HL-LHC}          & Cross-section (LO) {[}pb{]}    & 0.009758       & 61.25     & 8833    & 1.946     & 5.773$\times 10^{-5}$ \\
   &  Events w/ loose kin. cuts     & 2.92$\times 10^{4}$      & 1.84$\times 10^{8}$           & 2.65$\times 10^{10}$       & 5.83$\times 10^{6}$  &  1.73$\times 10^{2}$  & 2.67$\times 10^{10}$ \\
  & + tagging \& efficiencies        & 1.10$\times 10^{4}$           & 1.56$\times 10^{4}$       & 5.50$\times 10^{2}$    &  2.06$\times 10^{3}$     & 6.6$\times 10^1$      & 2.93$\times 10^4$         \\
  &  + kinematic cuts     & 4.55$\times 10^{3}$           & 3.96$\times 10^{3}$     & 1.75$\times 10^{2}$    &  5.48$\times 10^{2}$       & 4.3$\times 10^1$ & 9.28$\times 10^3$           \\
 & + $m_h$ windows & 1.59$\times 10^{2}$   &   5.90$\times 10^{1}$ & 8.01$\times 10^{0}$           & 3.27$\times 10^{1}$    &  2.2$\times 10^{1}$ &  2.81$\times 10^{2}$\\
  & + angular cuts & 1.29$\times 10^{1}$   &   6.51$\times 10^{0}$ & 1.43$\times 10^{0}$           & 2.19$\times 10^{0}$    &  7.7$\times 10^{0}$ &  3.07$\times 10^{1}$\\
      & $K$-factor adjusted & 1.75$\times10^1$  &  1.15$\times10^1$ & 2.15$\times10^0$      &3.37$\times10^0$    &  9.16$\times10^0$ &  4.37$\times 10^{1}$\\
\midrule
\multirow{4}{*}{Future Collider} & Cross-section (LO) {[}pb{]}    &0.09731       & 707    & 1.127$\times 10^5$    & 15.72 & 4.062$\times 10^{-4}$           \\
 &  Events w/ loose kin. cuts    & 2.91$\times 10^6$      & 2.12$\times 10^{10}$    & 3.38$\times 10^{12}$     & 4.72$\times 10^8$  & 1.22$\times 10^4$   & 3.040$\times 10^{12}$\\
& + tagging \& efficiencies   & 1.10$\times 10^6$      & 1.83$\times 10^{6}$    & 7.24$\times 10^{4}$     & 2.54$\times 10^5$  & 4.61$\times 10^3$     & 3.26$\times 10^{6}$ \\
&  + kinematic cuts    & 4.37$\times 10^5$      & 4.86$\times 10^{5}$    & 2.33$\times 10^{4}$     & 6.86$\times 10^{4}$  & 3.00$\times 10^{3}$    & 1.02$\times 10^6$    \\
 & + $m_h$ windows & 1.47$\times 10^{4}$   &   6.57$\times 10^{3}$ & 1.10$\times 10^{3}$           & 3.98$\times 10^{3}$    & 1.50$\times 10^{3}$ &  2.79$\times 10^{4}$\\
  & + angular cuts & 1.01$\times 10^{3}$   &   8.38$\times 10^{2}$ & 2.73$\times 10^{2}$           &2.94$\times 10^{2}$    &  5.92$\times 10^{2}$ &  3.01$\times 10^{3}$\\
    & $K$-factor adjusted & 1.37$\times10^3$ & 1.47$\times10^3$ & 4.10$\times10^2$         & 3.82$\times10^2$ &  7.04$\times10^2$ &  4.34$\times 10^{3}$\\
\bottomrule
\end{tabular} \end{small}
\caption{All background cross-sections are given in pb. The $jj\gamma\gamma$ background 
includes the $c\bar{c}\gamma\gamma$ background. The $pp\rightarrow Zh$ cross sections and event yields are for the full decay $pp\rightarrow Z(\rightarrow b\bar{b})h(\rightarrow\gamma\gamma)$. Cross sections are calculated at leading order (LO) with loose kinematic cuts. All rows except the last represent the LO event yields; the last row multiplies by the process-specific $K$-factor stated in the text to give the NLO event yields. }
\label{tab:background_cross_sections}
\end{table*}

\begin{figure*}[t]
\centering
\includegraphics[width=\textwidth]{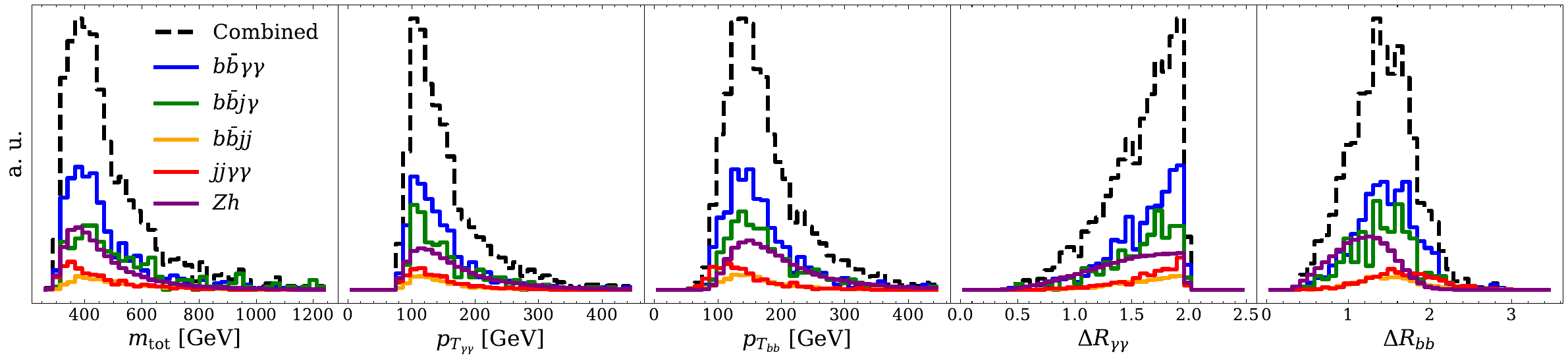}
\caption{Kinematic background distributions used in the shape analysis. We only show the HL-LHC predictions. The dashed line shows the total background from considering all processes in \Tab{tab:background_cross_sections}, while the solid lines represent each individual process, scaled to their level of contribution. The vertical axis is linearly scaled. We defer definition and motivation of these observables to \Sec{sec:evt_obs}.}
\label{fig:bkg_comparisons}
\end{figure*}

\subsection{Continuum backgrounds}
\label{sec:evt_bkg}

\begin{figure*}
    \centering
    \includegraphics[width = 0.9\textwidth]{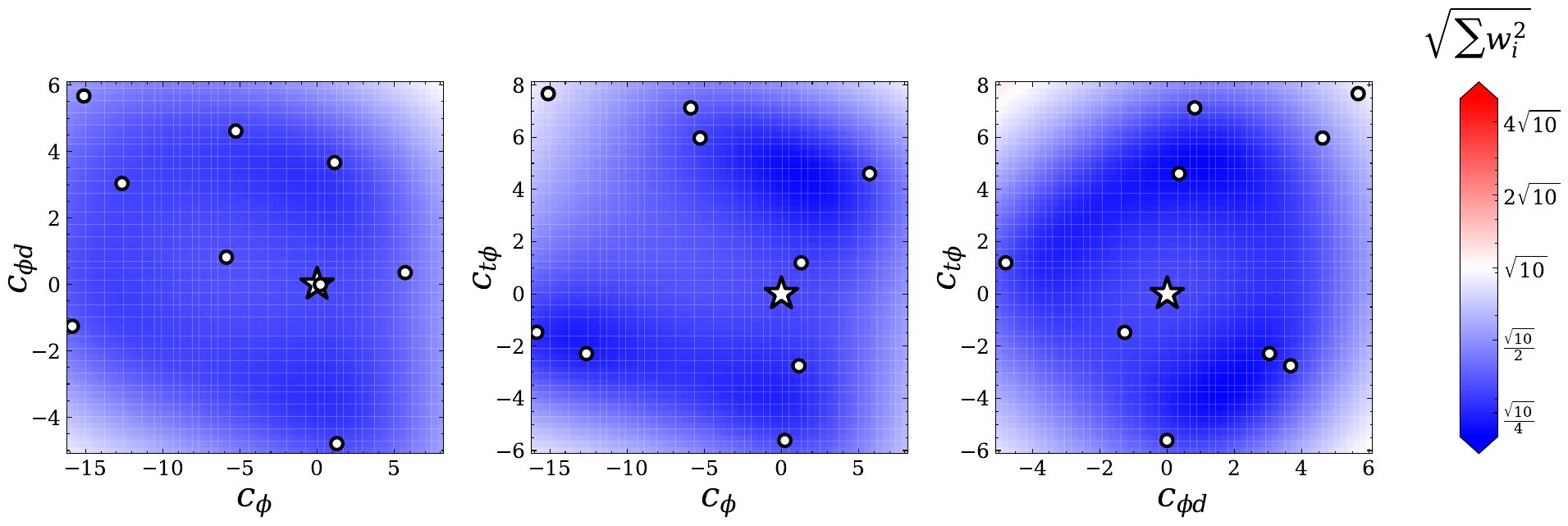}
    \caption{Squared morphing weights over the chosen parameter ranges for the set of Wilson coefficients $c = (\cp{}, \cdp{}, \ctp{})$.
    The SM point is denoted by a star, and the other 9 benchmarks given in \Tab{tab:generation_points} are denoted by circles.
    At each parameter point $c$, the weights needed to convert from 
    that point to each of the 10 benchmarks are computed; the squared sum is 
    then displayed. For numerical stability of the reweighting process, it is 
    desirable that $\sqrt{\sum 
    w_i^2} \lesssim \sqrt{10}$.}
    \label{fig:morphing_weights}
\end{figure*}

The main backgrounds for the $hh\rightarrow b\bar{b}\gamma\gamma$ channel can be divided into three categories:
\begin{enumerate}
    \item \textit{Continuum:} $gg\rightarrow b\bar{b}\gamma\gamma$ with 
    two real $b$ quarks and two real photons, but no intermediate resonances.
    \item \textit{Mistags}: a light jet mistagged as a photon, or a light or a $c$-jet 
    mistagged as a $b$-jet. Possible channels include: $b\bar{b}j\gamma$, $b\bar{b}jj$, $jj\gamma\gamma$, and $c\bar{c}\gamma\gamma$, all without an intermediate Higgs.
    \item \textit{Single-Higgs:} a single Higgs boson produced in association with other objects, including $b\bar{b}h$, $c\bar{c}h$, and $Zh$. 
\end{enumerate}

To estimate the importance of each background, we run 
\textsc{MadGraph} at LO for each of the SM processes in \Tab{tab:background_cross_sections} 
at leading order. Slightly looser kinematic selections to those from \Sec{sec:evt_sig} are enforced at 
the parton level during event generation, to speed up the analysis. 
When calculating the expected background yields, we do not run \textsc{Pythia}, but we simulate detector effects manually for computational 
tractability: $p_T$~smearing, jet efficiencies, photon efficiencies, and mistagging 
rates are carried out following the HL-LHC \textsc{Delphes} run card. 

We calculate the NLO yields by multiplying the LO yields for each process by its corresponding $K$-factor, given in \Tab{tab:k_factors}. In is important to highlight that the $K$-factors listed are for the total cross section. In reality, the $K$-factors are scale-dependent functions of $p_T$ and will modify the shapes of the distributions of kinematic observables. Since this shape modification is expected to be independent of the SMEFT Wilson coefficients, we do not consider this effect in this analysis.

\begin{table}[b!]
    \centering
    \begin{small} \begin{tabular}{lcc}
    \toprule
    Process & $K_\textrm{NLO/LO}$, 14 TeV & $K_\textrm{NLO/LO}$, 100 TeV \\ 
     \midrule
        $pp\rightarrow b\bar{b}\gamma\gamma$  & 1.36  \cite{Fah:2017wlf} (13 TeV)& (14 TeV value) \\
       $pp\rightarrow b\bar{b}j\gamma$ & 1.76 \cite{Kim:2024ppt} & (14 TeV value)\\
         $pp\rightarrow b\bar{b}jj$  & 1.50 \cite{Kim:2024ppt} & (14 TeV value)\\
        $pp\rightarrow jj\gamma\gamma$& 1.54 \cite{Kim:2024ppt}  & 1.3  \cite{Mangano:2016jyj} \\
      $pp\rightarrow Zh$&  1.19 \cite{Kim:2024ppt}  & (14 TeV value) \\

       \bottomrule
    \end{tabular} \end{small}
    \caption{$K_\textrm{NLO/LO}$ factors for the cross section for the designated background process. Where we cannot find a value in the literature for the 100 TeV $K$-factors, we use the 14 TeV value. For the 100 TeV $K$-factor for the $jj\gamma\gamma$ background, there is a cut $p_{T_j} > 50$ GeV.}
    \label{tab:k_factors}
\end{table}

Altogether, we generate $8 \times 10^6$ ($7.6 \times 10^6$) background events for the HL-LHC (100~TeV). This 
large number is necessary to train the parameterized classifier for the shape 
analysis, described in \Sec{sec:shape_analysis}. 
After the selections given in Sec.~\ref{sec:evt_sig}, we are left with 290k (183k) 
background events.\medskip

According to Tab.~\ref{tab:background_cross_sections},
the $b\bar{b}\gamma\gamma$ continuum and $Zh$ backgrounds contribute the most to the total background,  followed by the $b\bar{b}j\gamma$ process. The $b\bar{b}jj$ background
has by far the largest cross section, but is effectively reduced by the low 
$j\rightarrow\gamma$ mistag rate and the $m_h$-windows. The 
$b\bar{b}h$ and $c\bar{c}h$ backgrounds with their moderate cross sections, 
0.012 and 0.040 pb at the HL-LHC and 0.22 and 0.34 pb at 100~TeV, respectively, are completely removed by the $m_h$ window requirement and thus omitted from 
Tab.~\ref{tab:background_cross_sections}.
To simplify the analysis, we can look at the 
kinematic distributions of the combined and component backgrounds in 
\Fig{fig:bkg_comparisons}. We see that the shapes of the full background distributions are very close to those of the two most contributing processes $b\bar{b}\gamma\gamma$ and $b\bar{b}j\gamma$ (which in turn look similar to each other), justifying the computationally-motivated choice to only generate continuum $b\bar{b}\gamma\gamma$ events for the shape analysis background.

We additionally make the assumption that none of the continuum backgrounds is 
significantly affected by SMEFT modifications. 
While this is not exactly true, we have checked the 
effect of $\otp$ on the $Zh$-background and find that the change in 
the background prediction 
is negligible compared with that of the $hh$ signal. Therefore we do not consider SMEFT modifications to backgrounds in this analysis.

When comparing the number of expected background events with the number of expected 
signal events in \Tab{tab:event_yields}, we see that $S/B \sim 0.30$ at the HL-LHC 
and $S/B \sim 0.83$ at 100~TeV. Without the hard angular cuts, these would be 0.036
and 0.098, respectively. These values, as well as the background 
yields in \Tab{tab:background_cross_sections}, are in agreement with 
Ref.~\cite{Goncalves:2018qas}.

Finally, we acknowledge that our treatment of the continuum and mistag backgrounds are simplified and require more detailed studies with full simulations and/or data for improved accuracy.  In an actual analysis, these 
backgrounds would likely be estimated from data~\cite{CMS:2020tkr, Jia:2024nbv}. 
We prioritize showing a proof-of-concept of shape information
constraining SMEFT coefficients, rather than a highly-realistic collider 
analysis.


\subsection{Morphing through parameter space}
\label{sec:morphing}

\begin{figure*}[t]
    \centering
    \includegraphics[width = \textwidth]{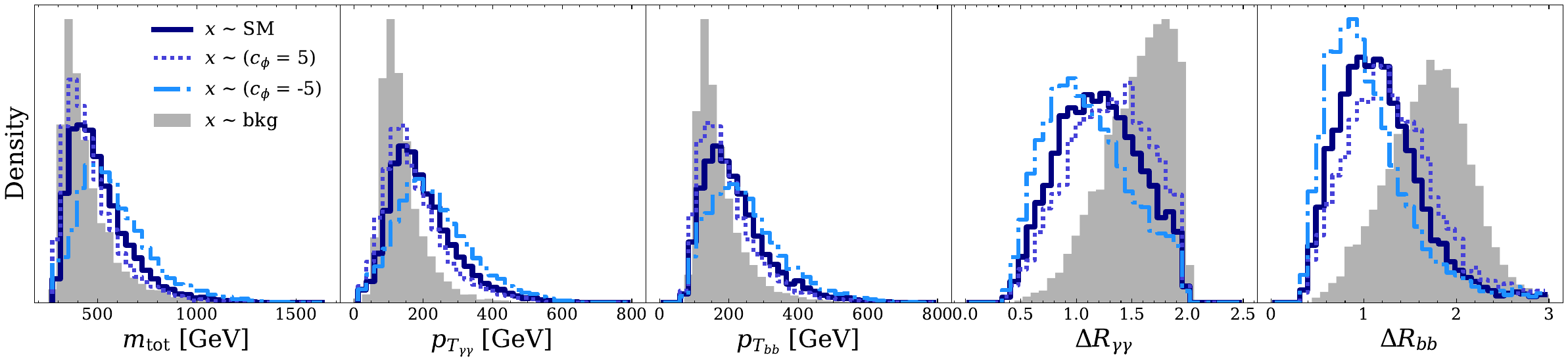}
    \caption{Kinematic signal and background distributions, including 
    two choices of $\cp$. We only show HL-LHC predictions. The vertical axis is linearly scaled.}
    \label{fig:core_5_features}
\end{figure*}

Parameter estimation with simulation-based inference compares 
data with simulations assuming different parameter values describing the underlying physics.  
Methods differ in how the real and synthetic data are compared (see 
Sec.~\ref{sec:analysis_procedure}), but they must all cover the full parameter space.  

\begin{table}[b!]
    \centering
    \begin{small} \begin{tabular}{crrr}
        \hline
        Generation Point & $c_\phi$ & $c_{\phi d}$ & $c_{t \phi}$ \\
        \hline
        1 (SM) & 0 & 0 & 0   \\
       2 & 5.710 & 0.354 & 4.604   \\
       3 & -5.873 & 0.817 & 7.124   \\
       4 & 1.135 & 3.664 & -2.754   \\
       5 & -12.638 & 3.035 & -2.288\\
       6 & 1.281 & -4.792 & 1.188   \\
       7 & -15.854 & -1.261 & -1.477   \\
       8 & -15.107 & 5.670 & 7.668   \\
       9 & -5.265 & 4.612 & 5.967   \\
       10 & 0.221 & -0.006 & -5.613   \\
       \hline
    \end{tabular} \end{small}
    \caption{Morphing basis points, from \textsc{MadMiner}. This basis minimizes the sum of the squared weights out of a set of $10^7$ random bases.}
    \label{tab:generation_points}
\end{table}

Generating new events at each parameter point is computationally 
prohibitive.  Instead, one re-uses events generated at one parameter point 
by leveraging the factorization between the parton-level physics governing 
Higgs production and decay and the long(er) distance-scale physics governing 
fragmentation and subsequent simulation steps. This scale-separation 
implies that the likelihood ratio between two parameter points is fully 
covered by an exchange of partonic matrix elements. 
We can reweight an event from one parameter value to another 
and in this way morph the set events for one parameter value to the set
for another parameter point. 

This idea is implemented in \textsc{MadMiner}~\cite{Brehmer:2019xox}. 
In particular, \textsc{MadMiner} expresses the matrix elements for a given 
process as a polynomial over chosen Wilson coefficients, simulates events 
at a number of parameter points comprising the ``morphing basis", and fits
the polynomial to the morphing basis to quickly generate event weights 
at an arbitrary parameter point. 
This works well when the new parameter point is close to the original one; 
when moving too far away, the weights can be far from unity, and the statistical power of the weighted sample is diminished~\cite{Chatterjee:2022oco}.
An alternative, future direction could be to use differentiable simulations 
instead of surrogate
models~\cite{Carrazza:2021zug,Heinrich:2022xfa,Nachman:2022jbj}.

In our case, the model parameter space is composed of three-dimensional vectors 
\begin{align}
 c =  (\cp, \cdp, \ctp) \in \mathbb{R}^3 \; .
\label{eq:def_c}
\end{align}
For these three Wilson coefficients, we define 
a polynomial up to squared terms, resulting in
10 fit parameters, which is also the standard procedure in 
global SMEFT analyses~\cite{Elmer:2023wtr}.
In our case, the corresponding 10 basis points are given in 
\Tab{tab:generation_points}. 
This 
set is chosen by \textsc{MadMiner} by minimizing the sum of the squared 
weights from $10^7$ random instantiations. The locations of the 
generation points and the corresponding squared 
morphing weights are shown 
in \Fig{fig:morphing_weights}. To generate this figure, we scan over a 
dense grid of points $c$, compute the weights needed to convert from that 
point to each of the 10 benchmarks, $w_i$, and display 
the sum $\sqrt{\sum_i w_i^2}$. Since higher weights are associated with 
larger uncertainties on the morphing basis, it is encouraging that the 
majority of the computed weights are $\lesssim 1$, or $\sqrt{\sum 
    w_i^2} \lesssim \sqrt{10}$.

\subsection{Kinematic Observables}
\label{sec:evt_obs}

To study the main differences between the SM Higgs signal, the SMEFT effects,
and the continuum backgrounds, we define a set of $N=5$ observables, for which 
the sensitivity to the Higgs self-couplings and SMEFT effects is well known:
(i) the reconstructed di-Higgs mass, (ii) the reconstructed transverse momentum
of the photonically decaying Higgs, (iii) the reconstructed transverse momentum 
of the hadronically decaying Higgs, (iv) the the angular separation between the two 
photons, and (v) the angular separation between the two $b$-jets from the Higgs decay,
\begin{align}
N=5 
\qquad 
\Big\{ \; 
 m_\text{tot} ,p_{T_{\gamma\gamma}}, p_{T_{bb}}, \Delta R_{\gamma\gamma}, \Delta R_{bb} 
 \; \Big\} \; .
\label{eq:def_obs}
\end{align}
The reconstructed di-Higgs mass combines sensitivity to threshold cancellations
with a test of the Higgs kinematics~\cite{Baur:2002rb,Goncalves:2018qas}, and the 
Higgs transverse momenta test the momentum dependence of the production process
and the top-loop threshold~\cite{Dolan:2012rv}. The angular separations 
are strongly correlated to the transverse momenta, but only for the signal and
not for the continuum backgrounds.

All five observables are shown for the SM Higgs signal, two choices of 
the Wilson coefficient $\cp$, and the continuum background in 
\Fig{fig:core_5_features}. The most striking feature is that the 
Higgs signals and the continuum background look very different, where the 
backgrounds contain much less energy per event and the reconstructed Higgs 
decays products are widely separated. 
Analyses searching for deviations from the SM $hh$ production and for deviations
from the SM signal and the continuum background will produce quite different results~\cite{Kling:2016lay}.

\section{Analyses}
\label{sec:analysis_procedure}

Independent of the representation of our data $D$, the analysis goal is to determine 
if the data is more consistent with the SM prediction $c=0$ or some finite values for the 
Wilson coefficients $c = (\cp, \cdp, \ctp)$ defined in Eq. \eqref{eq:def_c}.
The data can either be events or bins of an $N$-dimensional observable, 
$D=\{x_i\}_{i=1}^k$. For a fixed BSM model $c$, the Neyman-Pearson 
Lemma~\cite{neyman1933ix} states that the most powerful test statistic is the 
likelihood ratio mentioned already in \label{eq:def_llr}
\begin{align}
    q(c|D) =  -2 \log \frac{p(D|c)}{p(D|c=0)} \; ,
\label{eq:np}
\end{align}
where the factor of two and natural logarithm imply that 
in the Gaussian approximation, changes in $q$ by one unit correspond 
to one standard deviation for a 1-dimensional $c$. 


If a production cross section strongly depends on a set of model 
parameters, a natural but by no means optimal first step is a rate-only analysis.
This is the current approach used by CMS~\cite{CMS:2022dwd} and ATLAS~\cite{ATLAS:2022jtk}.
This case corresponds to $N=0$, and the probability density for $D$ simplifies 
to a comparison of Poisson distributions for a given number of $k$ events,
\begin{align}
    q_\text{rate}(c|D)
    &= -2 \; \log\left(\frac{\text{Pois}(k|c)}{\text{Pois}(k|c=0)}\right) \notag \\
%
%
&=-2\left(\overline{k}(0)-\overline{k}(c)+k\ln\left(\frac{\overline{k}(c)}{\overline{k}(0)}\right)\right)\,.
\end{align}
where $\overline{k}(c)=\sigma_\text{tot}(c)\times L$ is the predicted number of events for given parameters $c$.

\subsection{Shape analysis with classifiers}
\label{sec:shape_analysis}

Collider events are statistically independent, which means that the full 
log-likelihood ratio factorizes into a rate term and a shape term, where 
the shape term is a sum over events or observable bins, 
\begin{align}
    q(c|D)
    &=q_\text{rate}(c|D)
    -2 \sum_{i=1}^k \log \frac{p(x_i|c)}{p(x_i|c=0)} \notag \\
    &\equiv q_{c,\text{rate}}(c|D) + q_{c,\text{shape}}(c|D)\,.
\end{align}
We explore various scenarios, with $N$ up to 5. 
The corresponding observables are given in Eq. \eqref{eq:def_obs}, and their histograms 
are presented in \Fig{fig:core_5_features}.  
For our analysis, we use subsets of the five observables, namely\footnote{The $N=5$ case is presented in the Appendix.}
\begin{alignat}{7}
N&=1 \qquad 
 \Big\{ \; 
 m_\text{tot} 
 \; \Big\} \notag \\
N&=3 \qquad 
 \Big\{ \; 
 m_\text{tot} ,p_{T_{\gamma\gamma}}, p_{T_{bb}}
 \; \Big\} \; .
\label{eq:def_obs2}
\end{alignat}
Unlike in the rate-only case, we do not know $p(x_i|c)$ explicitly. 
To estimate the per-event likelihood ratio, we use the 
fact that trained classifiers $C(x) \in [0,1]$ learn a monotonic function of the density 
ratio (see e.g. Ref.~\cite{hastie01statisticallearning,sugiyama_suzuki_kanamori_2012,cranmer}). 
A calibrated classifier then becomes
%
%
%
\begin{align}
    \frac{C(x)}{1-C(x)}\approx \frac{p(x|c)}{p(x|c=0)}\,,
\label{eq:notparameterized}
\end{align}
%
and the baseline configuration is known to work well~\cite{Rizvi:2023mws}.  
Our key assumption here is that we can sample accurately and precisely from $p(x|c)$.  
Here, the 
simulation-based inference represents an ideal that may be achievable completely or partially 
by the time such an analysis is performed on data. 

In our case, we need to promote the likelihood ratio to be a function of $c$. This is accomplished 
by training a parameterized classifier~\cite{cranmer,Baldi:2016fzo} where $c$ is promoted 
to a feature, This means that  we train a classifier $C$ acting on $(x,c)$.  For training, 
$c$ is drawn from a prior $p_0(c)$ and then signal events are drawn from $p(x|c)$.  
For the background events, we assign values $c$ so that the marginal distribution 
of $c$ is $p_0$.  In this way, $c$ is not useful for the classifier and 
\begin{align}
    \frac{C(x,c)}{1-C(x,c)}\approx 
    \frac{p_\text{BSM}(x,c)}{p_\text{SM}(x,c)}=\frac{p(x|c)}{p(x|0)}\,.
\label{eq:parameterized}
\end{align}
This looks the same as Eq. \eqref{eq:notparameterized}, but now it is a continuous 
function of $c$.


In practice, we can further simplify the construction of $q$ if we assume that we can neglect 
quantum interference between the $hh$ signal and the continuum background.
For our specific analysis, this is ensured by the small Higgs width, which 
suppresses all interference contributions. 
In that case $p(x|c)$ can be approximated by a mixture model
\begin{align}
\label{eq:mm}
    p(x|c)\approx \mu(c) \, p_S(x|c) + (1-\mu(c)) \, p_B(x) \; ,
\end{align}
where the relative proportion $\mu(c)$ of $hh$ to the total number of events 
is known from Sec.~\ref{sec:morphing}, and  $p_{S,B}$ are the 
corresponding probability densities of the $hh$ and the continuum events. 
As noted above, we have verified that for the Wilson coefficients considered, the 
variation of $p_B$ with $c$ is negligible. Thus we can 
rewrite the log-likelihood ratio in terms of $p_S$ and 
$p_B$~\cite{cranmer}
\begin{align}
\frac{p(x|c)}{p(x|0)} &= \left[ \frac{\mu(0)\,p_S(x|0)}{\mu(c)\,p_S(x|c)} +  \frac{(1-\mu(0))\,p_B(x)}{\mu(c)\,p_S(x|c)}\right]^{-1} \notag \\
&\hspace{5mm}+ \left[ \frac{\mu(0)\,p_S(x|0)}{(1-\mu(c))\,  p_B(x)} + \frac{ 1-\mu(0)}{1-\mu(c)}\right]^{-1} \; .
\label{eq:decompose}
\end{align}
While this expanded form may look complicated, it allows us to
break down the problem into three easier problems. In particular, the first three terms in 
Eq. \eqref{eq:decompose} can be approximated with classifiers that each have an easier task than 
distinguishing samples drawn from $p(x|c)$ and $p(x|0)$ all at once:
\begin{enumerate}
    \item $\frac{p_S(x|0)}{p_S(x|c)}$ is learned by a parameterized classifier distinguishing SM from SMEFT $hh$ events; 
    \item $\frac{p_B(x)}{p_S(x|c)}$ is learned by a parameterized classifier distinguishing background from SMEFT $hh$ events;
    \item $\frac{p_S(x|0)}{p_B(x)}$ is learned by a non-parameterized classifier distinguishing SM $hh$ events from continuum background.
\end{enumerate}

\begin{figure*}[t]
\includegraphics[width=0.45\textwidth]{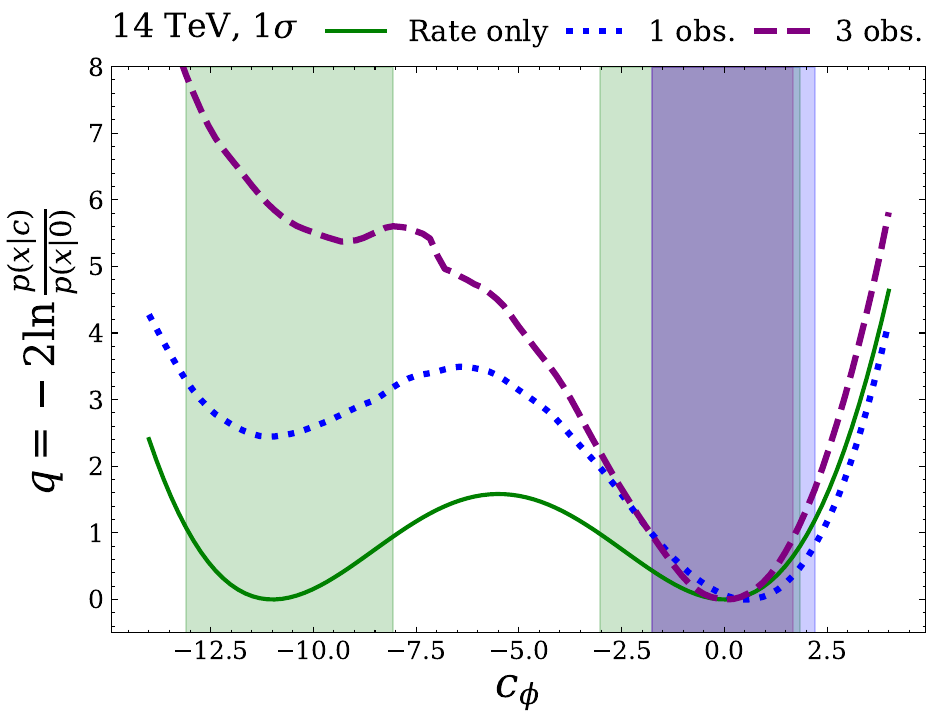} \hfill
\includegraphics[width=0.45\textwidth]{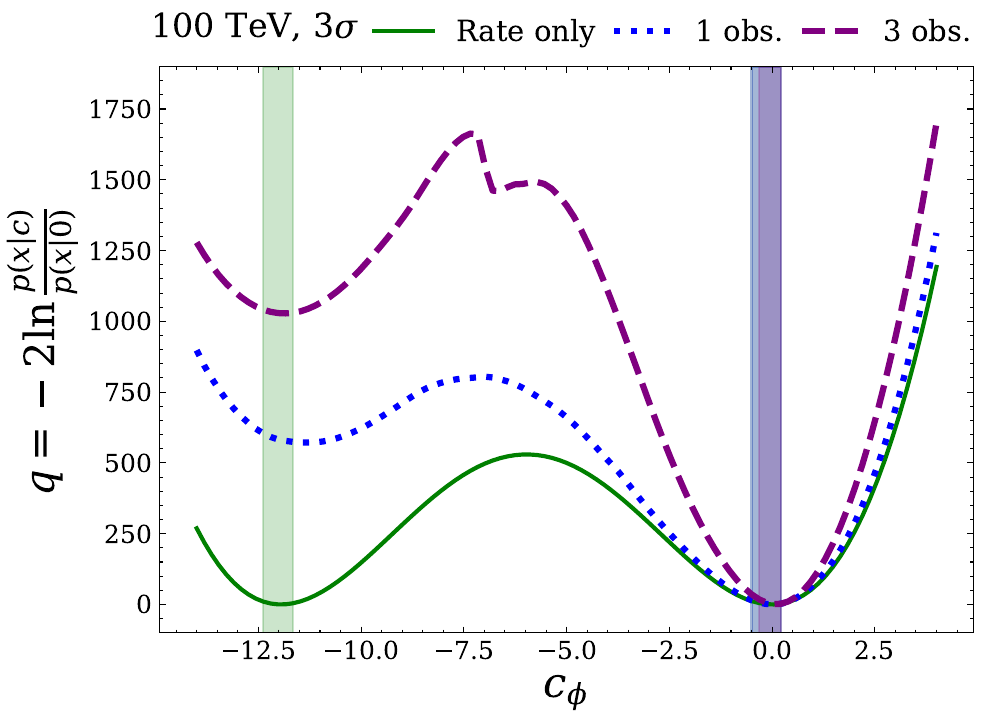} \\
\includegraphics[width=0.45\textwidth]{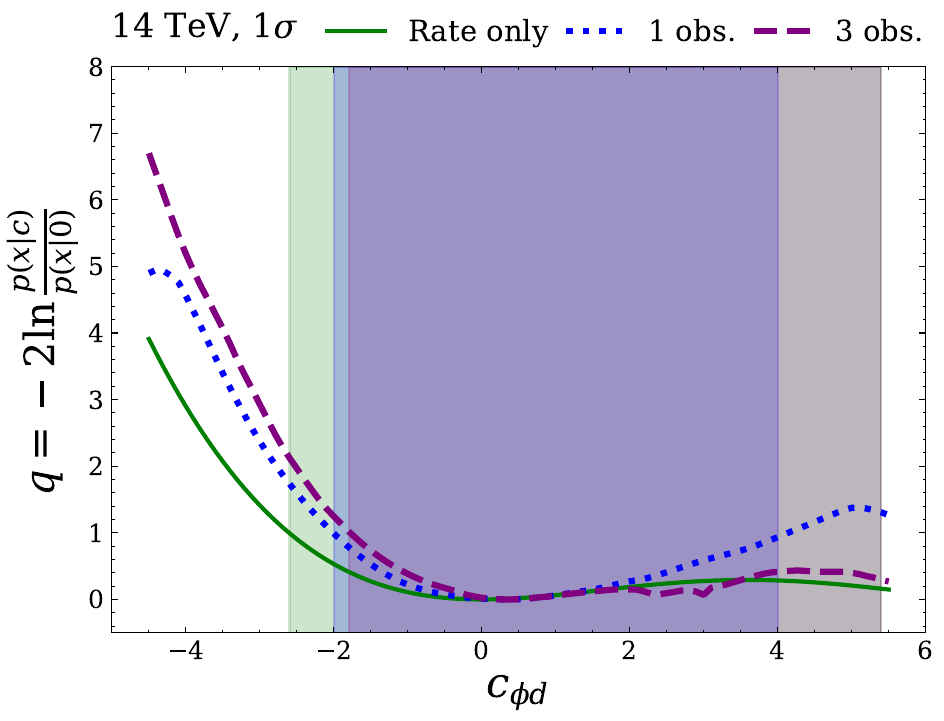} \hfill
\includegraphics[width=0.45\textwidth]{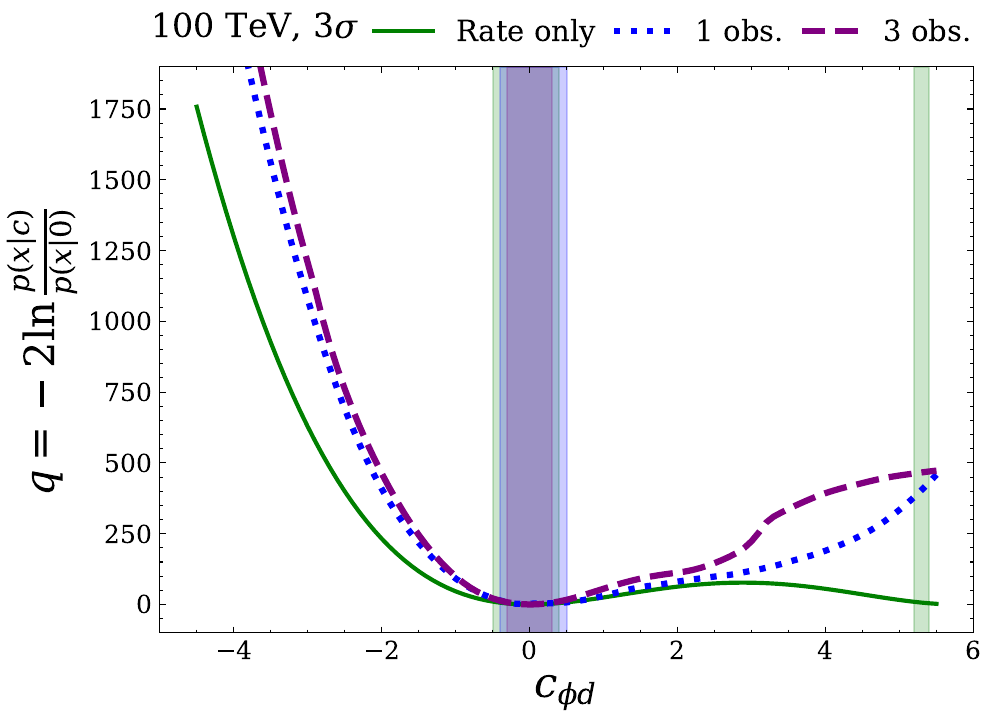} \\
\includegraphics[width=0.45\textwidth]{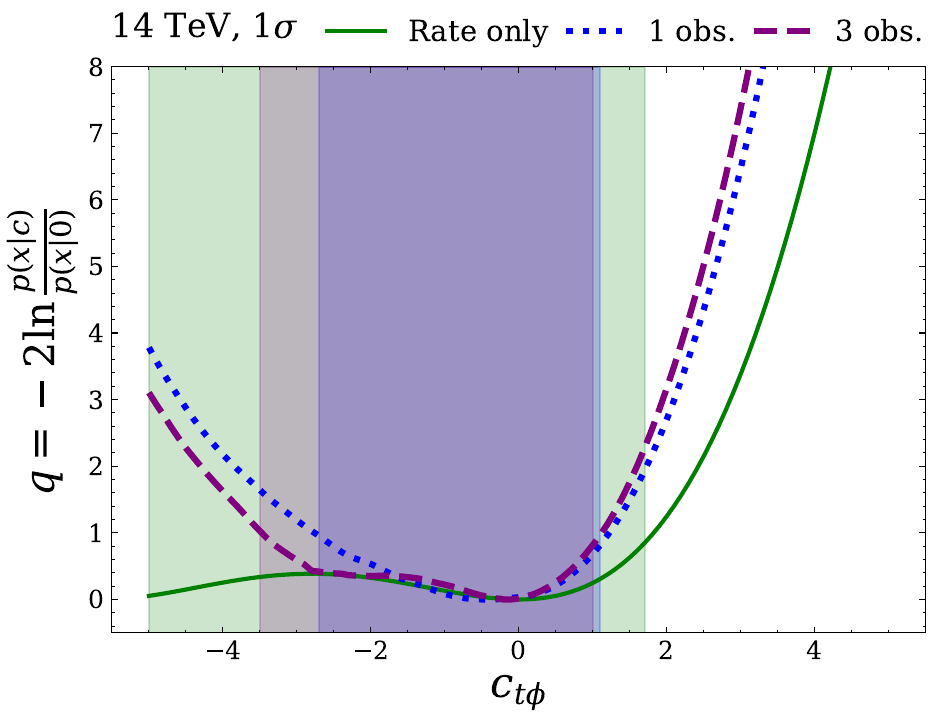} \hfill
\includegraphics[width=0.45\textwidth]{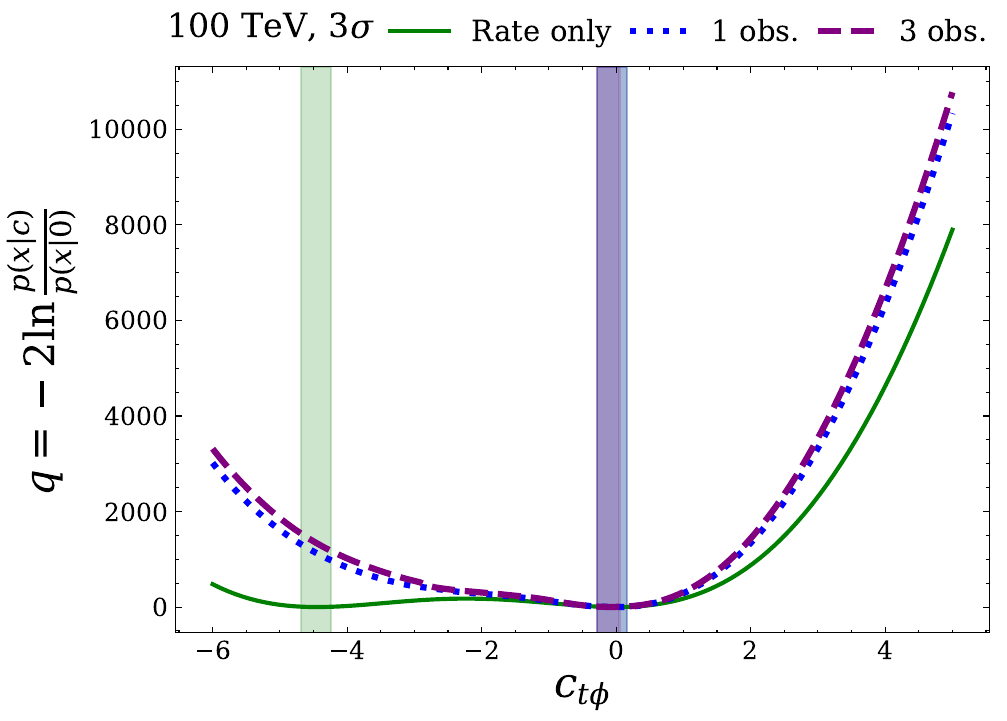}
\caption{Log-likelihood ratio test statistics in terms of 1$\sigma$ (for HL-LHC) or 
3$\sigma$ (for 100 TeV) confidence intervals for one
non-zero Wilson coefficient at a time. 
Data generation and test set size reflect the collider setup.}
\label{fig:q_results_1d}
\end{figure*}

\subsection{Training specifications}

Event generation for the parameterized classifier is done within the \textsc{MadMiner} 
framework, whose morphing feature allows for quick generation of events at arbitrary 
parameter points $c$. 

\begin{figure*}
\includegraphics[width=0.32\textwidth]{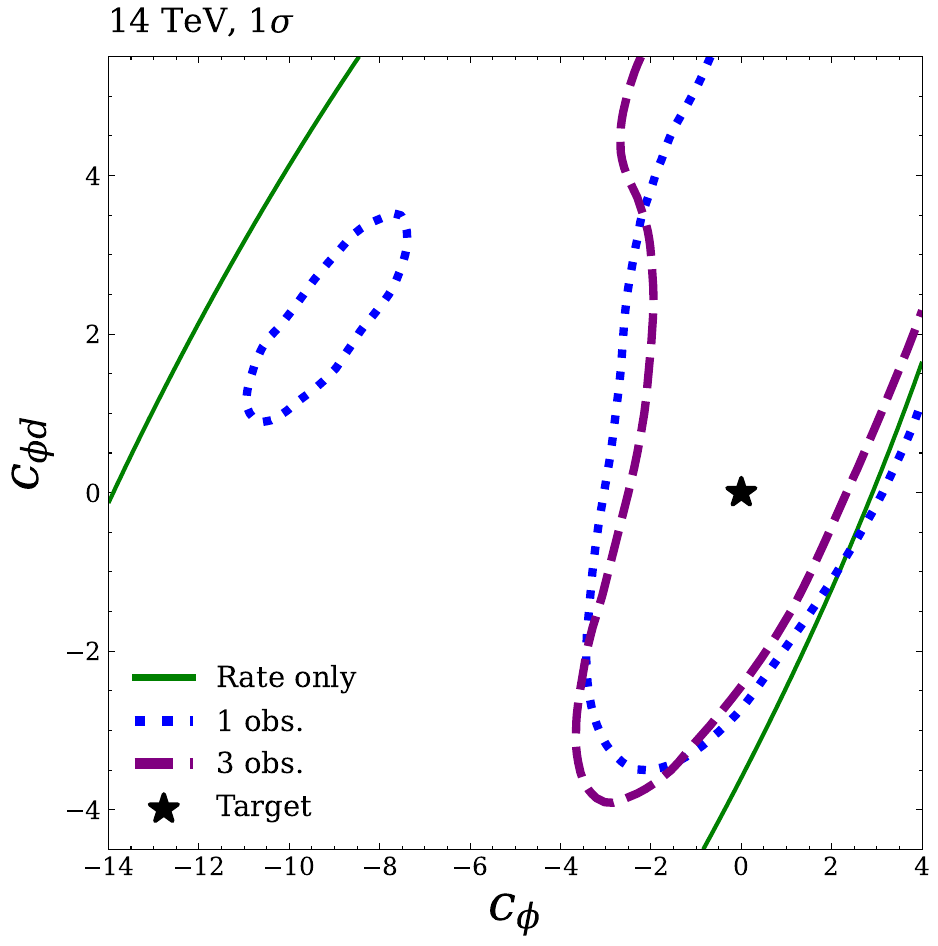}
\includegraphics[width=0.32\textwidth]{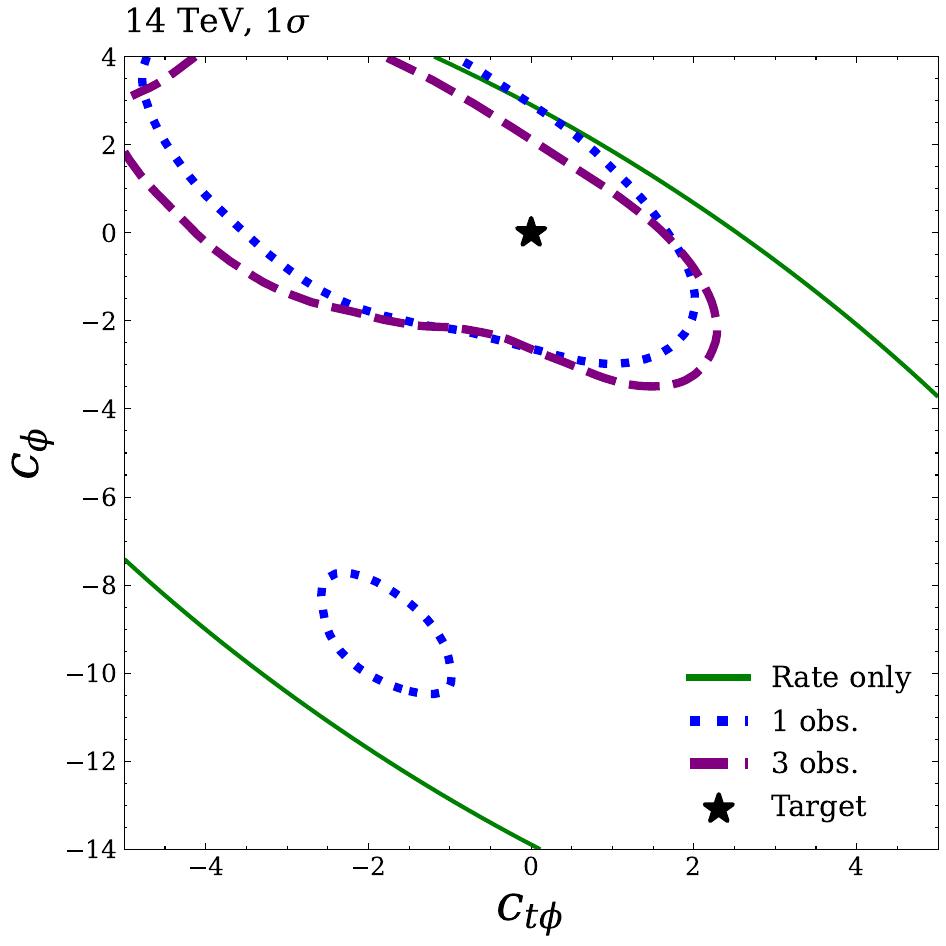}
\includegraphics[width=0.32\textwidth]{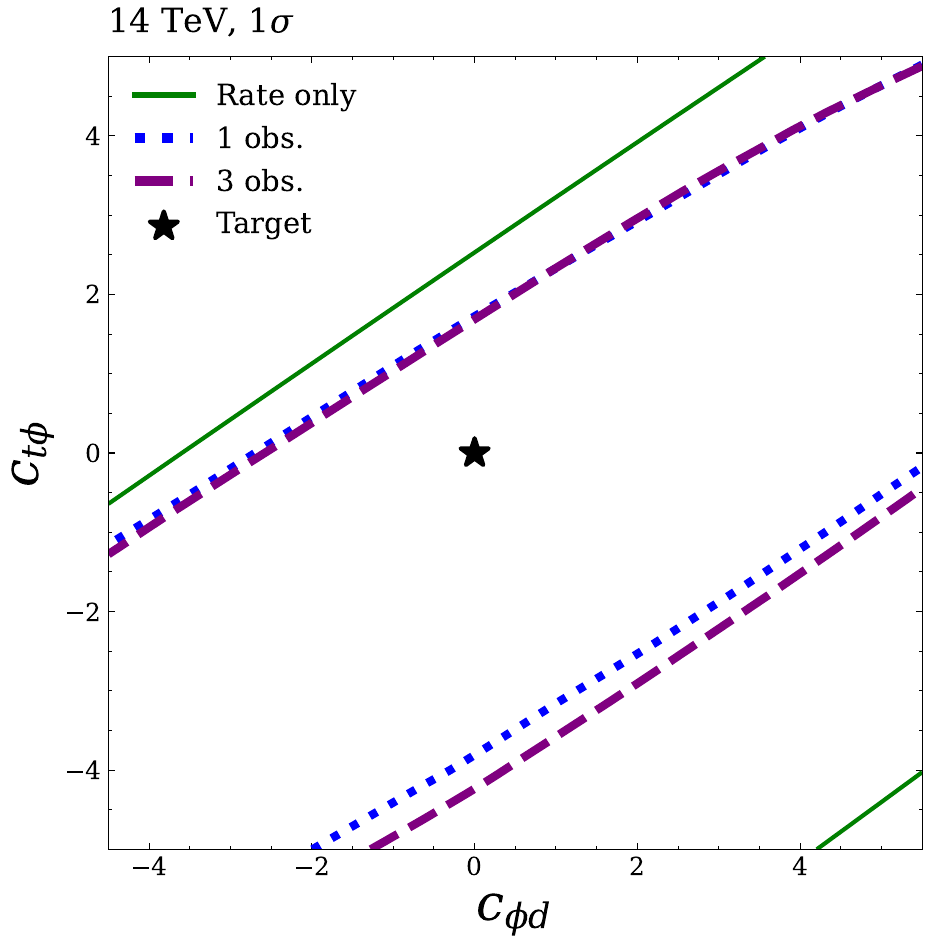} \\
\includegraphics[width=0.32\textwidth]{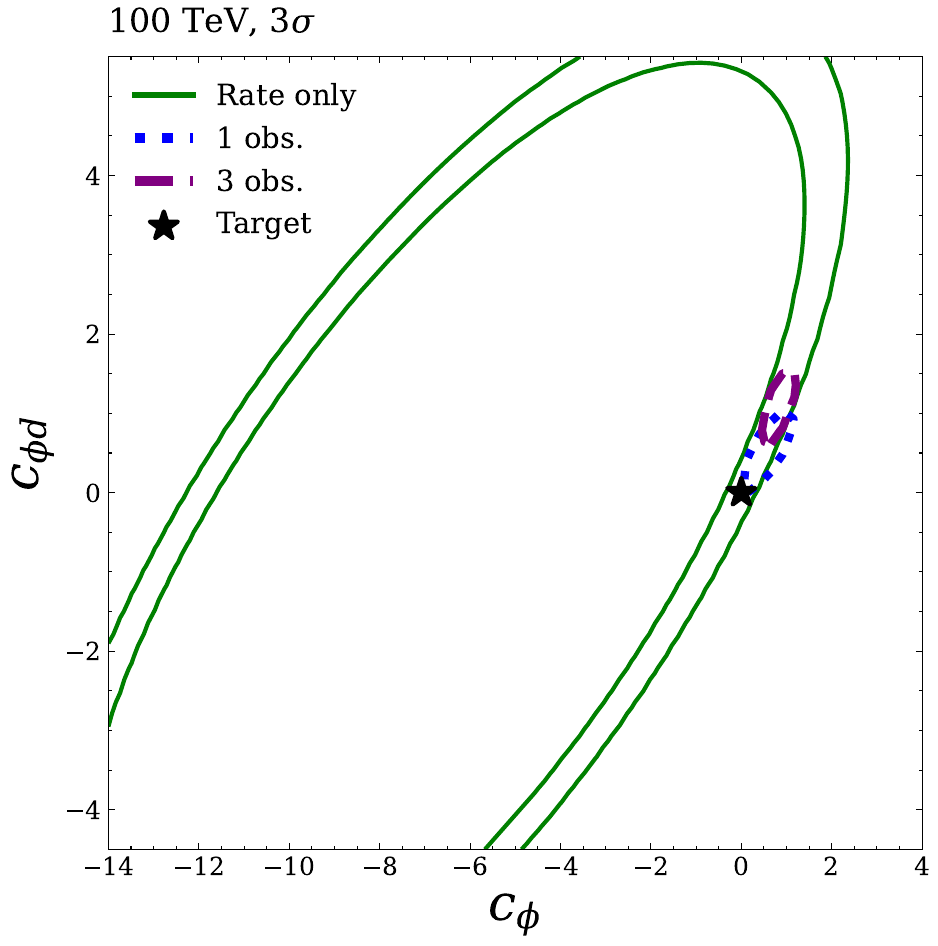} 
\includegraphics[width=0.32\textwidth]{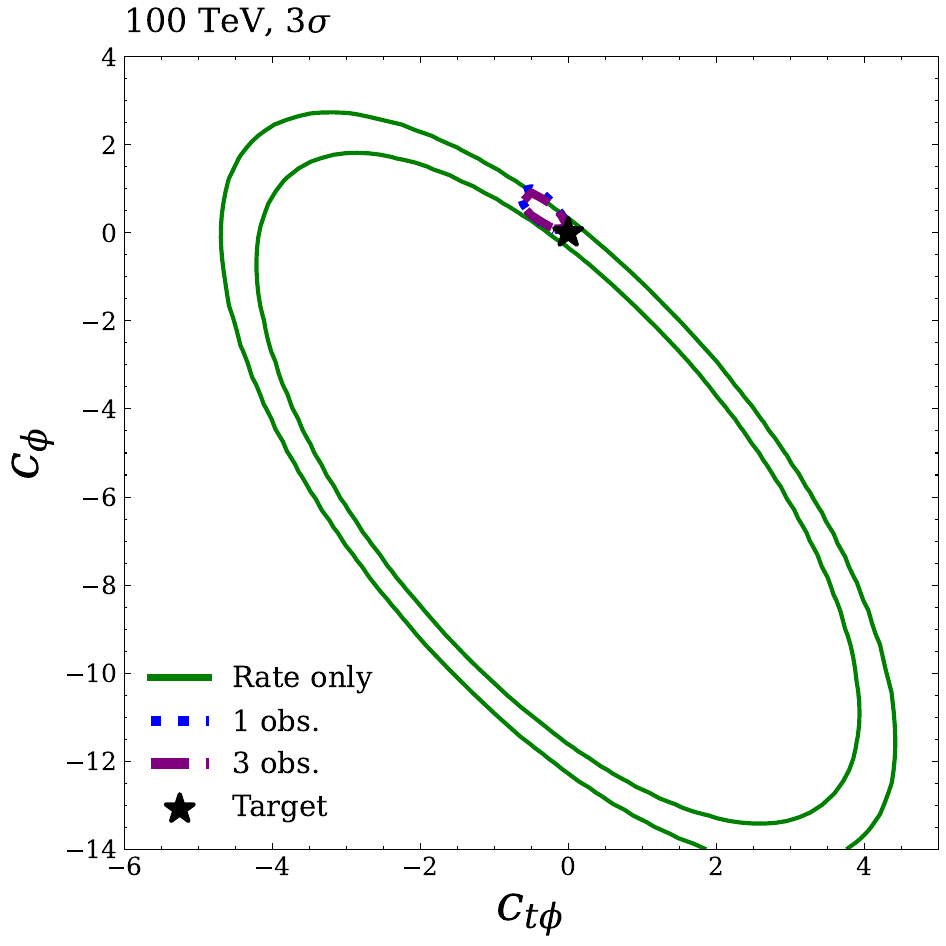} 
\includegraphics[width=0.32\textwidth]{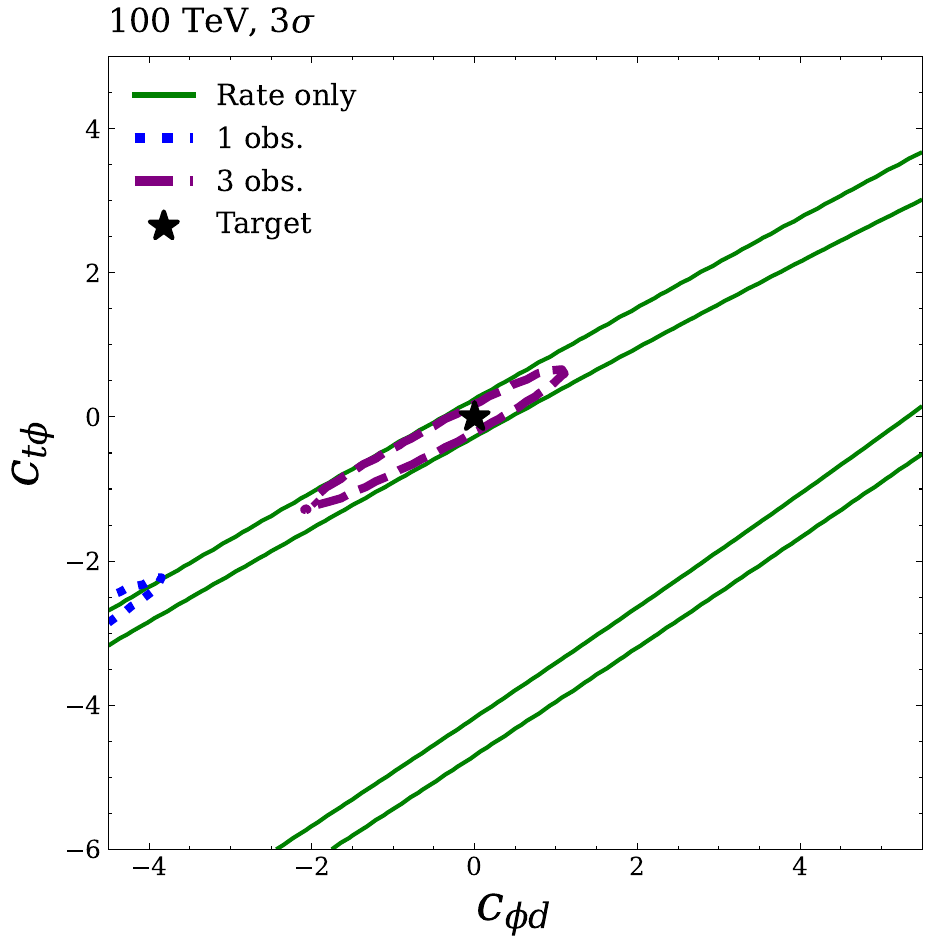}
\caption{Log-likelihood ratio test statistics in terms of 1$\sigma$ (for HL-LHC) or 
3$\sigma$ (for 100 TeV) confidence contours for two non-zero Wilson coefficients
at a time. 
Data generation and test set size reflect the collider setup.}
\label{fig:q_results_2d}
\end{figure*}

For classifier training, we generate sets of $10^7$ events each for the SM $hh$ signal, the BSM $hh$ signal, and the continuum background. For the BSM sample, we generate 
the events by uniformly sampling $1000$ values of $c$, which means $p_0$ in 
Sec.~\ref{sec:shape_analysis} is uniform. For the 1D coefficient tests which will be
shown in \Fig{fig:q_results_1d}, we only allow the single scanned Wilson coefficient 
to be non-zero; for the 2D coefficient tests, which will be shown in 
\Fig{fig:q_results_2d}, only the two scanned Wilson coefficients are non-zero. 
The non-zero coefficients cover the prior ranges 
$\cp \in (-14, +4)$, $\cdp \in (-4, +5)$, and $\ctp \in (-5, +7)$. For the test sets, 
we generate sets of events following the expected event yields 
from \Tab{tab:event_yields}.\medskip

All classifiers are parameterized as relatively small, dense neural networks consisting 
of 2 layers with 32 hidden nodes. We use a batch size of 1024, a weight decay 
$10^{-4}$, and an initial learning rate of $10^{-3}$. The learning rate reduces 
by half if the validation loss does not decrease for 5 epochs. We train for up to 1600 
epochs, stopping when the validation loss does not decrease for 20 epochs and evaluating 
the networks at the epoch of lowest validation loss. In practice, the classifiers trained on data for the 100~TeV collider often converged in fewer than 200 epochs. We use an 80:20 
training-validation split. All networks are implemented in 
\textsc{PyTorch}~\cite{NEURIPS2019_9015} and optimized with 
\textsc{Adam}~\cite{kingma2017adam}. All hyperparameters are optimized by 
manual tuning on a simplified version of the problem. This simplified problem refers to carrying out the 1D coefficient tests on pre-\textsc{Delphes} samples in the zero-background case (i.e. we just train the classifier that discriminates SM $hh$ signal from BSM $hh$ signal). Performance was fairly robust with respect to classifier architecture and training hyperparamters, although we did find that a longer early stopping parameter produced better results.


To mitigate the stochastic nature of the network training, we ensemble 
the outputs of five networks with identical architectures and different 
initial random number generator seeds. 

\section{Results}
\label{sec:results}

\subsection{Coefficient recovery for the SM}
\label{sec:results_SM}

We first assume that the observed data coresponds to the SM hypothesis.  
Using the pre-trained network from the previous section, we compute the log-likelihood 
ratio for a reference dataset
with all Wilson coefficients $\cp$, $\cdp$, and $\ctp$ set to zero. 

A 1D-scan over one 
Wilson coefficient at a time is shown in \Fig{fig:q_results_1d} for both, the HL-LHC 
and the 100~TeV collider setups.  The resulting central parameter value is given by 
the maximum log-likelihood, and the confidence intervals are determined based on the shape 
around the minimum. 
These 
uncertainties are indicated in \Fig{fig:q_results_1d} by vertical bands showing
1$\sigma$ confidence intervals for the HL-LHC (so the $y$-axis decreases by 
roughly one unit from the minimum),  and 3$\sigma$ for 100~TeV.

We consider three analysis methods: one rate-only analysis, one that 
incorporates shape information from the $m_{hh}$ only, and one that incorporates 
shape information from the $m_{hh}$, $p_{T_{bb}}$, and $p_{T_{\gamma \gamma}}$ 
kinematic distributions. We provide equivalent plots for the 1-dimensional test statistics in the Appendix for all five observables in \Fig{fig:q_results_1d_f5}, also including 
$\Delta R_{bb}$ and $\Delta R_{\gamma\gamma}$.

Starting with single Wilson coefficients in Fig.~\ref{fig:q_results_1d},
all likelihood minima are consistent with zero within the reported uncertainty,
and adding more kinematic information generally sharpens the peaks and leads to 
smaller uncertainty around the SM-minimum.
Away from the SM-minimum, the additional but incomplete kinematic information
can lead to features in the likelihood ratio dependence. 
Both effects are especially prominent for the classic trilinear 
Higgs coupling $\cp$ -- in the rate-only analysis, there remains a degeneracy 
in the test statistic at $\cp = 0$ and $\cp = -11$, which is resolved by 
incorporating shape information. As a matter of fact, the rate-only analysis 
leaves a degeneracy for all three Wilson coefficients, but this degeneracy 
is effectively resolved by including single-Higgs production.\medskip

The fundamental assumption of effective field theories is that any high-energy 
model will induce all Wilson coefficients compatible with its symmetry. This 
means any LHC signal will be affected by more than one operator, and the 
correlations between contributions from different operators will reflect the 
underlying theory. This motivates the variation of 
two Wilson coefficients at a time. The corresponding log-likelihoods for the rate-only, 1-observable, and 3-observable test statistics are shown 
in~\Fig{fig:q_results_2d}. Equivalent plots for the 5-observable test statistic are shown in the Appendix in \Fig{fig:q_results_2d_f5}.  We overlay the 1$\sigma$ and 3$\sigma$ confidence
regions for the HL-LHC and 100 TeV colliders, respectively. For the rate-only analysis, the  double peak distribution from Fig.~\ref{fig:q_results_1d} now becomes an ellipse or annulus.  For the statistically stable 100~TeV setup, adding kinematic information can indeed break the degeneracy and improve the confidence contours to small regions on the correct side of the rate-only ellipse.  While we show confidence contours for only a single realization of nature, the fact that the SM is not always contained within the confidence region for the 100 TeV machine is representative.  With the high statistics of the 100 TeV machine, the precision required on the likelihood ratio estimation is much stricter than for 14 TeV.  The confidence regions are qualitatively in the correct location, but achieving quantitatively precise results will require additional research (see Sec.~\ref{sec:shape_analysis} for some progress in other studies). 

Going beyond two dimensions, it is difficult to visualize the full space.  Since the neural networks are differentiable, it is possible to find the maximum likelihood estimate using gradient descent and the Hessian can provide an estimate of the confidence interval.  This is left to future studies to explore in more detail.

\begin{figure*}
\includegraphics[width=0.37\textwidth]{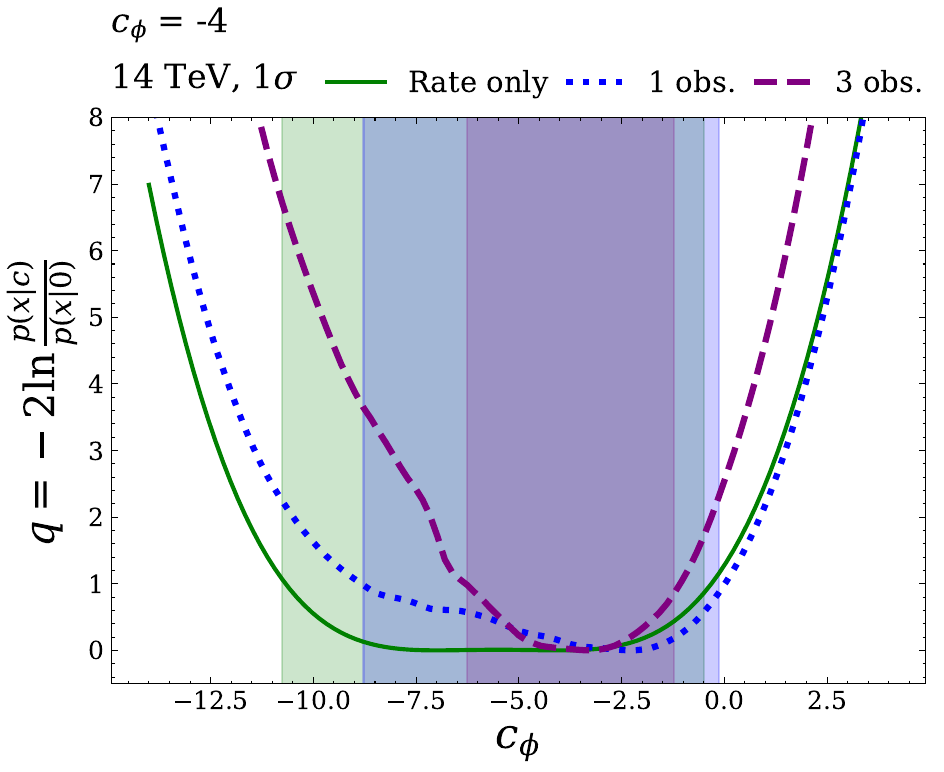} \hspace{2cm}
\includegraphics[width=0.37\textwidth]{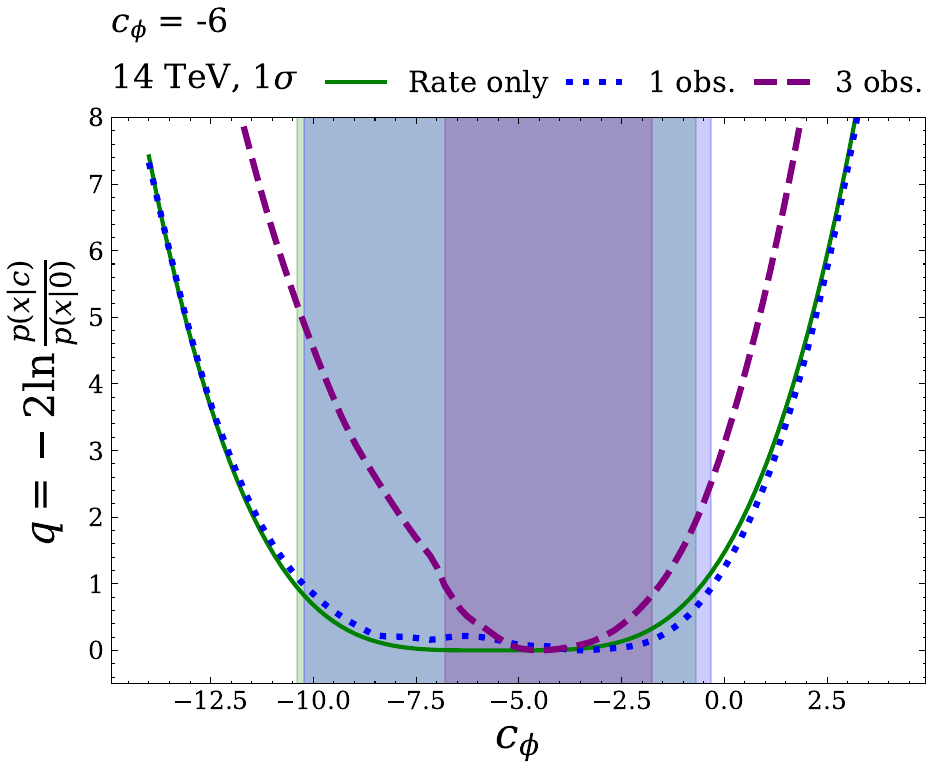}
\caption{Log-likelihood ratio test statistics in terms of 1$\sigma$ for the HL-LHC, for
one Wilson coefficient $\cp$ at a time. The central value of $\cp$ is chosen away from the 
SM.
Data generation and test set size reflect the collider setup.}
\label{fig:central_bsm}
\end{figure*}

\subsection{Coefficient recovery for BSM scenarios}

In order to explore the sensitivity away from the SM value of $c=0$, we also show the performance of recovering a 
non-zero $c$-vector. 
As a comprehensive scan of the accuracy and precision of the 
various approaches would be computationally demanding and difficult to visualize, 
we pick representative examples to study in detail.

In Fig.~\ref{fig:central_bsm}, we show 
the expected HL-LHC limits for the single Wilson coefficient
$\cp$ for assumed true values away from the SM. The 
almost-perfect cancellation of the triangle and box diagrams is
only true for a SM self coupling. This means that while
the $hh$ rate will increase for these points, the 
sensitivity will not be enhanced as much through the cancellation. 
Altogether, we find that the $1\sigma$ range for the 3-observable
analysis increases to 
\begin{alignat}{7}
\cp &= 0 \quad \text{assumed:} &\qquad \cp &\in [ -1.8,+1.7] \notag \\
\cp &= -4 \quad \text{assumed:} &\qquad \cp &\in [ -6.3,-1.2] \notag \\
\cp &= -6 \quad \text{assumed:} &\qquad \cp &\in [ -6.8,-1.8] \; .
\end{alignat}
In the HEFT basis assuming an EFT cutoff of 1 TeV and setting all other Wilson coefficients to zero, this is equivalent to
\begin{alignat}{7}
\kappa_\lambda &= +1 \quad \text{assumed:} &\qquad \kappa_\lambda &\in [ +0.2,+1.8] \notag \\
\kappa_\lambda &= +2.9 \quad \text{assumed:} &\qquad \kappa_\lambda &\in [ +1.6,+4.0] \notag \\
\kappa_\lambda &= +3.8 \quad \text{assumed:} &\qquad \kappa_\lambda &\in [ +1.8,+4.2] \; .
\end{alignat}
For both choices, the rate-only measurement would not be able to 
distinguish these parameter points from the SM.\medskip 

\begin{figure*}
\includegraphics[width=0.9\linewidth]{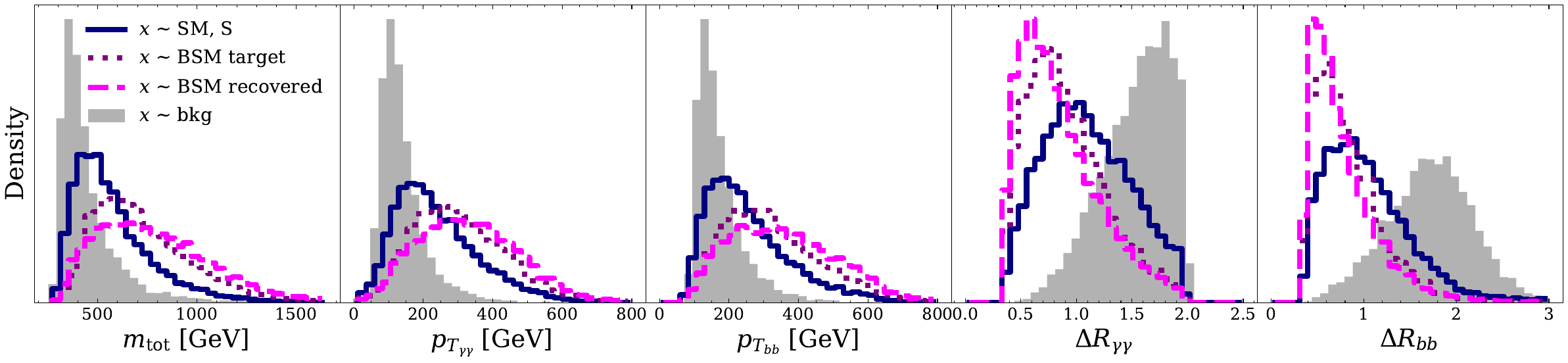}
\caption{Kinematic distributions from $b\bar{b}\gamma\gamma$ events from the 100 TeV collider setup. A test set generated with the Wilson coefficients $(\cp, \cdp, \ctp) = (-4, 0, 3)$, `BSM target', is close-to-degenerate in classifier output to a test set generated with $(0, 0, -5.5)$ `BSM recovered'. The vertical axis is linearly scaled.}
\label{fig:failure_features}
\end{figure*}

\begin{figure*}
\includegraphics[height=5.5cm]{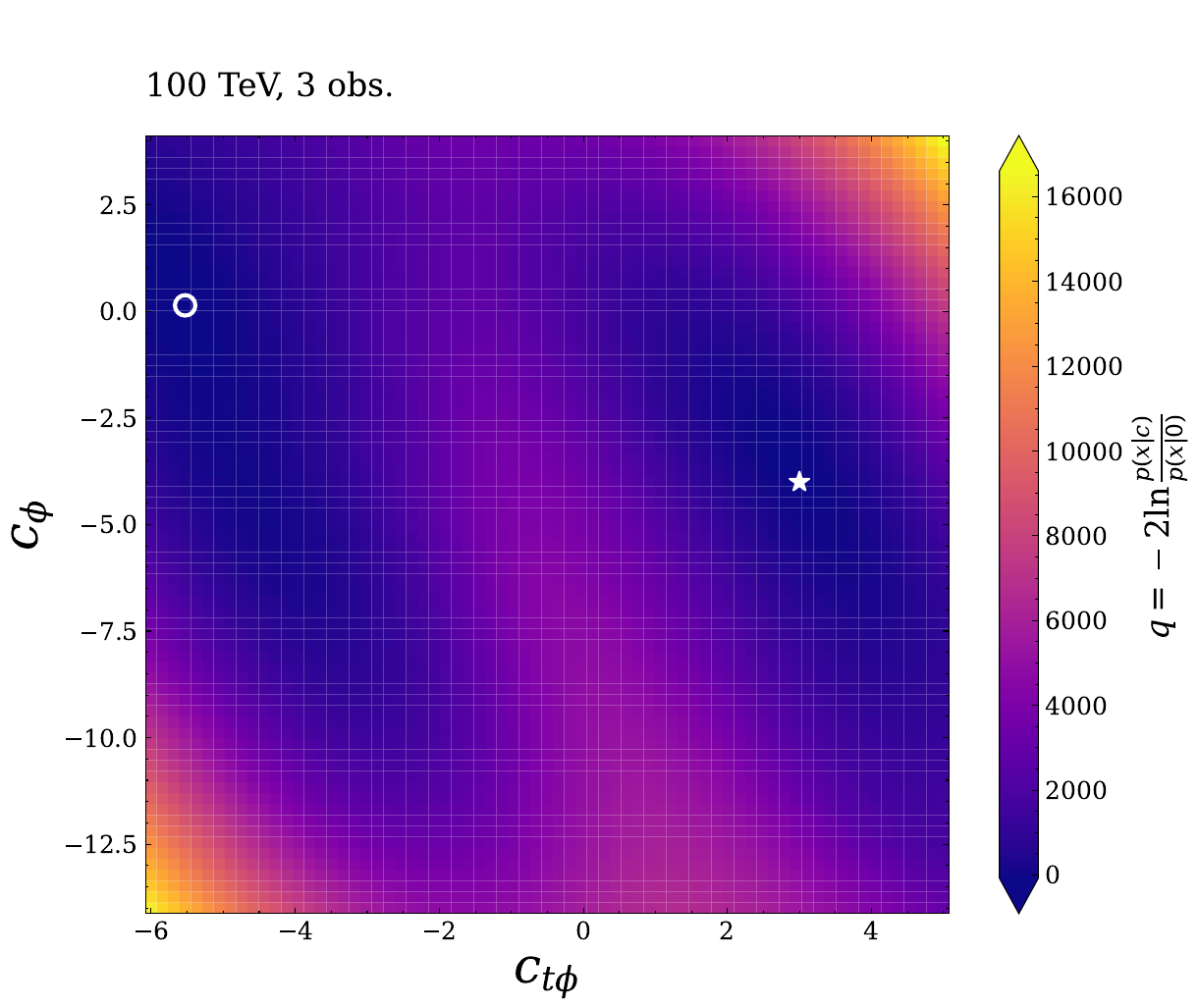}\hspace{1.5cm}
\includegraphics[height=5.1cm]{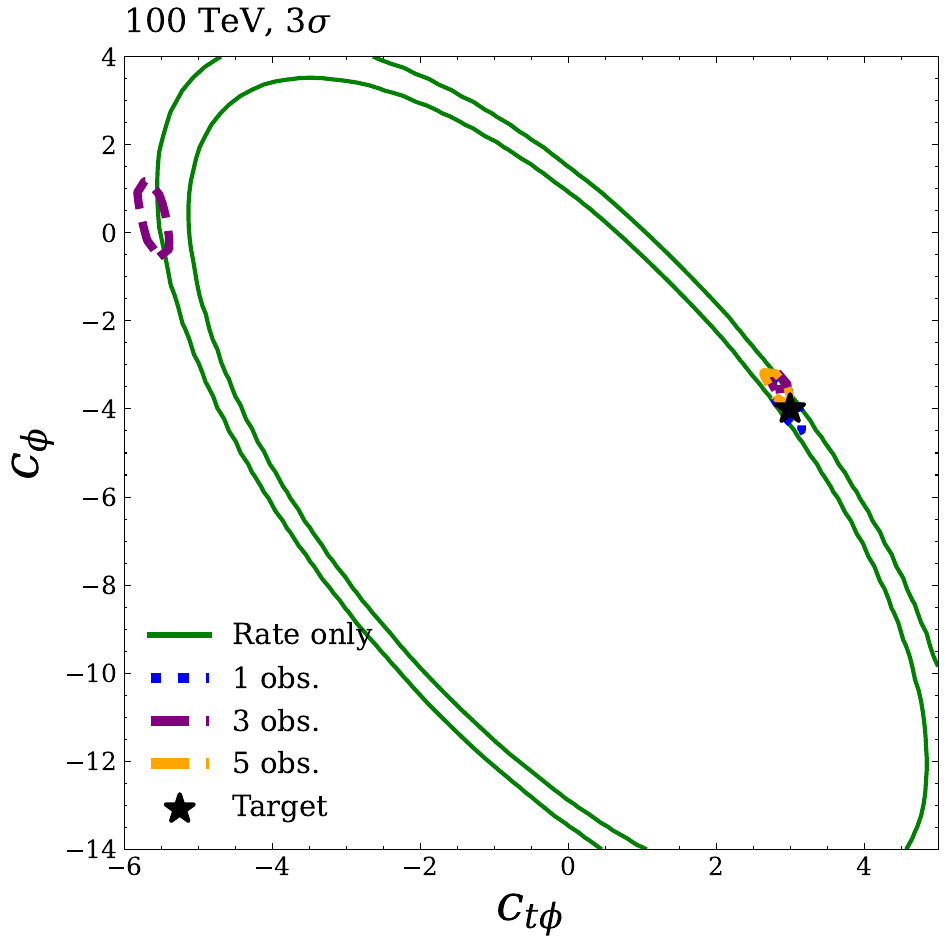}
\caption{Left: 2D test statistic for the 3-observable test. The generating Wilson coefficient vector $c = (-4, 0, 3)$ is denoted by a star; the test statistic is optimized at $c \approx (0, 0, -5.5)$, which is denoted by an open circle. Right: 3$\sigma$ confidence intervals for a variety of test statistics.}
\label{fig:q_results_poor1_100TeV}
\end{figure*}

Second, we can scan the likelihood landscape for a given 
set of events, for instance at the 100~TeV collider, and identify 
parameter points far away in model space but close in likelihood.
An example are the vectors of Wilson coefficients
\begin{align}
 (\cp, \cdp, \ctp)& =(0, 0, -5.5),  \notag \\
 (\cp, \cdp, \ctp)& =(-4, 0, 3), \; 
\label{eq:def_bsm}
\end{align}
which are indistinguishable from each other in a rate-only analysis and very 
similar in terms of our observables. In Fig.~\ref{fig:failure_features},
we see that the observables are clearly distinguishable from the SM, 
given the high assumed statistics of the 100~TeV collider, but not from each other. This similarity is reflected in a shape analysis, seen in \Fig{fig:q_results_poor1_100TeV}. Given a test set with the generating Wilson coefficient vector $c=(-4,0,3)$, the 3-observable test statistic is doubly-minimized, showing high likelihood for both the true underlying vector $c$ and the similarly-shaped  $ c = (0, 0, -5.5)$. The kinematic observables $m_\text{tot}$, $p_{T_{\gamma\gamma}}$, and $p_{T_{bb}}$ are indeed very similar to each other, especially when compared to the SM distributions; this similarity leads to the 
likelihood ratio degeneracy, which is somewhat broken when considering $\Delta R_{\gamma\gamma}$, which peaks in a distinct location for the true $c$ and recovered $c$ vectors. In fact, the 5-observable classifier is able to recover values of $c$ that are much closer to truth.

\subsection{Statistical properties and coverage}
\label{sec:stat}

\begin{figure*}
\includegraphics[width=0.32\textwidth]{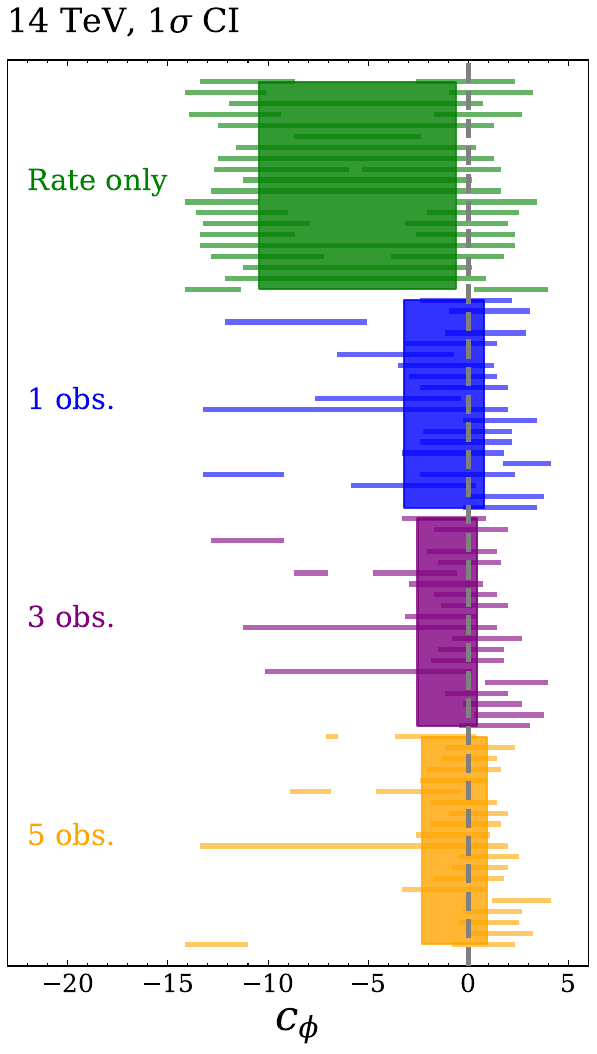}
\includegraphics[width=0.32\textwidth]{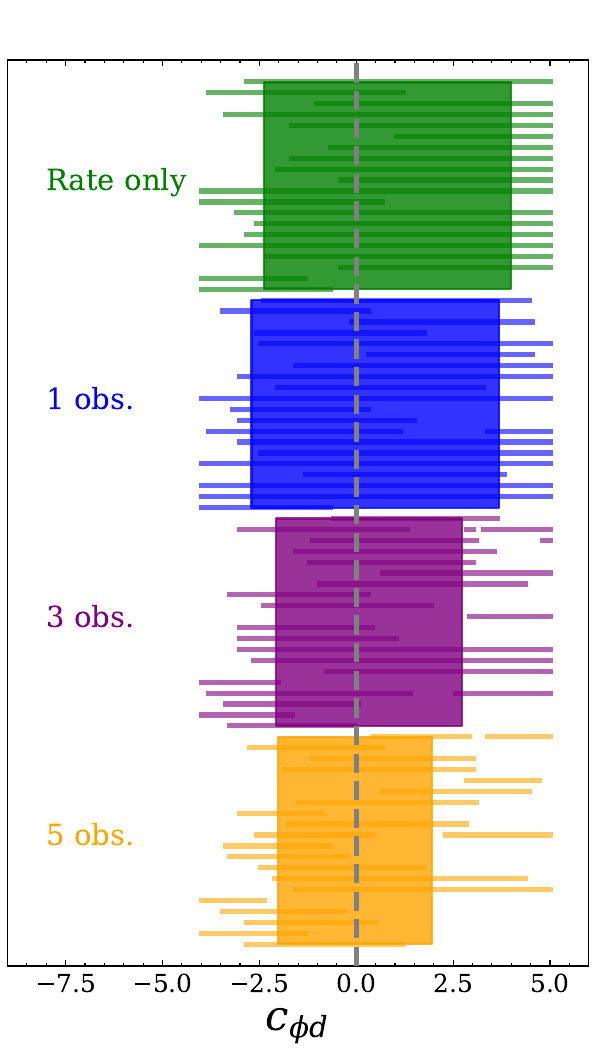}
\includegraphics[width=0.32\textwidth]{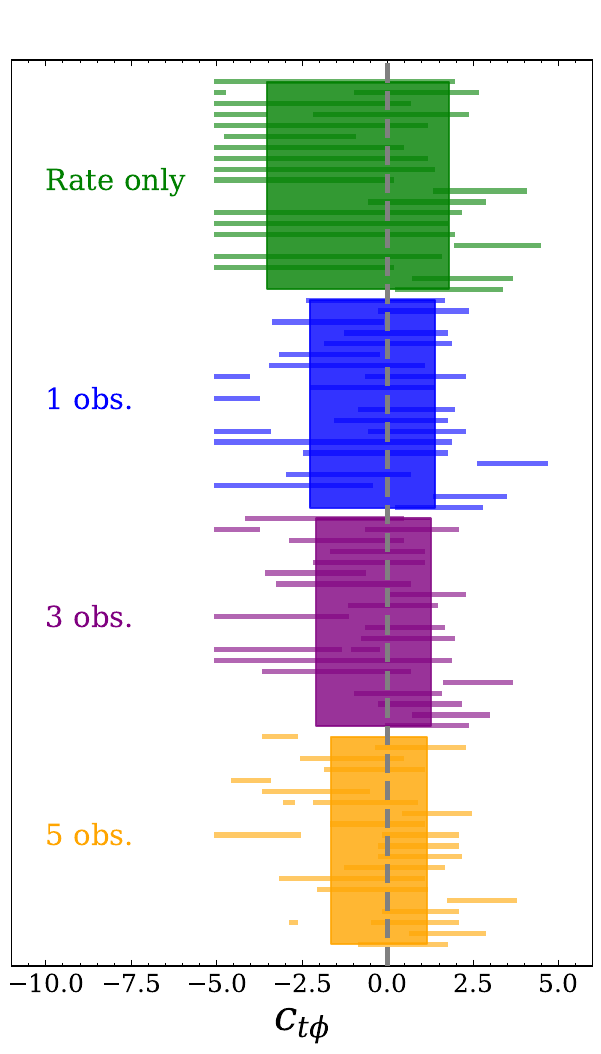}
\caption{Confidence intervals for 
log-likelihood ratio associated with a given Wilson coefficient. Data generation corresponds to the HL-LHC. Each horizontal band corresponds to a different 
event set. The shaded band shows the mean confidence interval for each test.}
\label{fig:coverage_results_14TeV}
\end{figure*}

In \Fig{fig:q_results_2d}, we see that for a single 
realization of nature, the confidence contours are not always ellipses and 
the SM is not always contained within the 
confidence region. This first of these is addressed with the higher statistics of the 100-TeV machine while the latter issue is more acute in this setting.
The problem is that for the HL-LHC, there are 
around 100 total events expected, with around 10 of them from the $hh$ signal.

This statistics limitation leads us to 
explore the accuracy and precision of the various 
parameter estimation approaches by computing confidence intervals for many synthetic 
datasets. Analyzing different datasets is of particular interest for the 
HL-LHC, where the small event yield may result in confidence intervals that
change significantly between different test event sets. 

For ease of visualization, we focus on the 1D confidence intervals testing one 
non-zero Wilson coefficient at a time. In \Fig{fig:coverage_results_14TeV}, we 
show expected 1$\sigma$ confidence intervals for each of our Wilson coefficients for
the HL-LHC. We use 20 different test event sets corresponding to the 
SM hypothesis, but with different signal and background events. 
For each test set, we also allow the sizes of the signal and background sets to vary, following Poisson distributions with means given by the last row of \Tab{tab:event_yields}.

The shaded bars in \Fig{fig:coverage_results_14TeV} denote the mean coverage interval across the test sets, 
for each of the analysis techniques --  rate-only, 1-observable, and 3-observable.  
We also include results from a 5-observable analysis
strategy, adding $\Delta R_{bb}$ and $\Delta R_{\gamma\gamma}$.
\medskip

Looking at the $\cp$ recovery at the HL-LHC, we see that for the rate-only 
analysis, $(90 \pm 7)$\% of the confidence intervals contain the correct value. 
For the 1-observable analysis, 
$(80 \pm 9)$\% contain the correct value; for the 3-observable analysis, $(80 \pm 9)$\%; 
and for the 5-observable analysis, $(85\pm 8)$\%. For standard 1$\sigma$ confidence intervals, 
we would expect around $68\%$ of the intervals to give  SM recovery. However, we almost always find that more than this percentage of the confidence intervals contain the SM, so our bounds are conservative. Given that for the rate-only analysis, the likelihood ratio is exact (i.e. not estimated with neural networks), this may be due to the non-Gaussian nature of the test statistic.

A similar recovery is also achieved for the other two Wilson coefficients: 
for $\cdp$, the confidence intervals for the rate-only, 1-observable, 3-observable, 
and 5-observable analyses return the SM value $(85 \pm 8)$\%, $(90 \pm 7)$\%,
$(80 \pm 9)$\%, and $(55 \pm 11)$\% of the time. For $\ctp$, the same numbers are
$(75 \pm 10)$\%, $(65 \pm 11)$\%,  $(70\pm 10)$\%, and $(70 \pm 10)$\% of the time.

A corresponding analysis can be carried out for the 100 TeV setup, where we evaluate 1D confidence intervals for 20 independent test sets. For $\cp$ recovery, all 3$\sigma$ confidence intervals for all test statistics contain the SM value; for $\cdp$, the 1-observable test statistic achieves SM recovery $(95\pm 5)$\% of the time and the other test statistics 100\% of the time; for $\ctp$, the rate-only test statistic returns the SM value 100\%, the 1-observable $(95\pm 5)$\%, and the 3- and 5-observable $(90\pm 7)$\% of the time.

We may further contrast the test statistic 
types. The mean confidence interval is narrower when derived from the 3-observable 
test statistic than when derived from the 1-observable test statistic, and the mean 
1-observable test statistic is narrower than that of the rate-only analysis. Further, the test statistics that make use of kinematic observable information more often resolve the likelihood degeneracy seen in the $\cp$ recovery that is left ambiguous for the rate-only test statistic. This is 
consistent with our earlier findings that including shape information for kinematic 
distributions can place tighter constraints on Wilson coefficient bounds than rate-only 
analyses can. In addition to the known $m_\text{tot}$, the set of 3 observables 
are informative and relevant for the coefficients $\cp$, $\cdp$, and $\ctp$. 

Finally, it is worth noting that for the HL-LHC, some test event sets
do not resolve the degeneracy for the $\cp$-coefficient, or resolve it incorrectly 
by choosing a large negative value for this coefficient. The degeneracy is always 
correctly resolved, to 3$\sigma$, for the 100 TeV collider. 

\section{Conclusions}
\label{sec:conclusions}

In this work, we have explored the use of neural simulation-based inference (nSBI) to enhance the sensitivity to searches for pair production of Higgs bosons.  As our example, we have simulated an analysis to place constraints on the SMEFT Wilson coefficients for a set of three dimension-6 operators associated with $hh$ production: $\cp$, $\cdp$, and $\ctp$.  We have considered two collider setups in this report: a HL-HLC-like setup with $\sqrt{s}$ = 14 TeV and 3ab$^{-1}$ of integrated luminosity, and a future hadron collider setup with $\sqrt{s}$ = 100 TeV to 30ab$^{-1}$. We have shown that through parameterized machine learning tools, we can augment more ``standard" cut-and-count analyses with per-event shape information to increase constraining power for these $hh$-relevant Wilson coefficients. 

In the idealized context of our study, we encountered a number of challenges that need to be addressed before these methods can be used in practice.  Most importantly, it is difficult to achieve the level of precision required to produce accurate and precise confidence regions near the global minimum of the likelihood landscape.  We have utilized a number of techniques to address this, such as factorizing the classifiers and using ensembling.  It would be interesting to explore additional proposals for improving the likelihood-ratio estimation and we hypothesize that additional methods are required, especially for the level of precision that will be afforded by future high-luminosity collider data. In addition, we assumed that simulation will be used to estimate the background.  It may be that this will be possible in the HL-LHC era, but the current state-of-the-art is data-driven background estimates.  It may be possible to combine such approaches with nSBI, which would be interesting to explore in the future.  Finally, we assumed that the signal and background are known with no systematic uncertainty.  In practice, such uncertainties can be directly folded into the analysis protocol, although profiling over a large number of nuisance parameters may be challenging.  

Going beyond the analysis presented here, it would also be interesting to explore how far we could push the dimensionality of the observable space and the parameter space.  There may also be gains possible from a dedicated study of the trade-offs between making restrictive selections and using per-event information for more events.  In particular, we could relax the preselection to reduce the starting significance of the signal, but then recover (and ideally, exceed) the sensitivity through the per-event likelihood estimation.  This approach will be limited in part by the ability of the neural networks to describe very low likelihood events.  While we have focused on $hh$ events, the tools and challenges are common to many nSBI analyses, and our study provides another important benchmark for refining and developing new methods.

\section*{Data and code availability}

All data used in this report is available on Zenodo at \url{https://zenodo.org/records/11222924}. The analysis code is available at \url{https://github.com/rmastand/nsbi_for_dihiggs}.

\section*{Acknowledgments}

BN and RM are supported by the U.S. Department of Energy (DOE), Office of Science under contract DE-AC02-05CH11231 and Grant No. 63038 from the John Templeton Foundation.  RM is additionally supported by Grant No. DGE 2146752 from the National Science Foundation Graduate Research Fellowship Program.
This research used resources of the National Energy Research Scientific Computing Center, a DOE Office of Science User Facility supported by the Office of Science of the U.S. Department of Energy under Contract No. DE-AC02-05CH11231 using NERSC award HEP-ERCAP0021099.  

TP is supported by the Baden-W\"urttemberg-Stiftung 
through the program \textit{Internationale
  Spitzenforschung}, project \textsl{Uncertainties --- Teaching AI its
  Limits} (BWST\_IF2020-010), the Deutsche Forschungsgemeinschaft (DFG, German Research
Foundation) under grant 396021762 -- TRR~257 \textsl{Particle Physics
  Phenomenology after the Higgs Discovery}, and through Germany's
Excellence Strategy EXC~2181/1 -- 390900948 (the \textsl{Heidelberg
  STRUCTURES Excellence Cluster}).  
\appendix

\section{Additional plots}

\begin{figure*}
\includegraphics[width=0.45\textwidth]{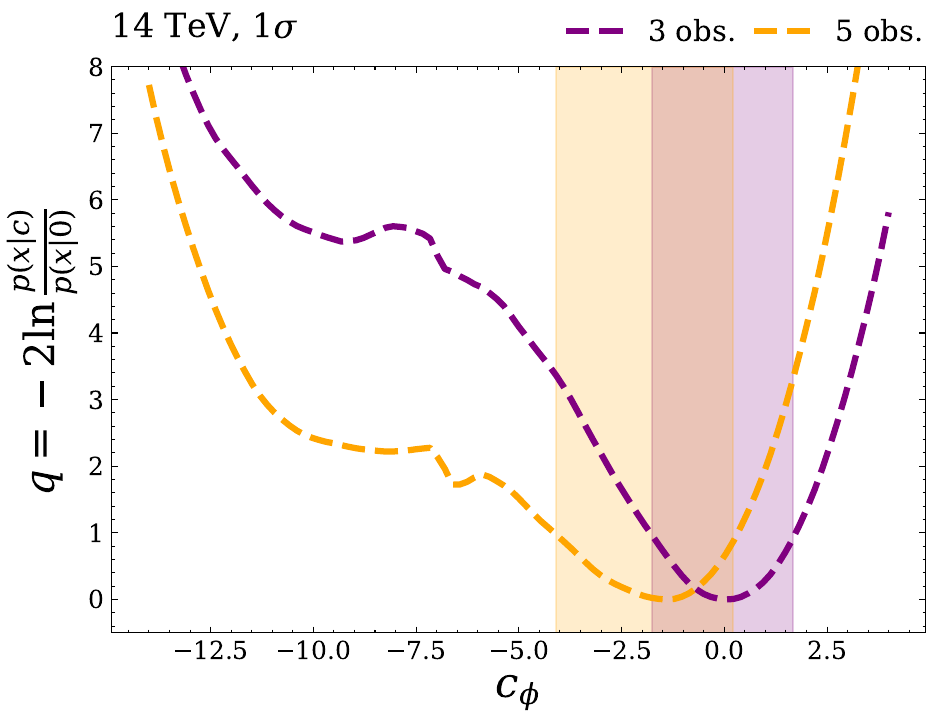} \hfill
\includegraphics[width=0.45\textwidth]{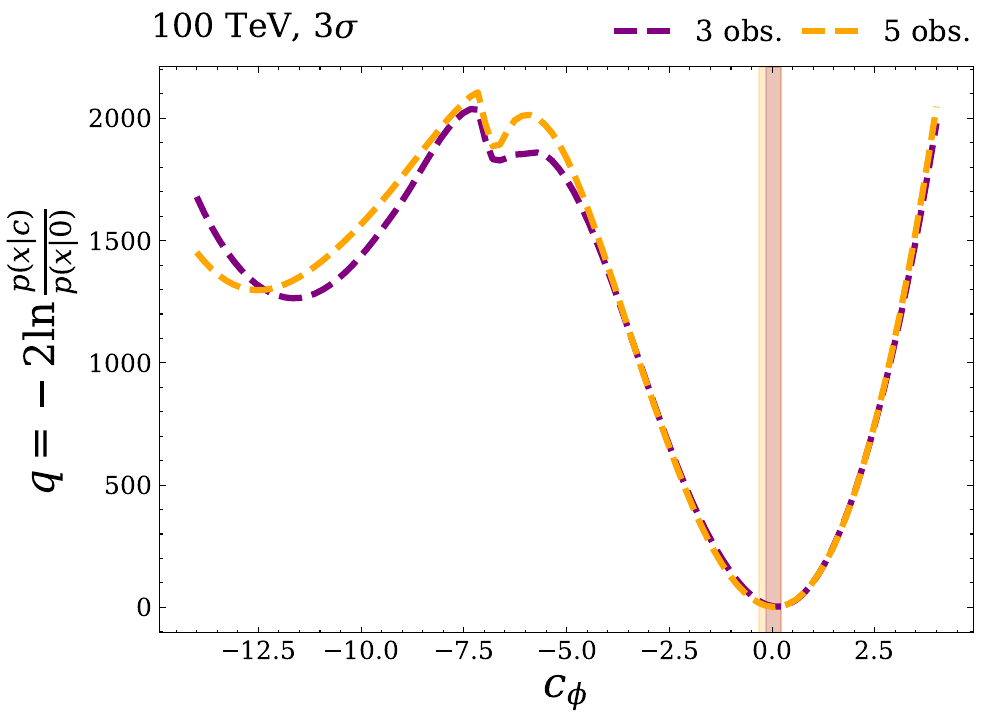} \\
\includegraphics[width=0.45\textwidth]{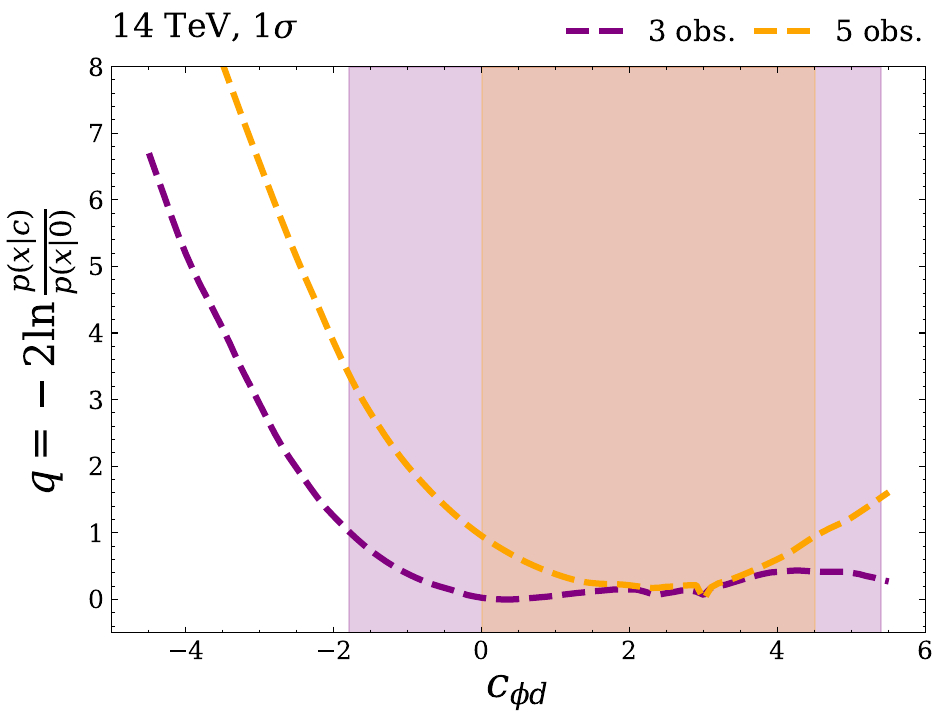} \hfill
\includegraphics[width=0.45\textwidth]{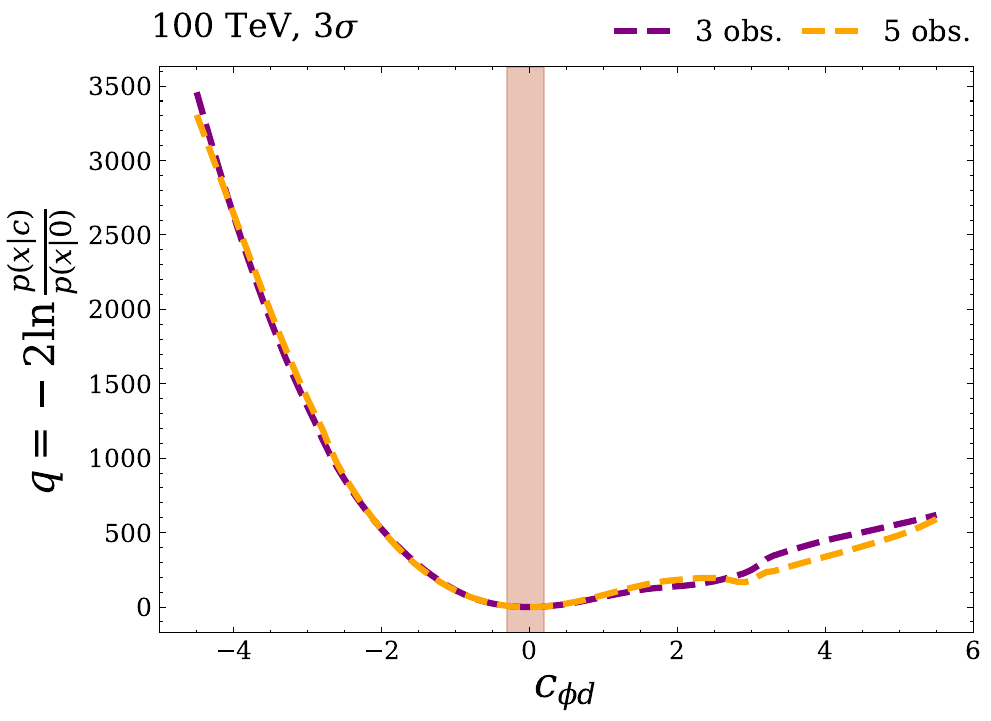} \\
\includegraphics[width=0.45\textwidth]{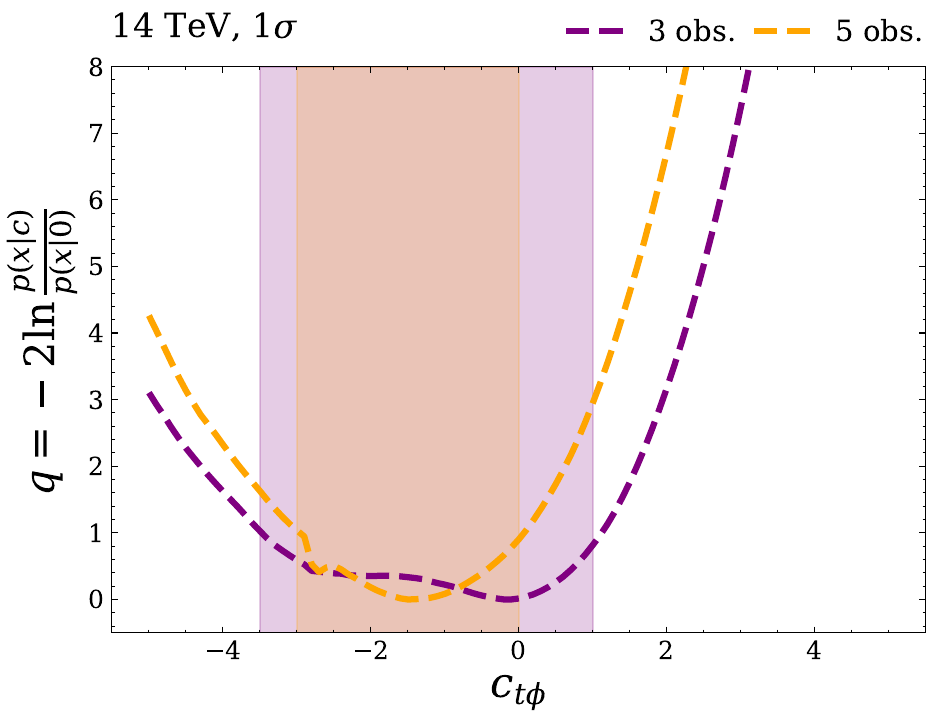} \hfill
\includegraphics[width=0.45\textwidth]{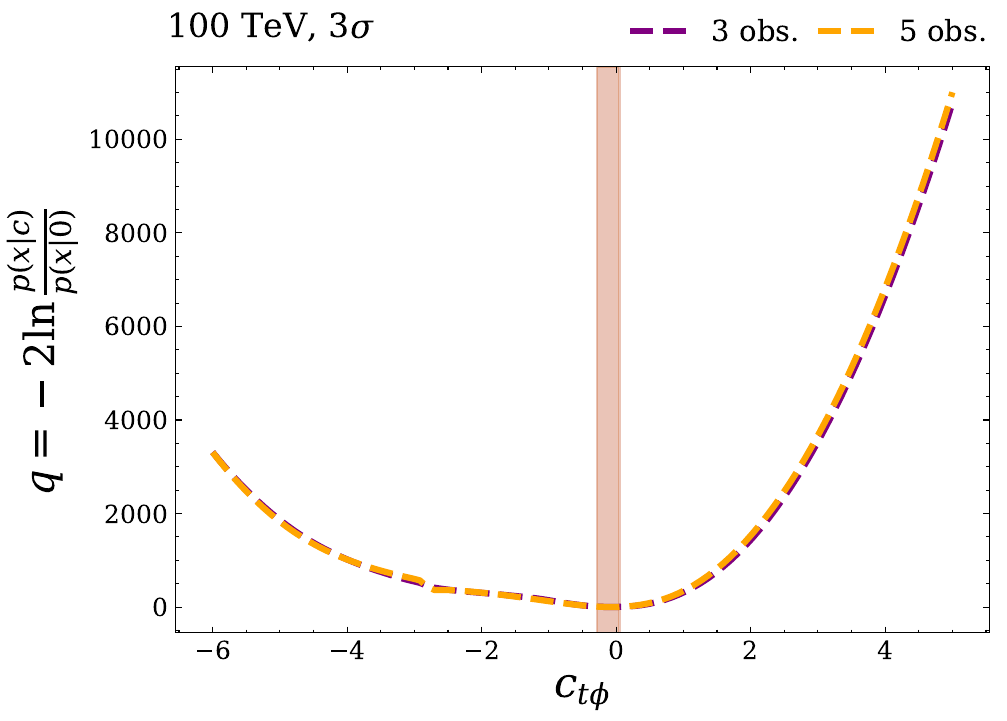}
\caption{Test statistics and 1$\sigma$ (for HL-LHC) or 3$\sigma$ (for 100 TeV) confidence intervals for the test statistic $q=-2\ln\frac{p(x|c)}{p(x|0)}$ associated with the given Wilson coefficient. Data generation and test set size reflects the labeled collider setup.}
\label{fig:q_results_1d_f5}
\end{figure*}

\begin{figure*}
\includegraphics[width=0.32\textwidth]{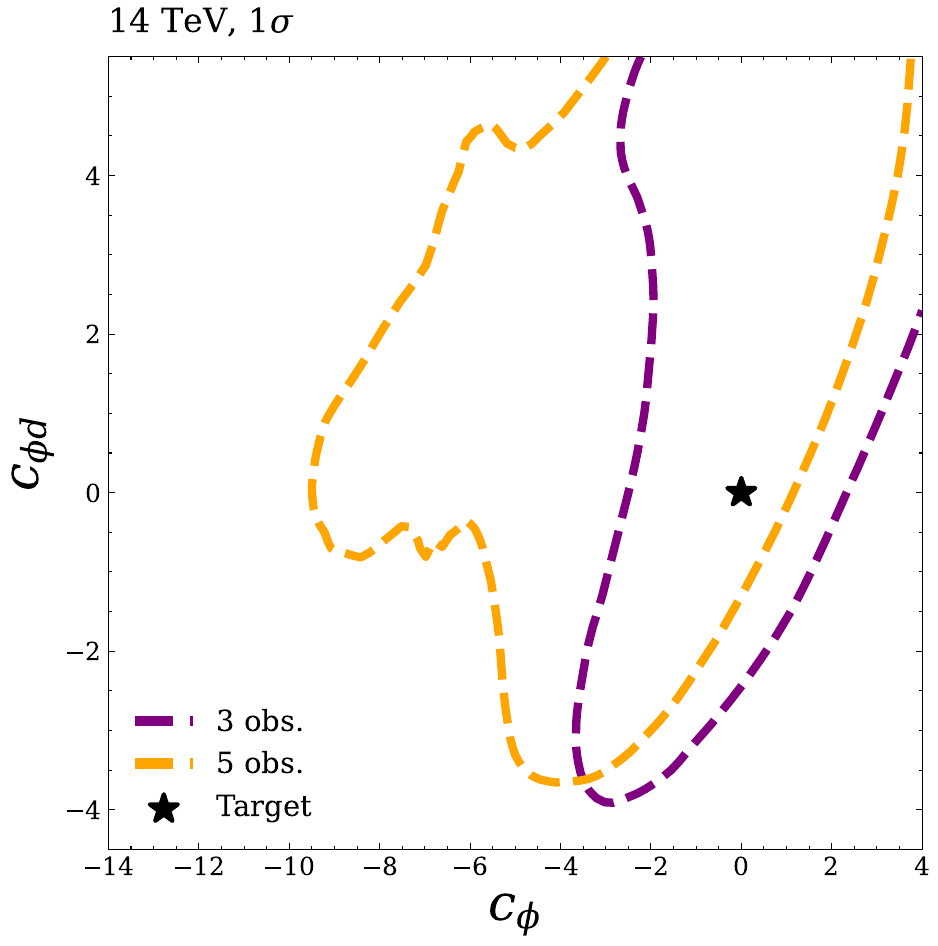} 
\includegraphics[width=0.32\textwidth]{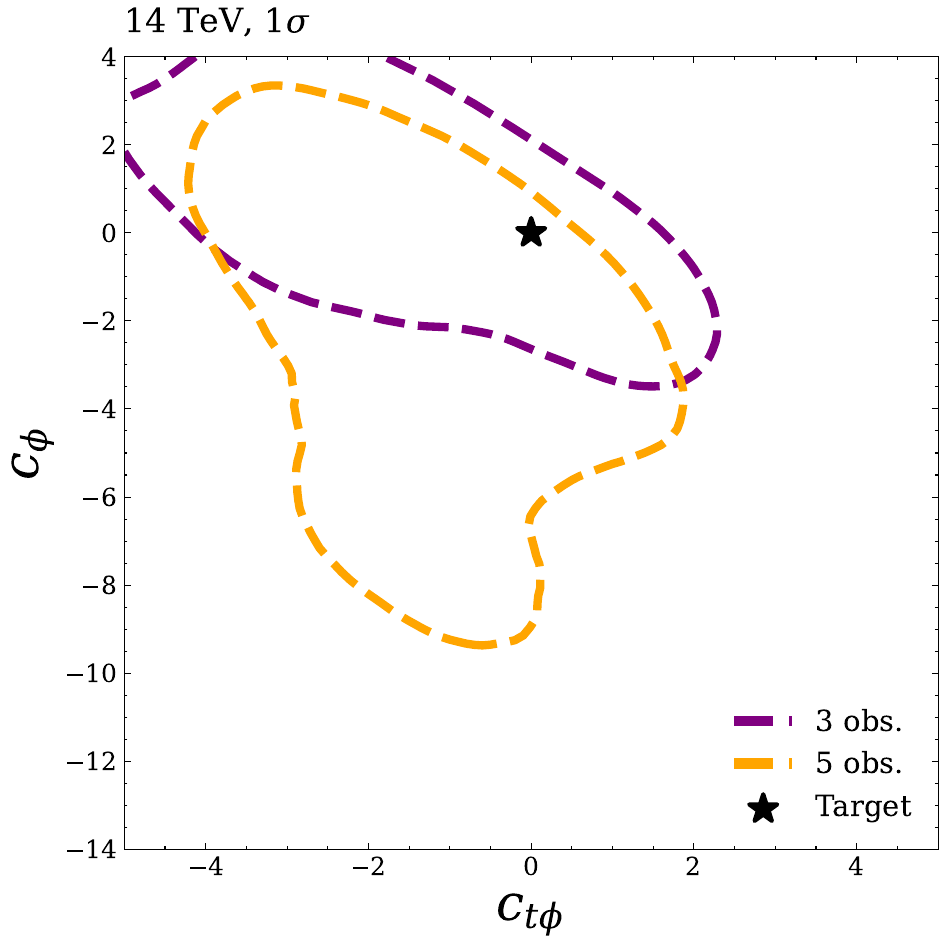} 
\includegraphics[width=0.32\textwidth]{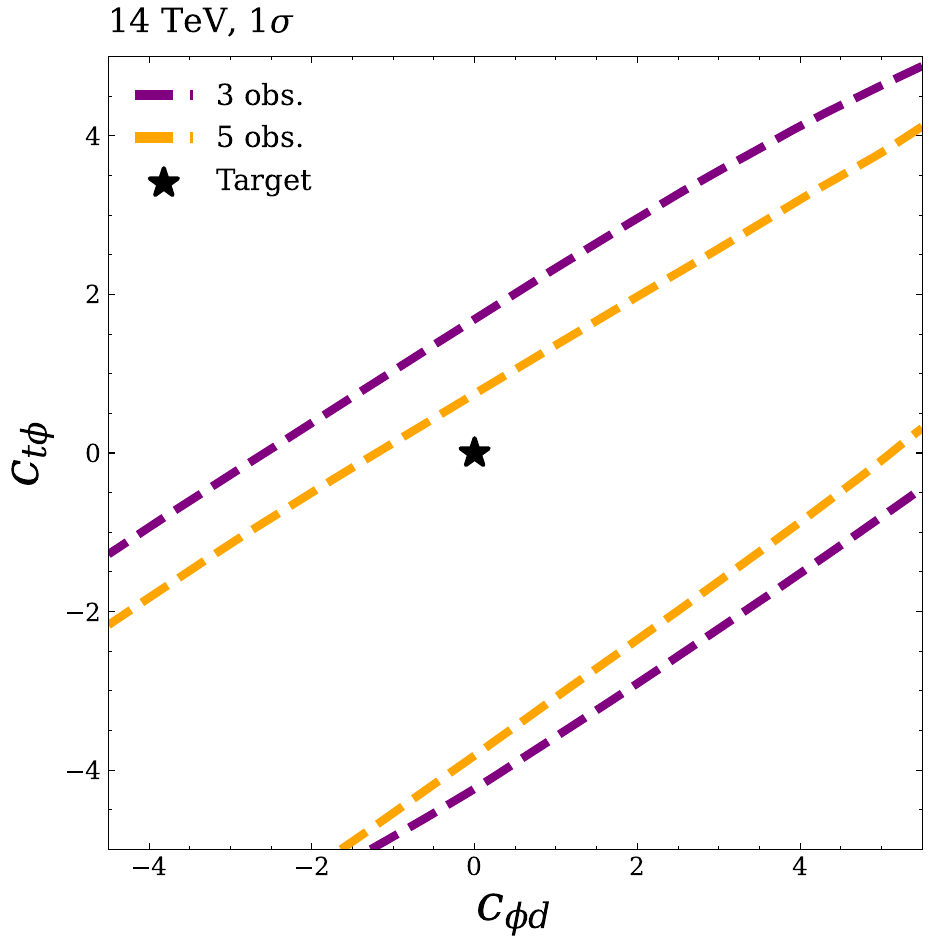} \\
\includegraphics[width=0.32\textwidth]{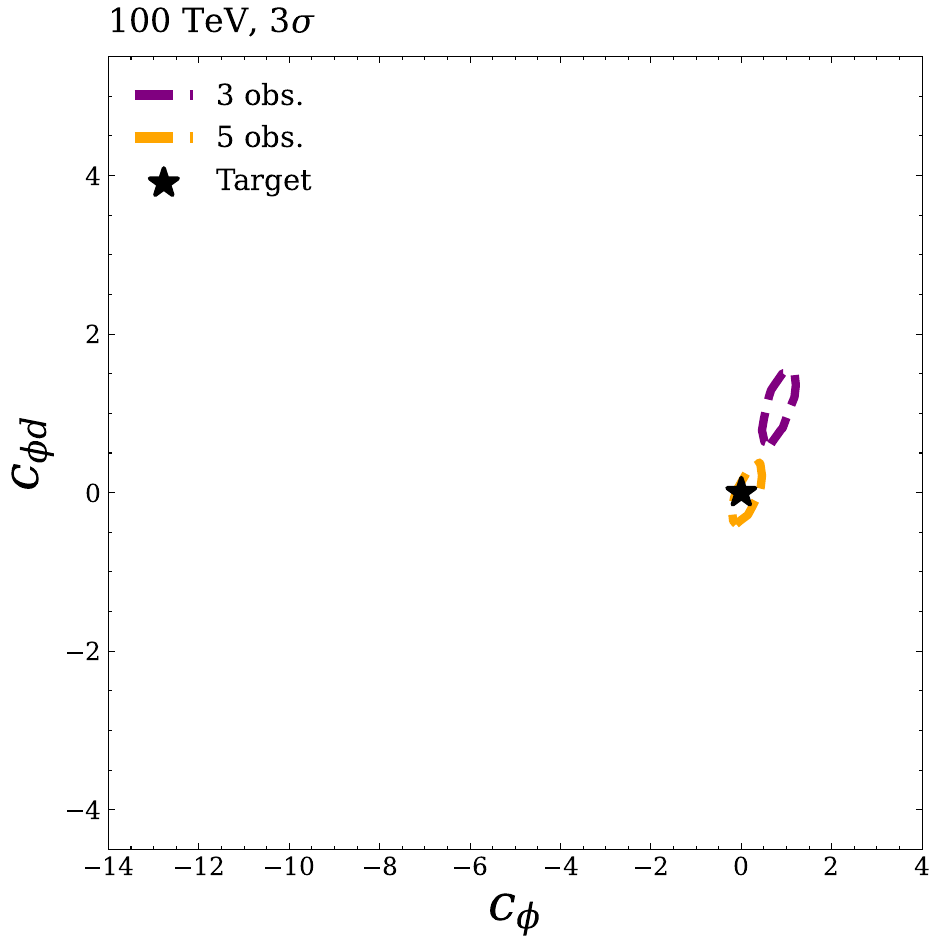} 
\includegraphics[width=0.32\textwidth]{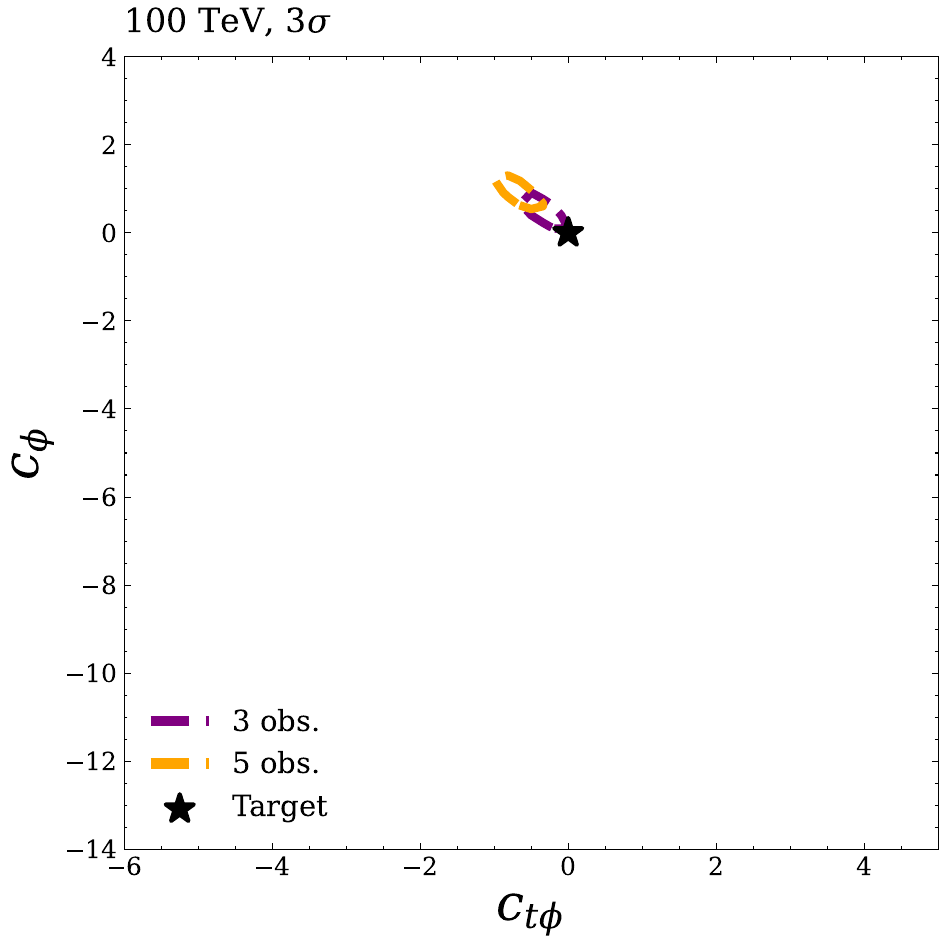} 
\includegraphics[width=0.32\textwidth]{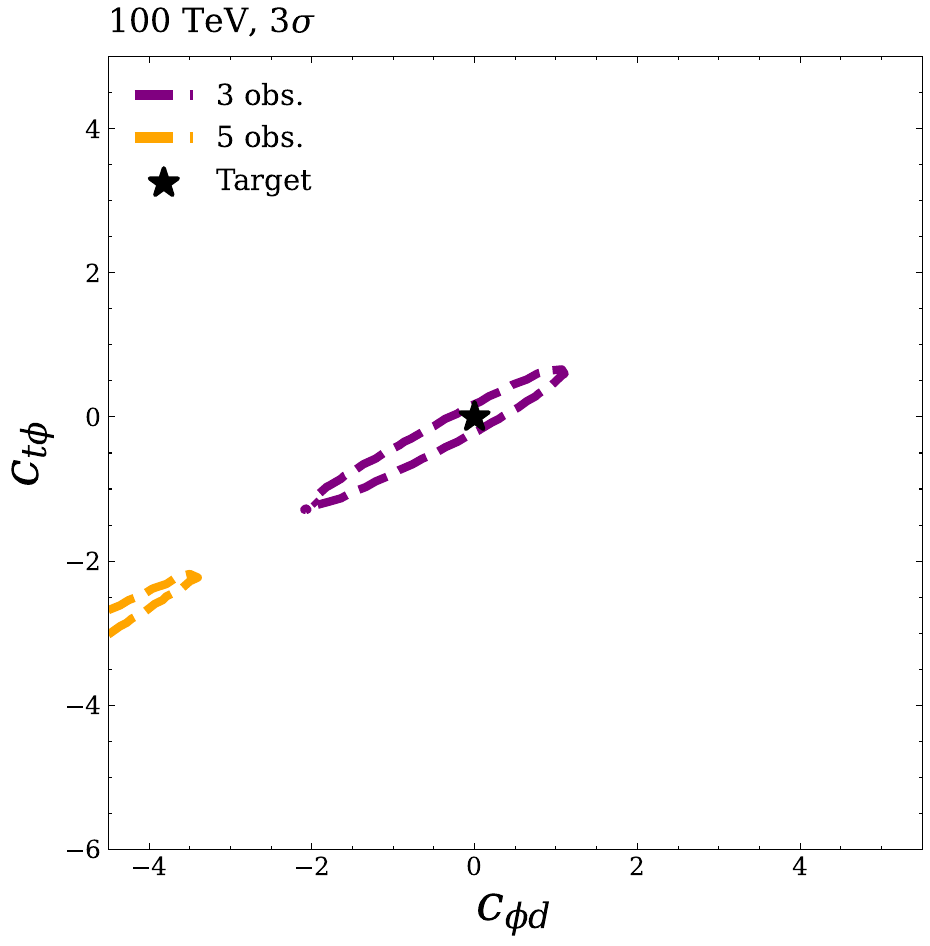}
\caption{Test statistics and 1$\sigma$ (for HL-LHC) or 3$\sigma$ (for 100 TeV) confidence contours for the test statistic $q=-2\ln\frac{p(x|c)}{p(x|0)}$ associated with the given Wilson coefficient. Data generation and test set size reflects the labeled collider setup.}
\label{fig:q_results_2d_f5}
\end{figure*}

In \Fig{fig:q_results_1d_f5}, we show the 1-dimensional test statistic and confidence intervals for the 3-observable ($m_\mathrm{tot}$, $p_{T_{bb}}$, $p_{T_{\gamma\gamma}}$) and 5-observable test statistics (+ $\Delta R_{bb}$, $\Delta R_{\gamma\gamma}$). This plot serves as an extension to \Fig{fig:q_results_1d}. In \Fig{fig:q_results_2d_f5}, we show the same for the 2-dimension test statistic and confidence internal, as an  extension to \Fig{fig:q_results_2d}. We generally find that including more observables allows for tighter constraints on the given Wilson coefficient -- this is unambiguous for the 100 TeV collider setup, while the 14 TeV setup does suffer from limited statistics (see, in particular, the $\cp$ limit for \Fig{fig:q_results_1d_f5}).

\begin{figure*}
\centering
\includegraphics[width=0.32\textwidth]{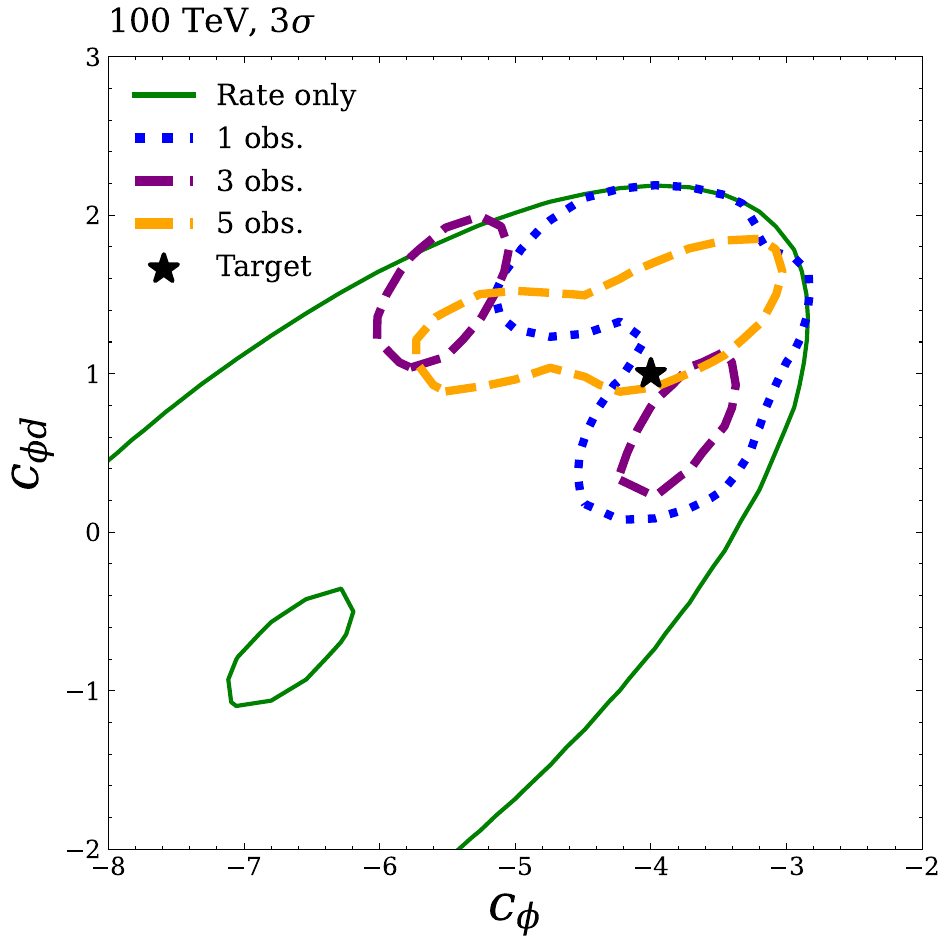} 
\hspace*{0.1\textwidth}
\includegraphics[width=0.32\textwidth]{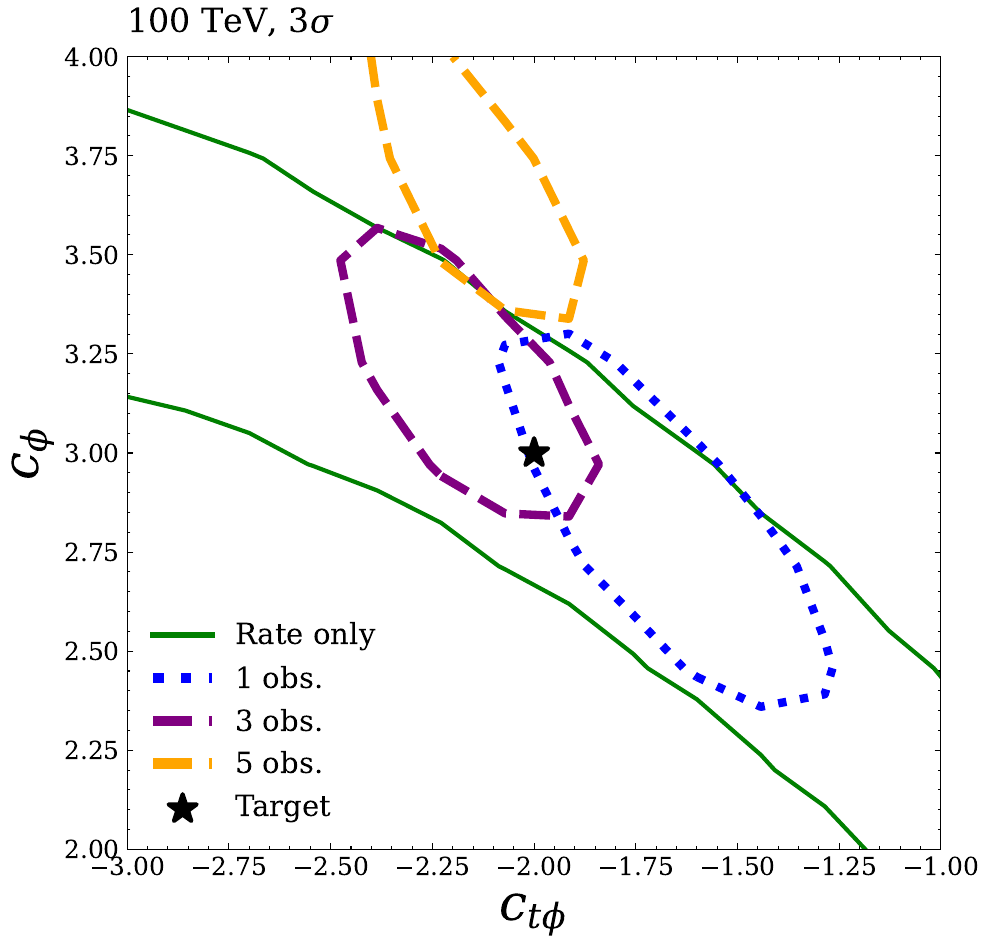}
\caption{Log-likelihood ratio test statistic in terms of 
$3\sigma$ confidence contours for two sets of Wilson coefficients with central values away 
from $c=0$. The generating values are denoted by a star.}
\label{fig:q_results_good1_100TeV}
\end{figure*}

In \Fig{fig:q_results_good1_100TeV}, we show two examples of 2D coefficient recovery for BSM test sets for the 100 TeV collider setup. All test statistics that make use of kinematic observables greatly reduce the confidence limit areas when compared to the rate-only analysis. However, the displacement of the recovered areas from zero demonstrates the challenges associated with the high-precision requirement of the 100 TeV collider.


\bibliography{tilman,main,old_refs}

\begin{thebibliography}{101}%
\makeatletter
\providecommand \@ifxundefined [1]{%
 \@ifx{#1\undefined}
}%
\providecommand \@ifnum [1]{%
 \ifnum #1\expandafter \@firstoftwo
 \else \expandafter \@secondoftwo
 \fi
}%
\providecommand \@ifx [1]{%
 \ifx #1\expandafter \@firstoftwo
 \else \expandafter \@secondoftwo
 \fi
}%
\providecommand \natexlab [1]{#1}%
\providecommand \enquote  [1]{``#1''}%
\providecommand \bibnamefont  [1]{#1}%
\providecommand \bibfnamefont [1]{#1}%
\providecommand \citenamefont [1]{#1}%
\providecommand \href@noop [0]{\@secondoftwo}%
\providecommand \href [0]{\begingroup \@sanitize@url \@href}%
\providecommand \@href[1]{\@@startlink{#1}\@@href}%
\providecommand \@@href[1]{\endgroup#1\@@endlink}%
\providecommand \@sanitize@url [0]{\catcode `\\12\catcode `\$12\catcode
  `\&12\catcode `\#12\catcode `\^12\catcode `\_12\catcode `\%12\relax}%
\providecommand \@@startlink[1]{}%
\providecommand \@@endlink[0]{}%
\providecommand \url  [0]{\begingroup\@sanitize@url \@url }%
\providecommand \@url [1]{\endgroup\@href {#1}{\urlprefix }}%
\providecommand \urlprefix  [0]{URL }%
\providecommand \Eprint [0]{\href }%
\providecommand \doibase [0]{https://doi.org/}%
\providecommand \selectlanguage [0]{\@gobble}%
\providecommand \bibinfo  [0]{\@secondoftwo}%
\providecommand \bibfield  [0]{\@secondoftwo}%
\providecommand \translation [1]{[#1]}%
\providecommand \BibitemOpen [0]{}%
\providecommand \bibitemStop [0]{}%
\providecommand \bibitemNoStop [0]{.\EOS\space}%
\providecommand \EOS [0]{\spacefactor3000\relax}%
\providecommand \BibitemShut  [1]{\csname bibitem#1\endcsname}%
\let\auto@bib@innerbib\@empty
\bibitem [{\citenamefont {Aad}\ \emph {et~al.}(2012)\citenamefont {Aad} \emph
  {et~al.}}]{Aad:2012tfa}%
  \BibitemOpen
  \bibfield  {author} {\bibinfo {author} {\bibfnamefont {G.}~\bibnamefont
  {Aad}} \emph {et~al.} (\bibinfo {collaboration} {ATLAS}),\ }\bibfield
  {title} {\bibinfo {title} {{Observation of a new particle in the search for
  the Standard Model Higgs boson with the ATLAS detector at the LHC}},\ }\href
  {https://doi.org/10.1016/j.physletb.2012.08.020} {\bibfield  {journal}
  {\bibinfo  {journal} {Phys. Lett. B}\ }\textbf {\bibinfo {volume} {716}},\
  \bibinfo {pages} {1} (\bibinfo {year} {2012})},\ \Eprint
  {https://arxiv.org/abs/1207.7214} {arXiv:1207.7214 [hep-ex]} \BibitemShut
  {NoStop}%
\bibitem [{\citenamefont {Chatrchyan}\ \emph {et~al.}(2012)\citenamefont
  {Chatrchyan} \emph {et~al.}}]{Chatrchyan:2012ufa}%
  \BibitemOpen
  \bibfield  {author} {\bibinfo {author} {\bibfnamefont {S.}~\bibnamefont
  {Chatrchyan}} \emph {et~al.} (\bibinfo {collaboration} {CMS}),\ }\bibfield
  {title} {\bibinfo {title} {{Observation of a New Boson at a Mass of 125 GeV
  with the CMS Experiment at the LHC}},\ }\href
  {https://doi.org/10.1016/j.physletb.2012.08.021} {\bibfield  {journal}
  {\bibinfo  {journal} {Phys. Lett. B}\ }\textbf {\bibinfo {volume} {716}},\
  \bibinfo {pages} {30} (\bibinfo {year} {2012})},\ \Eprint
  {https://arxiv.org/abs/1207.7235} {arXiv:1207.7235 [hep-ex]} \BibitemShut
  {NoStop}%
\bibitem [{\citenamefont {Isidori}\ \emph {et~al.}(2001)\citenamefont
  {Isidori}, \citenamefont {Ridolfi},\ and\ \citenamefont
  {Strumia}}]{Isidori:2001bm}%
  \BibitemOpen
  \bibfield  {author} {\bibinfo {author} {\bibfnamefont {G.}~\bibnamefont
  {Isidori}}, \bibinfo {author} {\bibfnamefont {G.}~\bibnamefont {Ridolfi}},\
  and\ \bibinfo {author} {\bibfnamefont {A.}~\bibnamefont {Strumia}},\
  }\bibfield  {title} {\bibinfo {title} {{On the metastability of the standard
  model vacuum}},\ }\href {https://doi.org/10.1016/S0550-3213(01)00302-9}
  {\bibfield  {journal} {\bibinfo  {journal} {Nucl. Phys. B}\ }\textbf
  {\bibinfo {volume} {609}},\ \bibinfo {pages} {387} (\bibinfo {year}
  {2001})},\ \Eprint {https://arxiv.org/abs/hep-ph/0104016}
  {arXiv:hep-ph/0104016} \BibitemShut {NoStop}%
\bibitem [{\citenamefont {Degrassi}\ \emph {et~al.}(2012)\citenamefont
  {Degrassi}, \citenamefont {Di~Vita}, \citenamefont {Elias-Miro},
  \citenamefont {Espinosa}, \citenamefont {Giudice}, \citenamefont {Isidori},\
  and\ \citenamefont {Strumia}}]{Degrassi:2012ry}%
  \BibitemOpen
  \bibfield  {author} {\bibinfo {author} {\bibfnamefont {G.}~\bibnamefont
  {Degrassi}}, \bibinfo {author} {\bibfnamefont {S.}~\bibnamefont {Di~Vita}},
  \bibinfo {author} {\bibfnamefont {J.}~\bibnamefont {Elias-Miro}}, \bibinfo
  {author} {\bibfnamefont {J.~R.}\ \bibnamefont {Espinosa}}, \bibinfo {author}
  {\bibfnamefont {G.~F.}\ \bibnamefont {Giudice}}, \bibinfo {author}
  {\bibfnamefont {G.}~\bibnamefont {Isidori}},\ and\ \bibinfo {author}
  {\bibfnamefont {A.}~\bibnamefont {Strumia}},\ }\bibfield  {title} {\bibinfo
  {title} {{Higgs mass and vacuum stability in the Standard Model at NNLO}},\
  }\href {https://doi.org/10.1007/JHEP08(2012)098} {\bibfield  {journal}
  {\bibinfo  {journal} {JHEP}\ }\textbf {\bibinfo {volume} {08}},\ \bibinfo
  {pages} {098}},\ \Eprint {https://arxiv.org/abs/1205.6497} {arXiv:1205.6497
  [hep-ph]} \BibitemShut {NoStop}%
\bibitem [{\citenamefont {Bednyakov}\ \emph {et~al.}(2015)\citenamefont
  {Bednyakov}, \citenamefont {Kniehl}, \citenamefont {Pikelner},\ and\
  \citenamefont {Veretin}}]{Bednyakov_2015}%
  \BibitemOpen
  \bibfield  {author} {\bibinfo {author} {\bibfnamefont {A.}~\bibnamefont
  {Bednyakov}}, \bibinfo {author} {\bibfnamefont {B.}~\bibnamefont {Kniehl}},
  \bibinfo {author} {\bibfnamefont {A.}~\bibnamefont {Pikelner}},\ and\
  \bibinfo {author} {\bibfnamefont {O.}~\bibnamefont {Veretin}},\ }\bibfield
  {title} {\bibinfo {title} {Stability of the electroweak vacuum: Gauge
  independence and advanced precision},\ }\bibfield  {journal} {\bibinfo
  {journal} {Physical Review Letters}\ }\textbf {\bibinfo {volume} {115}},\
  \href {https://doi.org/10.1103/physrevlett.115.201802}
  {10.1103/physrevlett.115.201802} (\bibinfo {year} {2015})\BibitemShut
  {NoStop}%
\bibitem [{\citenamefont {Grojean}\ \emph {et~al.}(2005)\citenamefont
  {Grojean}, \citenamefont {Servant},\ and\ \citenamefont
  {Wells}}]{Grojean:2004xa}%
  \BibitemOpen
  \bibfield  {author} {\bibinfo {author} {\bibfnamefont {C.}~\bibnamefont
  {Grojean}}, \bibinfo {author} {\bibfnamefont {G.}~\bibnamefont {Servant}},\
  and\ \bibinfo {author} {\bibfnamefont {J.~D.}\ \bibnamefont {Wells}},\
  }\bibfield  {title} {\bibinfo {title} {{First-order electroweak phase
  transition in the standard model with a low cutoff}},\ }\href
  {https://doi.org/10.1103/PhysRevD.71.036001} {\bibfield  {journal} {\bibinfo
  {journal} {Phys. Rev. D}\ }\textbf {\bibinfo {volume} {71}},\ \bibinfo
  {pages} {036001} (\bibinfo {year} {2005})},\ \Eprint
  {https://arxiv.org/abs/hep-ph/0407019} {arXiv:hep-ph/0407019} \BibitemShut
  {NoStop}%
\bibitem [{\citenamefont {Reichert}\ \emph {et~al.}(2018)\citenamefont
  {Reichert}, \citenamefont {Eichhorn}, \citenamefont {Gies}, \citenamefont
  {Pawlowski}, \citenamefont {Plehn},\ and\ \citenamefont
  {Scherer}}]{Reichert:2017puo}%
  \BibitemOpen
  \bibfield  {author} {\bibinfo {author} {\bibfnamefont {M.}~\bibnamefont
  {Reichert}}, \bibinfo {author} {\bibfnamefont {A.}~\bibnamefont {Eichhorn}},
  \bibinfo {author} {\bibfnamefont {H.}~\bibnamefont {Gies}}, \bibinfo {author}
  {\bibfnamefont {J.~M.}\ \bibnamefont {Pawlowski}}, \bibinfo {author}
  {\bibfnamefont {T.}~\bibnamefont {Plehn}},\ and\ \bibinfo {author}
  {\bibfnamefont {M.~M.}\ \bibnamefont {Scherer}},\ }\bibfield  {title}
  {\bibinfo {title} {{Probing baryogenesis through the Higgs boson
  self-coupling}},\ }\href {https://doi.org/10.1103/PhysRevD.97.075008}
  {\bibfield  {journal} {\bibinfo  {journal} {Phys. Rev. D}\ }\textbf {\bibinfo
  {volume} {97}},\ \bibinfo {pages} {075008} (\bibinfo {year} {2018})},\
  \Eprint {https://arxiv.org/abs/1711.00019} {arXiv:1711.00019 [hep-ph]}
  \BibitemShut {NoStop}%
\bibitem [{\citenamefont {Anisha}\ \emph {et~al.}(2022)\citenamefont {Anisha},
  \citenamefont {Biermann}, \citenamefont {Englert},\ and\ \citenamefont
  {M\"uhlleitner}}]{Anisha:2022hgv}%
  \BibitemOpen
  \bibfield  {author} {\bibinfo {author} {\bibnamefont {Anisha}}, \bibinfo
  {author} {\bibfnamefont {L.}~\bibnamefont {Biermann}}, \bibinfo {author}
  {\bibfnamefont {C.}~\bibnamefont {Englert}},\ and\ \bibinfo {author}
  {\bibfnamefont {M.}~\bibnamefont {M\"uhlleitner}},\ }\bibfield  {title}
  {\bibinfo {title} {{Two Higgs doublets, effective interactions and a strong
  first-order electroweak phase transition}},\ }\href
  {https://doi.org/10.1007/JHEP08(2022)091} {\bibfield  {journal} {\bibinfo
  {journal} {JHEP}\ }\textbf {\bibinfo {volume} {08}},\ \bibinfo {pages}
  {091}},\ \Eprint {https://arxiv.org/abs/2204.06966} {arXiv:2204.06966
  [hep-ph]} \BibitemShut {NoStop}%
\bibitem [{\citenamefont {Biek\"otter}\ \emph {et~al.}(2023)\citenamefont
  {Biek\"otter}, \citenamefont {Fuentes-Mart\'\i{}n}, \citenamefont {Galda},\
  and\ \citenamefont {Neubert}}]{Biekotter:2023mpd}%
  \BibitemOpen
  \bibfield  {author} {\bibinfo {author} {\bibfnamefont {A.}~\bibnamefont
  {Biek\"otter}}, \bibinfo {author} {\bibfnamefont {J.}~\bibnamefont
  {Fuentes-Mart\'\i{}n}}, \bibinfo {author} {\bibfnamefont {A.~M.}\
  \bibnamefont {Galda}},\ and\ \bibinfo {author} {\bibfnamefont
  {M.}~\bibnamefont {Neubert}},\ }\bibfield  {title} {\bibinfo {title} {{A
  global analysis of axion-like particle interactions using SMEFT fits}},\
  }\href {https://doi.org/10.1007/JHEP09(2023)120} {\bibfield  {journal}
  {\bibinfo  {journal} {JHEP}\ }\textbf {\bibinfo {volume} {09}},\ \bibinfo
  {pages} {120}},\ \Eprint {https://arxiv.org/abs/2307.10372} {arXiv:2307.10372
  [hep-ph]} \BibitemShut {NoStop}%
\bibitem [{\citenamefont {Workman}\ and\ \citenamefont
  {Others}(2022)}]{Workman:2022ynf}%
  \BibitemOpen
  \bibfield  {author} {\bibinfo {author} {\bibfnamefont {R.~L.}\ \bibnamefont
  {Workman}}\ and\ \bibinfo {author} {\bibnamefont {Others}} (\bibinfo
  {collaboration} {Particle Data Group}),\ }\bibfield  {title} {\bibinfo
  {title} {{Review of Particle Physics}},\ }\href
  {https://doi.org/10.1093/ptep/ptac097} {\bibfield  {journal} {\bibinfo
  {journal} {PTEP}\ }\textbf {\bibinfo {volume} {2022}},\ \bibinfo {pages}
  {083C01} (\bibinfo {year} {2022})}\BibitemShut {NoStop}%
\bibitem [{\citenamefont {Di~Vita}\ \emph {et~al.}(2017)\citenamefont
  {Di~Vita}, \citenamefont {Grojean}, \citenamefont {Panico}, \citenamefont
  {Riembau},\ and\ \citenamefont {Vantalon}}]{DiVita:2017eyz}%
  \BibitemOpen
  \bibfield  {author} {\bibinfo {author} {\bibfnamefont {S.}~\bibnamefont
  {Di~Vita}}, \bibinfo {author} {\bibfnamefont {C.}~\bibnamefont {Grojean}},
  \bibinfo {author} {\bibfnamefont {G.}~\bibnamefont {Panico}}, \bibinfo
  {author} {\bibfnamefont {M.}~\bibnamefont {Riembau}},\ and\ \bibinfo {author}
  {\bibfnamefont {T.}~\bibnamefont {Vantalon}},\ }\bibfield  {title} {\bibinfo
  {title} {{A global view on the Higgs self-coupling}},\ }\href
  {https://doi.org/10.1007/JHEP09(2017)069} {\bibfield  {journal} {\bibinfo
  {journal} {JHEP}\ }\textbf {\bibinfo {volume} {09}},\ \bibinfo {pages}
  {069}},\ \Eprint {https://arxiv.org/abs/1704.01953} {arXiv:1704.01953
  [hep-ph]} \BibitemShut {NoStop}%
\bibitem [{\citenamefont {Baur}\ \emph {et~al.}(2002)\citenamefont {Baur},
  \citenamefont {Plehn},\ and\ \citenamefont {Rainwater}}]{Baur:2002rb}%
  \BibitemOpen
  \bibfield  {author} {\bibinfo {author} {\bibfnamefont {U.}~\bibnamefont
  {Baur}}, \bibinfo {author} {\bibfnamefont {T.}~\bibnamefont {Plehn}},\ and\
  \bibinfo {author} {\bibfnamefont {D.~L.}\ \bibnamefont {Rainwater}},\
  }\bibfield  {title} {\bibinfo {title} {{Measuring the Higgs Boson Self
  Coupling at the LHC and Finite Top Mass Matrix Elements}},\ }\href
  {https://doi.org/10.1103/PhysRevLett.89.151801} {\bibfield  {journal}
  {\bibinfo  {journal} {Phys. Rev. Lett.}\ }\textbf {\bibinfo {volume} {89}},\
  \bibinfo {pages} {151801} (\bibinfo {year} {2002})},\ \Eprint
  {https://arxiv.org/abs/hep-ph/0206024} {arXiv:hep-ph/0206024} \BibitemShut
  {NoStop}%
\bibitem [{\citenamefont {Baur}\ \emph {et~al.}(2004)\citenamefont {Baur},
  \citenamefont {Plehn},\ and\ \citenamefont {Rainwater}}]{Baur:2003gp}%
  \BibitemOpen
  \bibfield  {author} {\bibinfo {author} {\bibfnamefont {U.}~\bibnamefont
  {Baur}}, \bibinfo {author} {\bibfnamefont {T.}~\bibnamefont {Plehn}},\ and\
  \bibinfo {author} {\bibfnamefont {D.~L.}\ \bibnamefont {Rainwater}},\
  }\bibfield  {title} {\bibinfo {title} {{Probing the Higgs selfcoupling at
  hadron colliders using rare decays}},\ }\href
  {https://doi.org/10.1103/PhysRevD.69.053004} {\bibfield  {journal} {\bibinfo
  {journal} {Phys. Rev. D}\ }\textbf {\bibinfo {volume} {69}},\ \bibinfo
  {pages} {053004} (\bibinfo {year} {2004})},\ \Eprint
  {https://arxiv.org/abs/hep-ph/0310056} {arXiv:hep-ph/0310056} \BibitemShut
  {NoStop}%
\bibitem [{\citenamefont {Chang}\ \emph {et~al.}(2020)\citenamefont {Chang},
  \citenamefont {Cheung}, \citenamefont {Lee},\ and\ \citenamefont
  {Park}}]{Chang:2019ncg}%
  \BibitemOpen
  \bibfield  {author} {\bibinfo {author} {\bibfnamefont {J.}~\bibnamefont
  {Chang}}, \bibinfo {author} {\bibfnamefont {K.}~\bibnamefont {Cheung}},
  \bibinfo {author} {\bibfnamefont {J.~S.}\ \bibnamefont {Lee}},\ and\ \bibinfo
  {author} {\bibfnamefont {J.}~\bibnamefont {Park}},\ }\bibfield  {title}
  {\bibinfo {title} {{Probing the trilinear Higgs boson self-coupling at the
  high-luminosity LHC via multivariate analysis}},\ }\href
  {https://doi.org/10.1103/PhysRevD.101.016004} {\bibfield  {journal} {\bibinfo
   {journal} {Phys. Rev. D}\ }\textbf {\bibinfo {volume} {101}},\ \bibinfo
  {pages} {016004} (\bibinfo {year} {2020})},\ \Eprint
  {https://arxiv.org/abs/1908.00753} {arXiv:1908.00753 [hep-ph]} \BibitemShut
  {NoStop}%
\bibitem [{\citenamefont {Zurbano~Fernandez}\ \emph {et~al.}(2020)\citenamefont
  {Zurbano~Fernandez} \emph {et~al.}}]{ZurbanoFernandez:2020cco}%
  \BibitemOpen
  \bibfield  {author} {\bibinfo {author} {\bibfnamefont {I.}~\bibnamefont
  {Zurbano~Fernandez}} \emph {et~al.},\ }\bibfield  {title} {\bibinfo {title}
  {{High-Luminosity Large Hadron Collider (HL-LHC): Technical design report}}\
  }\textbf {\bibinfo {volume} {10/2020}},\ \href
  {https://doi.org/10.23731/CYRM-2020-0010} {10.23731/CYRM-2020-0010} (\bibinfo
  {year} {2020})\BibitemShut {NoStop}%
\bibitem [{\citenamefont {Faus-Golfe}\ \emph {et~al.}(2022)\citenamefont
  {Faus-Golfe} \emph {et~al.}}]{Faus-Golfe:2022cgm}%
  \BibitemOpen
  \bibfield  {author} {\bibinfo {author} {\bibfnamefont {A.}~\bibnamefont
  {Faus-Golfe}} \emph {et~al.},\ }\bibfield  {title} {\bibinfo {title}
  {{Accelerators for Electroweak Physics and Higgs Boson Studies}},\
  }\href@noop {} {\  (\bibinfo {year} {2022})},\ \Eprint
  {https://arxiv.org/abs/2209.05827} {arXiv:2209.05827 [physics.acc-ph]}
  \BibitemShut {NoStop}%
\bibitem [{\citenamefont {Abada}\ \emph
  {et~al.}(2019{\natexlab{a}})\citenamefont {Abada} \emph
  {et~al.}}]{FCC:2018byv}%
  \BibitemOpen
  \bibfield  {author} {\bibinfo {author} {\bibfnamefont {A.}~\bibnamefont
  {Abada}} \emph {et~al.} (\bibinfo {collaboration} {FCC}),\ }\bibfield
  {title} {\bibinfo {title} {{FCC Physics Opportunities}: {Future Circular
  Collider Conceptual Design Report Volume 1}},\ }\href
  {https://doi.org/10.1140/epjc/s10052-019-6904-3} {\bibfield  {journal}
  {\bibinfo  {journal} {Eur. Phys. J. C}\ }\textbf {\bibinfo {volume} {79}},\
  \bibinfo {pages} {474} (\bibinfo {year} {2019}{\natexlab{a}})}\BibitemShut
  {NoStop}%
\bibitem [{\citenamefont {Buchmuller}\ and\ \citenamefont
  {Wyler}(1986)}]{Buchmuller:1985jz}%
  \BibitemOpen
  \bibfield  {author} {\bibinfo {author} {\bibfnamefont {W.}~\bibnamefont
  {Buchmuller}}\ and\ \bibinfo {author} {\bibfnamefont {D.}~\bibnamefont
  {Wyler}},\ }\bibfield  {title} {\bibinfo {title} {{Effective Lagrangian
  Analysis of New Interactions and Flavor Conservation}},\ }\href
  {https://doi.org/10.1016/0550-3213(86)90262-2} {\bibfield  {journal}
  {\bibinfo  {journal} {Nucl. Phys. B}\ }\textbf {\bibinfo {volume} {268}},\
  \bibinfo {pages} {621} (\bibinfo {year} {1986})}\BibitemShut {NoStop}%
\bibitem [{\citenamefont {Grzadkowski}\ \emph {et~al.}(2010)\citenamefont
  {Grzadkowski}, \citenamefont {Iskrzynski}, \citenamefont {Misiak},\ and\
  \citenamefont {Rosiek}}]{Grzadkowski:2010es}%
  \BibitemOpen
  \bibfield  {author} {\bibinfo {author} {\bibfnamefont {B.}~\bibnamefont
  {Grzadkowski}}, \bibinfo {author} {\bibfnamefont {M.}~\bibnamefont
  {Iskrzynski}}, \bibinfo {author} {\bibfnamefont {M.}~\bibnamefont {Misiak}},\
  and\ \bibinfo {author} {\bibfnamefont {J.}~\bibnamefont {Rosiek}},\
  }\bibfield  {title} {\bibinfo {title} {{Dimension-Six Terms in the Standard
  Model Lagrangian}},\ }\href {https://doi.org/10.1007/JHEP10(2010)085}
  {\bibfield  {journal} {\bibinfo  {journal} {JHEP}\ }\textbf {\bibinfo
  {volume} {10}},\ \bibinfo {pages} {085}},\ \Eprint
  {https://arxiv.org/abs/1008.4884} {arXiv:1008.4884 [hep-ph]} \BibitemShut
  {NoStop}%
\bibitem [{\citenamefont {Brivio}\ and\ \citenamefont
  {Trott}(2019)}]{Brivio:2017vri}%
  \BibitemOpen
  \bibfield  {author} {\bibinfo {author} {\bibfnamefont {I.}~\bibnamefont
  {Brivio}}\ and\ \bibinfo {author} {\bibfnamefont {M.}~\bibnamefont {Trott}},\
  }\bibfield  {title} {\bibinfo {title} {{The Standard Model as an Effective
  Field Theory}},\ }\href {https://doi.org/10.1016/j.physrep.2018.11.002}
  {\bibfield  {journal} {\bibinfo  {journal} {Phys. Rept.}\ }\textbf {\bibinfo
  {volume} {793}},\ \bibinfo {pages} {1} (\bibinfo {year} {2019})},\ \Eprint
  {https://arxiv.org/abs/1706.08945} {arXiv:1706.08945 [hep-ph]} \BibitemShut
  {NoStop}%
\bibitem [{\citenamefont {Feruglio}(1993)}]{Feruglio:1992wf}%
  \BibitemOpen
  \bibfield  {author} {\bibinfo {author} {\bibfnamefont {F.}~\bibnamefont
  {Feruglio}},\ }\bibfield  {title} {\bibinfo {title} {{The Chiral approach to
  the electroweak interactions}},\ }\href
  {https://doi.org/10.1142/S0217751X93001946} {\bibfield  {journal} {\bibinfo
  {journal} {Int. J. Mod. Phys. A}\ }\textbf {\bibinfo {volume} {8}},\ \bibinfo
  {pages} {4937} (\bibinfo {year} {1993})},\ \Eprint
  {https://arxiv.org/abs/hep-ph/9301281} {arXiv:hep-ph/9301281} \BibitemShut
  {NoStop}%
\bibitem [{\citenamefont {Alonso}\ \emph {et~al.}(2013)\citenamefont {Alonso},
  \citenamefont {Gavela}, \citenamefont {Merlo}, \citenamefont {Rigolin},\ and\
  \citenamefont {Yepes}}]{Alonso:2012px}%
  \BibitemOpen
  \bibfield  {author} {\bibinfo {author} {\bibfnamefont {R.}~\bibnamefont
  {Alonso}}, \bibinfo {author} {\bibfnamefont {M.~B.}\ \bibnamefont {Gavela}},
  \bibinfo {author} {\bibfnamefont {L.}~\bibnamefont {Merlo}}, \bibinfo
  {author} {\bibfnamefont {S.}~\bibnamefont {Rigolin}},\ and\ \bibinfo {author}
  {\bibfnamefont {J.}~\bibnamefont {Yepes}},\ }\bibfield  {title} {\bibinfo
  {title} {{The Effective Chiral Lagrangian for a Light Dynamical ''Higgs
  Particle''}},\ }\href {https://doi.org/10.1016/j.physletb.2013.04.037}
  {\bibfield  {journal} {\bibinfo  {journal} {Phys. Lett. B}\ }\textbf
  {\bibinfo {volume} {722}},\ \bibinfo {pages} {330} (\bibinfo {year}
  {2013})},\ \bibinfo {note} {[Erratum: Phys.Lett.B 726, 926 (2013)]},\ \Eprint
  {https://arxiv.org/abs/1212.3305} {arXiv:1212.3305 [hep-ph]} \BibitemShut
  {NoStop}%
\bibitem [{\citenamefont {Buchalla}\ \emph {et~al.}(2014)\citenamefont
  {Buchalla}, \citenamefont {Cat\`a},\ and\ \citenamefont
  {Krause}}]{Buchalla:2013rka}%
  \BibitemOpen
  \bibfield  {author} {\bibinfo {author} {\bibfnamefont {G.}~\bibnamefont
  {Buchalla}}, \bibinfo {author} {\bibfnamefont {O.}~\bibnamefont {Cat\`a}},\
  and\ \bibinfo {author} {\bibfnamefont {C.}~\bibnamefont {Krause}},\
  }\bibfield  {title} {\bibinfo {title} {{Complete Electroweak Chiral
  Lagrangian with a Light Higgs at NLO}},\ }\href
  {https://doi.org/10.1016/j.nuclphysb.2014.01.018} {\bibfield  {journal}
  {\bibinfo  {journal} {Nucl. Phys. B}\ }\textbf {\bibinfo {volume} {880}},\
  \bibinfo {pages} {552} (\bibinfo {year} {2014})},\ \bibinfo {note} {[Erratum:
  Nucl.Phys.B 913, 475--478 (2016)]},\ \Eprint
  {https://arxiv.org/abs/1307.5017} {arXiv:1307.5017 [hep-ph]} \BibitemShut
  {NoStop}%
\bibitem [{\citenamefont {Gon\c{c}alves}\ \emph {et~al.}(2018)\citenamefont
  {Gon\c{c}alves}, \citenamefont {Han}, \citenamefont {Kling}, \citenamefont
  {Plehn},\ and\ \citenamefont {Takeuchi}}]{Goncalves:2018qas}%
  \BibitemOpen
  \bibfield  {author} {\bibinfo {author} {\bibfnamefont {D.}~\bibnamefont
  {Gon\c{c}alves}}, \bibinfo {author} {\bibfnamefont {T.}~\bibnamefont {Han}},
  \bibinfo {author} {\bibfnamefont {F.}~\bibnamefont {Kling}}, \bibinfo
  {author} {\bibfnamefont {T.}~\bibnamefont {Plehn}},\ and\ \bibinfo {author}
  {\bibfnamefont {M.}~\bibnamefont {Takeuchi}},\ }\bibfield  {title} {\bibinfo
  {title} {{Higgs boson pair production at future hadron colliders: From
  kinematics to dynamics}},\ }\href
  {https://doi.org/10.1103/PhysRevD.97.113004} {\bibfield  {journal} {\bibinfo
  {journal} {Phys. Rev. D}\ }\textbf {\bibinfo {volume} {97}},\ \bibinfo
  {pages} {113004} (\bibinfo {year} {2018})},\ \Eprint
  {https://arxiv.org/abs/1802.04319} {arXiv:1802.04319 [hep-ph]} \BibitemShut
  {NoStop}%
\bibitem [{\citenamefont {Chang}\ \emph {et~al.}(2019)\citenamefont {Chang},
  \citenamefont {Cheung}, \citenamefont {Lee}, \citenamefont {Lu},\ and\
  \citenamefont {Park}}]{Chang:2018uwu}%
  \BibitemOpen
  \bibfield  {author} {\bibinfo {author} {\bibfnamefont {J.}~\bibnamefont
  {Chang}}, \bibinfo {author} {\bibfnamefont {K.}~\bibnamefont {Cheung}},
  \bibinfo {author} {\bibfnamefont {J.~S.}\ \bibnamefont {Lee}}, \bibinfo
  {author} {\bibfnamefont {C.-T.}\ \bibnamefont {Lu}},\ and\ \bibinfo {author}
  {\bibfnamefont {J.}~\bibnamefont {Park}},\ }\bibfield  {title} {\bibinfo
  {title} {{Higgs-boson-pair production
  H(\textrightarrow{}bb\textasciimacron{})H(\textrightarrow{}\ensuremath{\gamma}\ensuremath{\gamma})
  from gluon fusion at the HL-LHC and HL-100 TeV hadron collider}},\ }\href
  {https://doi.org/10.1103/PhysRevD.100.096001} {\bibfield  {journal} {\bibinfo
   {journal} {Phys. Rev. D}\ }\textbf {\bibinfo {volume} {100}},\ \bibinfo
  {pages} {096001} (\bibinfo {year} {2019})},\ \Eprint
  {https://arxiv.org/abs/1804.07130} {arXiv:1804.07130 [hep-ph]} \BibitemShut
  {NoStop}%
\bibitem [{\citenamefont {Chang}\ and\ \citenamefont
  {Luty}(2020)}]{Chang:2019vez}%
  \BibitemOpen
  \bibfield  {author} {\bibinfo {author} {\bibfnamefont {S.}~\bibnamefont
  {Chang}}\ and\ \bibinfo {author} {\bibfnamefont {M.~A.}\ \bibnamefont
  {Luty}},\ }\bibfield  {title} {\bibinfo {title} {{The Higgs Trilinear
  Coupling and the Scale of New Physics}},\ }\href
  {https://doi.org/10.1007/JHEP03(2020)140} {\bibfield  {journal} {\bibinfo
  {journal} {JHEP}\ }\textbf {\bibinfo {volume} {03}},\ \bibinfo {pages}
  {140}},\ \Eprint {https://arxiv.org/abs/1902.05556} {arXiv:1902.05556
  [hep-ph]} \BibitemShut {NoStop}%
\bibitem [{\citenamefont {Agrawal}\ \emph {et~al.}(2020)\citenamefont
  {Agrawal}, \citenamefont {Saha}, \citenamefont {Xu}, \citenamefont {Yu},\
  and\ \citenamefont {Yuan}}]{Agrawal:2019bpm}%
  \BibitemOpen
  \bibfield  {author} {\bibinfo {author} {\bibfnamefont {P.}~\bibnamefont
  {Agrawal}}, \bibinfo {author} {\bibfnamefont {D.}~\bibnamefont {Saha}},
  \bibinfo {author} {\bibfnamefont {L.-X.}\ \bibnamefont {Xu}}, \bibinfo
  {author} {\bibfnamefont {J.-H.}\ \bibnamefont {Yu}},\ and\ \bibinfo {author}
  {\bibfnamefont {C.~P.}\ \bibnamefont {Yuan}},\ }\bibfield  {title} {\bibinfo
  {title} {{Determining the shape of the Higgs potential at future
  colliders}},\ }\href {https://doi.org/10.1103/PhysRevD.101.075023} {\bibfield
   {journal} {\bibinfo  {journal} {Phys. Rev. D}\ }\textbf {\bibinfo {volume}
  {101}},\ \bibinfo {pages} {075023} (\bibinfo {year} {2020})},\ \Eprint
  {https://arxiv.org/abs/1907.02078} {arXiv:1907.02078 [hep-ph]} \BibitemShut
  {NoStop}%
\bibitem [{\citenamefont {Mangano}\ \emph {et~al.}(2020)\citenamefont
  {Mangano}, \citenamefont {Ortona},\ and\ \citenamefont
  {Selvaggi}}]{Mangano:2020sao}%
  \BibitemOpen
  \bibfield  {author} {\bibinfo {author} {\bibfnamefont {M.~L.}\ \bibnamefont
  {Mangano}}, \bibinfo {author} {\bibfnamefont {G.}~\bibnamefont {Ortona}},\
  and\ \bibinfo {author} {\bibfnamefont {M.}~\bibnamefont {Selvaggi}},\
  }\bibfield  {title} {\bibinfo {title} {{Measuring the Higgs self-coupling via
  Higgs-pair production at a 100 TeV p-p collider}},\ }\href
  {https://doi.org/10.1140/epjc/s10052-020-08595-3} {\bibfield  {journal}
  {\bibinfo  {journal} {Eur. Phys. J. C}\ }\textbf {\bibinfo {volume} {80}},\
  \bibinfo {pages} {1030} (\bibinfo {year} {2020})},\ \Eprint
  {https://arxiv.org/abs/2004.03505} {arXiv:2004.03505 [hep-ph]} \BibitemShut
  {NoStop}%
\bibitem [{\citenamefont {Chai}\ \emph {et~al.}(2023)\citenamefont {Chai},
  \citenamefont {Yu},\ and\ \citenamefont {Zhang}}]{Chai:2022zeq}%
  \BibitemOpen
  \bibfield  {author} {\bibinfo {author} {\bibfnamefont {K.}~\bibnamefont
  {Chai}}, \bibinfo {author} {\bibfnamefont {J.-H.}\ \bibnamefont {Yu}},\ and\
  \bibinfo {author} {\bibfnamefont {H.}~\bibnamefont {Zhang}},\ }\bibfield
  {title} {\bibinfo {title} {{Investigating Higgs self-interaction through
  di-Higgs plus jet production at a 100~TeV hadron collider}},\ }\href
  {https://doi.org/10.1103/PhysRevD.107.055031} {\bibfield  {journal} {\bibinfo
   {journal} {Phys. Rev. D}\ }\textbf {\bibinfo {volume} {107}},\ \bibinfo
  {pages} {055031} (\bibinfo {year} {2023})},\ \Eprint
  {https://arxiv.org/abs/2210.14929} {arXiv:2210.14929 [hep-ph]} \BibitemShut
  {NoStop}%
\bibitem [{\citenamefont {Da~Rold}\ \emph {et~al.}(2023)\citenamefont
  {Da~Rold}, \citenamefont {Epele}, \citenamefont {Medina}, \citenamefont
  {Mileo},\ and\ \citenamefont {Szynkman}}]{DaRold:2023hst}%
  \BibitemOpen
  \bibfield  {author} {\bibinfo {author} {\bibfnamefont {L.}~\bibnamefont
  {Da~Rold}}, \bibinfo {author} {\bibfnamefont {M.}~\bibnamefont {Epele}},
  \bibinfo {author} {\bibfnamefont {A.~D.}\ \bibnamefont {Medina}}, \bibinfo
  {author} {\bibfnamefont {N.~I.}\ \bibnamefont {Mileo}},\ and\ \bibinfo
  {author} {\bibfnamefont {A.}~\bibnamefont {Szynkman}},\ }\bibfield  {title}
  {\bibinfo {title} {{Double Higgs production at the HL-LHC: probing a
  loop-enhanced model with kinematical distributions}},\ }\href@noop {} {\
  (\bibinfo {year} {2023})},\ \Eprint {https://arxiv.org/abs/2312.13149}
  {arXiv:2312.13149 [hep-ph]} \BibitemShut {NoStop}%
\bibitem [{\citenamefont {Papaefstathiou}\ and\ \citenamefont
  {Tetlalmatzi-Xolocotzi}(2023)}]{Papaefstathiou:2023uum}%
  \BibitemOpen
  \bibfield  {author} {\bibinfo {author} {\bibfnamefont {A.}~\bibnamefont
  {Papaefstathiou}}\ and\ \bibinfo {author} {\bibfnamefont {G.}~\bibnamefont
  {Tetlalmatzi-Xolocotzi}},\ }\bibfield  {title} {\bibinfo {title}
  {{Multi-Higgs Boson Production with Anomalous Interactions at Current and
  Future Proton Colliders}},\ }\href@noop {} {\  (\bibinfo {year} {2023})},\
  \Eprint {https://arxiv.org/abs/2312.13562} {arXiv:2312.13562 [hep-ph]}
  \BibitemShut {NoStop}%
\bibitem [{\citenamefont {Tumasyan}\ \emph {et~al.}(2022)\citenamefont
  {Tumasyan} \emph {et~al.}}]{CMS:2022dwd}%
  \BibitemOpen
  \bibfield  {author} {\bibinfo {author} {\bibfnamefont {A.}~\bibnamefont
  {Tumasyan}} \emph {et~al.} (\bibinfo {collaboration} {CMS}),\ }\bibfield
  {title} {\bibinfo {title} {{A portrait of the Higgs boson by the CMS
  experiment ten years after the discovery.}},\ }\href
  {https://doi.org/10.1038/s41586-022-04892-x} {\bibfield  {journal} {\bibinfo
  {journal} {Nature}\ }\textbf {\bibinfo {volume} {607}},\ \bibinfo {pages}
  {60} (\bibinfo {year} {2022})},\ \Eprint {https://arxiv.org/abs/2207.00043}
  {arXiv:2207.00043 [hep-ex]} \BibitemShut {NoStop}%
\bibitem [{\citenamefont {Aad}\ \emph {et~al.}(2023)\citenamefont {Aad} \emph
  {et~al.}}]{ATLAS:2022jtk}%
  \BibitemOpen
  \bibfield  {author} {\bibinfo {author} {\bibfnamefont {G.}~\bibnamefont
  {Aad}} \emph {et~al.} (\bibinfo {collaboration} {ATLAS}),\ }\bibfield
  {title} {\bibinfo {title} {{Constraints on the Higgs boson self-coupling from
  single- and double-Higgs production with the ATLAS detector using pp
  collisions at s=13 TeV}},\ }\href
  {https://doi.org/10.1016/j.physletb.2023.137745} {\bibfield  {journal}
  {\bibinfo  {journal} {Phys. Lett. B}\ }\textbf {\bibinfo {volume} {843}},\
  \bibinfo {pages} {137745} (\bibinfo {year} {2023})},\ \Eprint
  {https://arxiv.org/abs/2211.01216} {arXiv:2211.01216 [hep-ex]} \BibitemShut
  {NoStop}%
\bibitem [{\citenamefont {Kling}\ \emph {et~al.}(2017)\citenamefont {Kling},
  \citenamefont {Plehn},\ and\ \citenamefont {Schichtel}}]{Kling:2016lay}%
  \BibitemOpen
  \bibfield  {author} {\bibinfo {author} {\bibfnamefont {F.}~\bibnamefont
  {Kling}}, \bibinfo {author} {\bibfnamefont {T.}~\bibnamefont {Plehn}},\ and\
  \bibinfo {author} {\bibfnamefont {P.}~\bibnamefont {Schichtel}},\ }\bibfield
  {title} {\bibinfo {title} {{Maximizing the significance in Higgs boson pair
  analyses}},\ }\href {https://doi.org/10.1103/PhysRevD.95.035026} {\bibfield
  {journal} {\bibinfo  {journal} {Phys. Rev. D}\ }\textbf {\bibinfo {volume}
  {95}},\ \bibinfo {pages} {035026} (\bibinfo {year} {2017})},\ \Eprint
  {https://arxiv.org/abs/1607.07441} {arXiv:1607.07441 [hep-ph]} \BibitemShut
  {NoStop}%
\bibitem [{\citenamefont {Cranmer}\ \emph {et~al.}(2020)\citenamefont
  {Cranmer}, \citenamefont {Brehme},\ and\ \citenamefont {Louppe}}]{sbi_pnas}%
  \BibitemOpen
  \bibfield  {author} {\bibinfo {author} {\bibfnamefont {K.}~\bibnamefont
  {Cranmer}}, \bibinfo {author} {\bibfnamefont {J.}~\bibnamefont {Brehme}},\
  and\ \bibinfo {author} {\bibfnamefont {G.}~\bibnamefont {Louppe}},\ }\href
  {https://doi.org/10.1073/pnas.1912789117} {\bibinfo {title} {The frontier of
  simulation-based inference}} (\bibinfo {year} {2020})\BibitemShut {NoStop}%
\bibitem [{\citenamefont {Carvalho}\ \emph {et~al.}(2016)\citenamefont
  {Carvalho}, \citenamefont {Dall’Osso}, \citenamefont {Dorigo},
  \citenamefont {Goertz}, \citenamefont {Gottardo},\ and\ \citenamefont
  {Tosi}}]{Carvalho_2016}%
  \BibitemOpen
  \bibfield  {author} {\bibinfo {author} {\bibfnamefont {A.}~\bibnamefont
  {Carvalho}}, \bibinfo {author} {\bibfnamefont {M.}~\bibnamefont
  {Dall’Osso}}, \bibinfo {author} {\bibfnamefont {T.}~\bibnamefont {Dorigo}},
  \bibinfo {author} {\bibfnamefont {F.}~\bibnamefont {Goertz}}, \bibinfo
  {author} {\bibfnamefont {C.~A.}\ \bibnamefont {Gottardo}},\ and\ \bibinfo
  {author} {\bibfnamefont {M.}~\bibnamefont {Tosi}},\ }\bibfield  {title}
  {\bibinfo {title} {Higgs pair production: choosing benchmarks with cluster
  analysis},\ }\href {https://doi.org/10.1007/jhep04(2016)126} {\bibfield
  {journal} {\bibinfo  {journal} {Journal of High Energy Physics}\ }\textbf
  {\bibinfo {volume} {2016}},\ \bibinfo {pages} {1–28} (\bibinfo {year}
  {2016})}\BibitemShut {NoStop}%
\bibitem [{\citenamefont {Buchalla}\ \emph {et~al.}(2018)\citenamefont
  {Buchalla}, \citenamefont {Capozi}, \citenamefont {Celis}, \citenamefont
  {Heinrich},\ and\ \citenamefont {Scyboz}}]{Buchalla:2018yce}%
  \BibitemOpen
  \bibfield  {author} {\bibinfo {author} {\bibfnamefont {G.}~\bibnamefont
  {Buchalla}}, \bibinfo {author} {\bibfnamefont {M.}~\bibnamefont {Capozi}},
  \bibinfo {author} {\bibfnamefont {A.}~\bibnamefont {Celis}}, \bibinfo
  {author} {\bibfnamefont {G.}~\bibnamefont {Heinrich}},\ and\ \bibinfo
  {author} {\bibfnamefont {L.}~\bibnamefont {Scyboz}},\ }\bibfield  {title}
  {\bibinfo {title} {{Higgs boson pair production in non-linear Effective Field
  Theory with full $m_t$-dependence at NLO QCD}},\ }\href
  {https://doi.org/10.1007/JHEP09(2018)057} {\bibfield  {journal} {\bibinfo
  {journal} {JHEP}\ }\textbf {\bibinfo {volume} {09}},\ \bibinfo {pages}
  {057}},\ \Eprint {https://arxiv.org/abs/1806.05162} {arXiv:1806.05162
  [hep-ph]} \BibitemShut {NoStop}%
\bibitem [{\citenamefont {Capozi}\ and\ \citenamefont
  {Heinrich}(2020)}]{Capozi:2019xsi}%
  \BibitemOpen
  \bibfield  {author} {\bibinfo {author} {\bibfnamefont {M.}~\bibnamefont
  {Capozi}}\ and\ \bibinfo {author} {\bibfnamefont {G.}~\bibnamefont
  {Heinrich}},\ }\bibfield  {title} {\bibinfo {title} {{Exploring anomalous
  couplings in Higgs boson pair production through shape analysis}},\ }\href
  {https://doi.org/10.1007/JHEP03(2020)091} {\bibfield  {journal} {\bibinfo
  {journal} {JHEP}\ }\textbf {\bibinfo {volume} {03}},\ \bibinfo {pages}
  {091}},\ \Eprint {https://arxiv.org/abs/1908.08923} {arXiv:1908.08923
  [hep-ph]} \BibitemShut {NoStop}%
\bibitem [{\citenamefont {Alasfar}\ \emph {et~al.}(2023)\citenamefont {Alasfar}
  \emph {et~al.}}]{Alasfar:2023xpc}%
  \BibitemOpen
  \bibfield  {author} {\bibinfo {author} {\bibfnamefont {L.}~\bibnamefont
  {Alasfar}} \emph {et~al.},\ }\bibfield  {title} {\bibinfo {title} {{Effective
  Field Theory descriptions of Higgs boson pair production}},\ }\href@noop {}
  {\  (\bibinfo {year} {2023})},\ \Eprint {https://arxiv.org/abs/2304.01968}
  {arXiv:2304.01968 [hep-ph]} \BibitemShut {NoStop}%
\bibitem [{\citenamefont {Brehmer}\ \emph {et~al.}(2020)\citenamefont
  {Brehmer}, \citenamefont {Kling}, \citenamefont {Espejo},\ and\ \citenamefont
  {Cranmer}}]{Brehmer:2019xox}%
  \BibitemOpen
  \bibfield  {author} {\bibinfo {author} {\bibfnamefont {J.}~\bibnamefont
  {Brehmer}}, \bibinfo {author} {\bibfnamefont {F.}~\bibnamefont {Kling}},
  \bibinfo {author} {\bibfnamefont {I.}~\bibnamefont {Espejo}},\ and\ \bibinfo
  {author} {\bibfnamefont {K.}~\bibnamefont {Cranmer}},\ }\bibfield  {title}
  {\bibinfo {title} {{MadMiner: Machine learning-based inference for particle
  physics}},\ }\href {https://doi.org/10.1007/s41781-020-0035-2} {\bibfield
  {journal} {\bibinfo  {journal} {Comput. Softw. Big Sci.}\ }\textbf {\bibinfo
  {volume} {4}},\ \bibinfo {pages} {3} (\bibinfo {year} {2020})},\ \Eprint
  {https://arxiv.org/abs/1907.10621} {arXiv:1907.10621 [hep-ph]} \BibitemShut
  {NoStop}%
\bibitem [{\citenamefont {Kong}\ \emph {et~al.}(2022)\citenamefont {Kong},
  \citenamefont {Matchev}, \citenamefont {Mrenna},\ and\ \citenamefont
  {Shyamsundar}}]{Kong:2022rnd}%
  \BibitemOpen
  \bibfield  {author} {\bibinfo {author} {\bibfnamefont {K.}~\bibnamefont
  {Kong}}, \bibinfo {author} {\bibfnamefont {K.~T.}\ \bibnamefont {Matchev}},
  \bibinfo {author} {\bibfnamefont {S.}~\bibnamefont {Mrenna}},\ and\ \bibinfo
  {author} {\bibfnamefont {P.}~\bibnamefont {Shyamsundar}},\ }\bibfield
  {title} {\bibinfo {title} {{New Machine Learning Techniques for
  Simulation-Based Inference: InferoStatic Nets, Kernel Score Estimation, and
  Kernel Likelihood Ratio Estimation}},\ }\href@noop {} {\  (\bibinfo {year}
  {2022})},\ \Eprint {https://arxiv.org/abs/2210.01680} {arXiv:2210.01680
  [stat.ML]} \BibitemShut {NoStop}%
\bibitem [{\citenamefont {Gomez~Ambrosio}\ \emph {et~al.}(2023)\citenamefont
  {Gomez~Ambrosio}, \citenamefont {ter Hoeve}, \citenamefont {Madigan},
  \citenamefont {Rojo},\ and\ \citenamefont {Sanz}}]{GomezAmbrosio:2022mpm}%
  \BibitemOpen
  \bibfield  {author} {\bibinfo {author} {\bibfnamefont {R.}~\bibnamefont
  {Gomez~Ambrosio}}, \bibinfo {author} {\bibfnamefont {J.}~\bibnamefont {ter
  Hoeve}}, \bibinfo {author} {\bibfnamefont {M.}~\bibnamefont {Madigan}},
  \bibinfo {author} {\bibfnamefont {J.}~\bibnamefont {Rojo}},\ and\ \bibinfo
  {author} {\bibfnamefont {V.}~\bibnamefont {Sanz}},\ }\bibfield  {title}
  {\bibinfo {title} {{Unbinned multivariate observables for global SMEFT
  analyses from machine learning}},\ }\href
  {https://doi.org/10.1007/JHEP03(2023)033} {\bibfield  {journal} {\bibinfo
  {journal} {JHEP}\ }\textbf {\bibinfo {volume} {03}},\ \bibinfo {pages}
  {033}},\ \Eprint {https://arxiv.org/abs/2211.02058} {arXiv:2211.02058
  [hep-ph]} \BibitemShut {NoStop}%
\bibitem [{\citenamefont {Brehmer}\ \emph {et~al.}(2019)\citenamefont
  {Brehmer}, \citenamefont {Dawson}, \citenamefont {Homiller}, \citenamefont
  {Kling},\ and\ \citenamefont {Plehn}}]{Brehmer:2019gmn}%
  \BibitemOpen
  \bibfield  {author} {\bibinfo {author} {\bibfnamefont {J.}~\bibnamefont
  {Brehmer}}, \bibinfo {author} {\bibfnamefont {S.}~\bibnamefont {Dawson}},
  \bibinfo {author} {\bibfnamefont {S.}~\bibnamefont {Homiller}}, \bibinfo
  {author} {\bibfnamefont {F.}~\bibnamefont {Kling}},\ and\ \bibinfo {author}
  {\bibfnamefont {T.}~\bibnamefont {Plehn}},\ }\bibfield  {title} {\bibinfo
  {title} {{Benchmarking simplified template cross sections in $WH$
  production}},\ }\href {https://doi.org/10.1007/JHEP11(2019)034} {\bibfield
  {journal} {\bibinfo  {journal} {JHEP}\ }\textbf {\bibinfo {volume} {11}},\
  \bibinfo {pages} {034}},\ \Eprint {https://arxiv.org/abs/1908.06980}
  {arXiv:1908.06980 [hep-ph]} \BibitemShut {NoStop}%
\bibitem [{\citenamefont {Ellis}\ \emph
  {et~al.}(2021{\natexlab{a}})\citenamefont {Ellis}, \citenamefont {Madigan},
  \citenamefont {Mimasu}, \citenamefont {Sanz},\ and\ \citenamefont
  {You}}]{Ellis:2020unq}%
  \BibitemOpen
  \bibfield  {author} {\bibinfo {author} {\bibfnamefont {J.}~\bibnamefont
  {Ellis}}, \bibinfo {author} {\bibfnamefont {M.}~\bibnamefont {Madigan}},
  \bibinfo {author} {\bibfnamefont {K.}~\bibnamefont {Mimasu}}, \bibinfo
  {author} {\bibfnamefont {V.}~\bibnamefont {Sanz}},\ and\ \bibinfo {author}
  {\bibfnamefont {T.}~\bibnamefont {You}},\ }\bibfield  {title} {\bibinfo
  {title} {{Top, Higgs, Diboson and Electroweak Fit to the Standard Model
  Effective Field Theory}},\ }\href {https://doi.org/10.1007/JHEP04(2021)279}
  {\bibfield  {journal} {\bibinfo  {journal} {JHEP}\ }\textbf {\bibinfo
  {volume} {04}},\ \bibinfo {pages} {279}},\ \Eprint
  {https://arxiv.org/abs/2012.02779} {arXiv:2012.02779 [hep-ph]} \BibitemShut
  {NoStop}%
\bibitem [{\citenamefont {Ethier}\ \emph {et~al.}(2021)\citenamefont {Ethier},
  \citenamefont {Magni}, \citenamefont {Maltoni}, \citenamefont {Mantani},
  \citenamefont {Nocera}, \citenamefont {Rojo}, \citenamefont {Slade},
  \citenamefont {Vryonidou},\ and\ \citenamefont {Zhang}}]{Ethier:2021bye}%
  \BibitemOpen
  \bibfield  {author} {\bibinfo {author} {\bibfnamefont {J.~J.}\ \bibnamefont
  {Ethier}}, \bibinfo {author} {\bibfnamefont {G.}~\bibnamefont {Magni}},
  \bibinfo {author} {\bibfnamefont {F.}~\bibnamefont {Maltoni}}, \bibinfo
  {author} {\bibfnamefont {L.}~\bibnamefont {Mantani}}, \bibinfo {author}
  {\bibfnamefont {E.~R.}\ \bibnamefont {Nocera}}, \bibinfo {author}
  {\bibfnamefont {J.}~\bibnamefont {Rojo}}, \bibinfo {author} {\bibfnamefont
  {E.}~\bibnamefont {Slade}}, \bibinfo {author} {\bibfnamefont
  {E.}~\bibnamefont {Vryonidou}},\ and\ \bibinfo {author} {\bibfnamefont
  {C.}~\bibnamefont {Zhang}} (\bibinfo {collaboration} {SMEFiT}),\ }\bibfield
  {title} {\bibinfo {title} {{Combined SMEFT interpretation of Higgs, diboson,
  and top quark data from the LHC}},\ }\href
  {https://doi.org/10.1007/JHEP11(2021)089} {\bibfield  {journal} {\bibinfo
  {journal} {JHEP}\ }\textbf {\bibinfo {volume} {11}},\ \bibinfo {pages}
  {089}},\ \Eprint {https://arxiv.org/abs/2105.00006} {arXiv:2105.00006
  [hep-ph]} \BibitemShut {NoStop}%
\bibitem [{\citenamefont {Elmer}\ \emph {et~al.}(2023)\citenamefont {Elmer},
  \citenamefont {Madigan}, \citenamefont {Plehn},\ and\ \citenamefont
  {Schmal}}]{Elmer:2023wtr}%
  \BibitemOpen
  \bibfield  {author} {\bibinfo {author} {\bibfnamefont {N.}~\bibnamefont
  {Elmer}}, \bibinfo {author} {\bibfnamefont {M.}~\bibnamefont {Madigan}},
  \bibinfo {author} {\bibfnamefont {T.}~\bibnamefont {Plehn}},\ and\ \bibinfo
  {author} {\bibfnamefont {N.}~\bibnamefont {Schmal}},\ }\bibfield  {title}
  {\bibinfo {title} {{Staying on Top of SMEFT-Likelihood Analyses}},\
  }\href@noop {} {\  (\bibinfo {year} {2023})},\ \Eprint
  {https://arxiv.org/abs/2312.12502} {arXiv:2312.12502 [hep-ph]} \BibitemShut
  {NoStop}%
\bibitem [{\citenamefont {Borowka}\ \emph {et~al.}(2019)\citenamefont
  {Borowka}, \citenamefont {Duhr}, \citenamefont {Maltoni}, \citenamefont
  {Pagani}, \citenamefont {Shivaji},\ and\ \citenamefont
  {Zhao}}]{Borowka:2018pxx}%
  \BibitemOpen
  \bibfield  {author} {\bibinfo {author} {\bibfnamefont {S.}~\bibnamefont
  {Borowka}}, \bibinfo {author} {\bibfnamefont {C.}~\bibnamefont {Duhr}},
  \bibinfo {author} {\bibfnamefont {F.}~\bibnamefont {Maltoni}}, \bibinfo
  {author} {\bibfnamefont {D.}~\bibnamefont {Pagani}}, \bibinfo {author}
  {\bibfnamefont {A.}~\bibnamefont {Shivaji}},\ and\ \bibinfo {author}
  {\bibfnamefont {X.}~\bibnamefont {Zhao}},\ }\bibfield  {title} {\bibinfo
  {title} {{Probing the scalar potential via double Higgs boson production at
  hadron colliders}},\ }\href {https://doi.org/10.1007/JHEP04(2019)016}
  {\bibfield  {journal} {\bibinfo  {journal} {JHEP}\ }\textbf {\bibinfo
  {volume} {04}},\ \bibinfo {pages} {016}},\ \Eprint
  {https://arxiv.org/abs/1811.12366} {arXiv:1811.12366 [hep-ph]} \BibitemShut
  {NoStop}%
\bibitem [{\citenamefont {Murphy}(2020)}]{Murphy:2020rsh}%
  \BibitemOpen
  \bibfield  {author} {\bibinfo {author} {\bibfnamefont {C.~W.}\ \bibnamefont
  {Murphy}},\ }\bibfield  {title} {\bibinfo {title} {{Dimension-8 operators in
  the Standard Model Effective Field Theory}},\ }\href
  {https://doi.org/10.1007/JHEP10(2020)174} {\bibfield  {journal} {\bibinfo
  {journal} {JHEP}\ }\textbf {\bibinfo {volume} {10}},\ \bibinfo {pages}
  {174}},\ \Eprint {https://arxiv.org/abs/2005.00059} {arXiv:2005.00059
  [hep-ph]} \BibitemShut {NoStop}%
\bibitem [{\citenamefont {Biek\"otter}\ \emph {et~al.}(2019)\citenamefont
  {Biek\"otter}, \citenamefont {Gon\c{c}alves}, \citenamefont {Plehn},
  \citenamefont {Takeuchi},\ and\ \citenamefont {Zerwas}}]{Biekotter:2018jzu}%
  \BibitemOpen
  \bibfield  {author} {\bibinfo {author} {\bibfnamefont {A.}~\bibnamefont
  {Biek\"otter}}, \bibinfo {author} {\bibfnamefont {D.}~\bibnamefont
  {Gon\c{c}alves}}, \bibinfo {author} {\bibfnamefont {T.}~\bibnamefont
  {Plehn}}, \bibinfo {author} {\bibfnamefont {M.}~\bibnamefont {Takeuchi}},\
  and\ \bibinfo {author} {\bibfnamefont {D.}~\bibnamefont {Zerwas}},\
  }\bibfield  {title} {\bibinfo {title} {{The global Higgs picture at 27
  TeV}},\ }\href {https://doi.org/10.21468/SciPostPhys.6.2.024} {\bibfield
  {journal} {\bibinfo  {journal} {SciPost Phys.}\ }\textbf {\bibinfo {volume}
  {6}},\ \bibinfo {pages} {024} (\bibinfo {year} {2019})},\ \Eprint
  {https://arxiv.org/abs/1811.08401} {arXiv:1811.08401 [hep-ph]} \BibitemShut
  {NoStop}%
\bibitem [{\citenamefont {Alwall}\ \emph {et~al.}(2014)\citenamefont {Alwall},
  \citenamefont {Frederix}, \citenamefont {Frixione}, \citenamefont {Hirschi},
  \citenamefont {Maltoni}, \citenamefont {Mattelaer}, \citenamefont {Shao},
  \citenamefont {Stelzer}, \citenamefont {Torrielli},\ and\ \citenamefont
  {Zaro}}]{Alwall_2014}%
  \BibitemOpen
  \bibfield  {author} {\bibinfo {author} {\bibfnamefont {J.}~\bibnamefont
  {Alwall}}, \bibinfo {author} {\bibfnamefont {R.}~\bibnamefont {Frederix}},
  \bibinfo {author} {\bibfnamefont {S.}~\bibnamefont {Frixione}}, \bibinfo
  {author} {\bibfnamefont {V.}~\bibnamefont {Hirschi}}, \bibinfo {author}
  {\bibfnamefont {F.}~\bibnamefont {Maltoni}}, \bibinfo {author} {\bibfnamefont
  {O.}~\bibnamefont {Mattelaer}}, \bibinfo {author} {\bibfnamefont {H.-S.}\
  \bibnamefont {Shao}}, \bibinfo {author} {\bibfnamefont {T.}~\bibnamefont
  {Stelzer}}, \bibinfo {author} {\bibfnamefont {P.}~\bibnamefont {Torrielli}},\
  and\ \bibinfo {author} {\bibfnamefont {M.}~\bibnamefont {Zaro}},\ }\bibfield
  {title} {\bibinfo {title} {The automated computation of tree-level and
  next-to-leading order differential cross sections, and their matching to
  parton shower simulations},\ }\bibfield  {journal} {\bibinfo  {journal}
  {Journal of High Energy Physics}\ }\textbf {\bibinfo {volume} {2014}},\ \href
  {https://doi.org/10.1007/jhep07(2014)079} {10.1007/jhep07(2014)079} (\bibinfo
  {year} {2014})\BibitemShut {NoStop}%
\bibitem [{\citenamefont {Degrande}\ \emph {et~al.}(2021)\citenamefont
  {Degrande}, \citenamefont {Durieux}, \citenamefont {Maltoni}, \citenamefont
  {Mimasu}, \citenamefont {Vryonidou},\ and\ \citenamefont
  {Zhang}}]{Degrande:2020evl}%
  \BibitemOpen
  \bibfield  {author} {\bibinfo {author} {\bibfnamefont {C.}~\bibnamefont
  {Degrande}}, \bibinfo {author} {\bibfnamefont {G.}~\bibnamefont {Durieux}},
  \bibinfo {author} {\bibfnamefont {F.}~\bibnamefont {Maltoni}}, \bibinfo
  {author} {\bibfnamefont {K.}~\bibnamefont {Mimasu}}, \bibinfo {author}
  {\bibfnamefont {E.}~\bibnamefont {Vryonidou}},\ and\ \bibinfo {author}
  {\bibfnamefont {C.}~\bibnamefont {Zhang}},\ }\bibfield  {title} {\bibinfo
  {title} {{Automated one-loop computations in the standard model effective
  field theory}},\ }\href {https://doi.org/10.1103/PhysRevD.103.096024}
  {\bibfield  {journal} {\bibinfo  {journal} {Phys. Rev. D}\ }\textbf {\bibinfo
  {volume} {103}},\ \bibinfo {pages} {096024} (\bibinfo {year} {2021})},\
  \Eprint {https://arxiv.org/abs/2008.11743} {arXiv:2008.11743 [hep-ph]}
  \BibitemShut {NoStop}%
\bibitem [{\citenamefont {Plehn}\ \emph {et~al.}(1996)\citenamefont {Plehn},
  \citenamefont {Spira},\ and\ \citenamefont {Zerwas}}]{Plehn:1996wb}%
  \BibitemOpen
  \bibfield  {author} {\bibinfo {author} {\bibfnamefont {T.}~\bibnamefont
  {Plehn}}, \bibinfo {author} {\bibfnamefont {M.}~\bibnamefont {Spira}},\ and\
  \bibinfo {author} {\bibfnamefont {P.~M.}\ \bibnamefont {Zerwas}},\ }\bibfield
   {title} {\bibinfo {title} {{Pair production of neutral Higgs particles in
  gluon-gluon collisions}},\ }\href
  {https://doi.org/10.1016/0550-3213(96)00418-X} {\bibfield  {journal}
  {\bibinfo  {journal} {Nucl. Phys. B}\ }\textbf {\bibinfo {volume} {479}},\
  \bibinfo {pages} {46} (\bibinfo {year} {1996})},\ \bibinfo {note} {[Erratum:
  Nucl.Phys.B 531, 655--655 (1998)]},\ \Eprint
  {https://arxiv.org/abs/hep-ph/9603205} {arXiv:hep-ph/9603205} \BibitemShut
  {NoStop}%
\bibitem [{\citenamefont {Plehn}(2015)}]{Plehn:2015dqa}%
  \BibitemOpen
  \bibfield  {author} {\bibinfo {author} {\bibfnamefont {T.}~\bibnamefont
  {Plehn}},\ }\href {https://doi.org/10.1007/978-3-319-05942-6} {\emph
  {\bibinfo {title} {{Lectures on LHC Physics}}}}\ (\bibinfo {year}
  {2015})\BibitemShut {NoStop}%
\bibitem [{\citenamefont {Buschmann}\ \emph {et~al.}(2015)\citenamefont
  {Buschmann}, \citenamefont {Goncalves}, \citenamefont {Kuttimalai},
  \citenamefont {Schonherr}, \citenamefont {Krauss},\ and\ \citenamefont
  {Plehn}}]{Buschmann:2014sia}%
  \BibitemOpen
  \bibfield  {author} {\bibinfo {author} {\bibfnamefont {M.}~\bibnamefont
  {Buschmann}}, \bibinfo {author} {\bibfnamefont {D.}~\bibnamefont
  {Goncalves}}, \bibinfo {author} {\bibfnamefont {S.}~\bibnamefont
  {Kuttimalai}}, \bibinfo {author} {\bibfnamefont {M.}~\bibnamefont
  {Schonherr}}, \bibinfo {author} {\bibfnamefont {F.}~\bibnamefont {Krauss}},\
  and\ \bibinfo {author} {\bibfnamefont {T.}~\bibnamefont {Plehn}},\ }\bibfield
   {title} {\bibinfo {title} {{Mass Effects in the Higgs-Gluon Coupling:
  Boosted vs Off-Shell Production}},\ }\href
  {https://doi.org/10.1007/JHEP02(2015)038} {\bibfield  {journal} {\bibinfo
  {journal} {JHEP}\ }\textbf {\bibinfo {volume} {02}},\ \bibinfo {pages}
  {038}},\ \Eprint {https://arxiv.org/abs/1410.5806} {arXiv:1410.5806 [hep-ph]}
  \BibitemShut {NoStop}%
\bibitem [{\citenamefont {Ellis}\ \emph
  {et~al.}(2021{\natexlab{b}})\citenamefont {Ellis}, \citenamefont {Madigan},
  \citenamefont {Mimasu}, \citenamefont {Sanz},\ and\ \citenamefont
  {You}}]{Ellis_2021}%
  \BibitemOpen
  \bibfield  {author} {\bibinfo {author} {\bibfnamefont {J.}~\bibnamefont
  {Ellis}}, \bibinfo {author} {\bibfnamefont {M.}~\bibnamefont {Madigan}},
  \bibinfo {author} {\bibfnamefont {K.}~\bibnamefont {Mimasu}}, \bibinfo
  {author} {\bibfnamefont {V.}~\bibnamefont {Sanz}},\ and\ \bibinfo {author}
  {\bibfnamefont {T.}~\bibnamefont {You}},\ }\bibfield  {title} {\bibinfo
  {title} {Top, higgs, diboson and electroweak fit to the standard model
  effective field theory},\ }\bibfield  {journal} {\bibinfo  {journal} {Journal
  of High Energy Physics}\ }\textbf {\bibinfo {volume} {2021}},\ \href
  {https://doi.org/10.1007/jhep04(2021)279} {10.1007/jhep04(2021)279} (\bibinfo
  {year} {2021}{\natexlab{b}})\BibitemShut {NoStop}%
\bibitem [{\citenamefont {Brivio}\ \emph {et~al.}(2024)\citenamefont {Brivio},
  \citenamefont {Bruggisser}, \citenamefont {Elmer}, \citenamefont {Geoffray},
  \citenamefont {Luchmann},\ and\ \citenamefont {Plehn}}]{Brivio:2022hrb}%
  \BibitemOpen
  \bibfield  {author} {\bibinfo {author} {\bibfnamefont {I.}~\bibnamefont
  {Brivio}}, \bibinfo {author} {\bibfnamefont {S.}~\bibnamefont {Bruggisser}},
  \bibinfo {author} {\bibfnamefont {N.}~\bibnamefont {Elmer}}, \bibinfo
  {author} {\bibfnamefont {E.}~\bibnamefont {Geoffray}}, \bibinfo {author}
  {\bibfnamefont {M.}~\bibnamefont {Luchmann}},\ and\ \bibinfo {author}
  {\bibfnamefont {T.}~\bibnamefont {Plehn}},\ }\bibfield  {title} {\bibinfo
  {title} {{To profile or to marginalize - A SMEFT case study}},\ }\href
  {https://doi.org/10.21468/SciPostPhys.16.1.035} {\bibfield  {journal}
  {\bibinfo  {journal} {SciPost Phys.}\ }\textbf {\bibinfo {volume} {16}},\
  \bibinfo {pages} {035} (\bibinfo {year} {2024})},\ \Eprint
  {https://arxiv.org/abs/2208.08454} {arXiv:2208.08454 [hep-ph]} \BibitemShut
  {NoStop}%
\bibitem [{\citenamefont {Baur}\ \emph {et~al.}(2003)\citenamefont {Baur},
  \citenamefont {Plehn},\ and\ \citenamefont {Rainwater}}]{Baur:2003gpa}%
  \BibitemOpen
  \bibfield  {author} {\bibinfo {author} {\bibfnamefont {U.}~\bibnamefont
  {Baur}}, \bibinfo {author} {\bibfnamefont {T.}~\bibnamefont {Plehn}},\ and\
  \bibinfo {author} {\bibfnamefont {D.~L.}\ \bibnamefont {Rainwater}},\
  }\bibfield  {title} {\bibinfo {title} {{Examining the Higgs boson potential
  at lepton and hadron colliders: A Comparative analysis}},\ }\href
  {https://doi.org/10.1103/PhysRevD.68.033001} {\bibfield  {journal} {\bibinfo
  {journal} {Phys. Rev. D}\ }\textbf {\bibinfo {volume} {68}},\ \bibinfo
  {pages} {033001} (\bibinfo {year} {2003})},\ \Eprint
  {https://arxiv.org/abs/hep-ph/0304015} {arXiv:hep-ph/0304015} \BibitemShut
  {NoStop}%
\bibitem [{\citenamefont {Dolan}\ \emph {et~al.}(2012)\citenamefont {Dolan},
  \citenamefont {Englert},\ and\ \citenamefont {Spannowsky}}]{Dolan:2012rv}%
  \BibitemOpen
  \bibfield  {author} {\bibinfo {author} {\bibfnamefont {M.~J.}\ \bibnamefont
  {Dolan}}, \bibinfo {author} {\bibfnamefont {C.}~\bibnamefont {Englert}},\
  and\ \bibinfo {author} {\bibfnamefont {M.}~\bibnamefont {Spannowsky}},\
  }\bibfield  {title} {\bibinfo {title} {{Higgs self-coupling measurements at
  the LHC}},\ }\href {https://doi.org/10.1007/JHEP10(2012)112} {\bibfield
  {journal} {\bibinfo  {journal} {JHEP}\ }\textbf {\bibinfo {volume} {10}},\
  \bibinfo {pages} {112}},\ \Eprint {https://arxiv.org/abs/1206.5001}
  {arXiv:1206.5001 [hep-ph]} \BibitemShut {NoStop}%
\bibitem [{\citenamefont {Baglio}\ \emph {et~al.}(2013)\citenamefont {Baglio},
  \citenamefont {Djouadi}, \citenamefont {Gr\"ober}, \citenamefont
  {M\"uhlleitner}, \citenamefont {Quevillon},\ and\ \citenamefont
  {Spira}}]{Baglio:2012np}%
  \BibitemOpen
  \bibfield  {author} {\bibinfo {author} {\bibfnamefont {J.}~\bibnamefont
  {Baglio}}, \bibinfo {author} {\bibfnamefont {A.}~\bibnamefont {Djouadi}},
  \bibinfo {author} {\bibfnamefont {R.}~\bibnamefont {Gr\"ober}}, \bibinfo
  {author} {\bibfnamefont {M.~M.}\ \bibnamefont {M\"uhlleitner}}, \bibinfo
  {author} {\bibfnamefont {J.}~\bibnamefont {Quevillon}},\ and\ \bibinfo
  {author} {\bibfnamefont {M.}~\bibnamefont {Spira}},\ }\bibfield  {title}
  {\bibinfo {title} {{The measurement of the Higgs self-coupling at the LHC:
  theoretical status}},\ }\href {https://doi.org/10.1007/JHEP04(2013)151}
  {\bibfield  {journal} {\bibinfo  {journal} {JHEP}\ }\textbf {\bibinfo
  {volume} {04}},\ \bibinfo {pages} {151}},\ \Eprint
  {https://arxiv.org/abs/1212.5581} {arXiv:1212.5581 [hep-ph]} \BibitemShut
  {NoStop}%
\bibitem [{\citenamefont {Abada}\ \emph
  {et~al.}(2019{\natexlab{b}})\citenamefont {Abada} \emph {et~al.}}]{FCC:2019}%
  \BibitemOpen
  \bibfield  {author} {\bibinfo {author} {\bibfnamefont {A.}~\bibnamefont
  {Abada}} \emph {et~al.} (\bibinfo {collaboration} {FCC}),\ }\bibfield
  {title} {\bibinfo {title} {{FCC-hh: The Hadron Collider}},\ }\href
  {https://doi.org/10.1140/epjst/e2019-900087-0} {\bibfield  {journal}
  {\bibinfo  {journal} {Eur. Phys. J. Spec. Top.}\ }\textbf {\bibinfo {volume}
  {228}},\ \bibinfo {pages} {755} (\bibinfo {year}
  {2019}{\natexlab{b}})}\BibitemShut {NoStop}%
\bibitem [{\citenamefont {Dawson}\ \emph {et~al.}(1998)\citenamefont {Dawson},
  \citenamefont {Dittmaier},\ and\ \citenamefont {Spira}}]{Dawson:1998py}%
  \BibitemOpen
  \bibfield  {author} {\bibinfo {author} {\bibfnamefont {S.}~\bibnamefont
  {Dawson}}, \bibinfo {author} {\bibfnamefont {S.}~\bibnamefont {Dittmaier}},\
  and\ \bibinfo {author} {\bibfnamefont {M.}~\bibnamefont {Spira}},\ }\bibfield
   {title} {\bibinfo {title} {{Neutral Higgs boson pair production at hadron
  colliders: QCD corrections}},\ }\href
  {https://doi.org/10.1103/PhysRevD.58.115012} {\bibfield  {journal} {\bibinfo
  {journal} {Phys. Rev. D}\ }\textbf {\bibinfo {volume} {58}},\ \bibinfo
  {pages} {115012} (\bibinfo {year} {1998})},\ \Eprint
  {https://arxiv.org/abs/hep-ph/9805244} {arXiv:hep-ph/9805244} \BibitemShut
  {NoStop}%
\bibitem [{\citenamefont {Grigo}\ \emph {et~al.}(2013)\citenamefont {Grigo},
  \citenamefont {Hoff}, \citenamefont {Melnikov},\ and\ \citenamefont
  {Steinhauser}}]{Grigo:2013rya}%
  \BibitemOpen
  \bibfield  {author} {\bibinfo {author} {\bibfnamefont {J.}~\bibnamefont
  {Grigo}}, \bibinfo {author} {\bibfnamefont {J.}~\bibnamefont {Hoff}},
  \bibinfo {author} {\bibfnamefont {K.}~\bibnamefont {Melnikov}},\ and\
  \bibinfo {author} {\bibfnamefont {M.}~\bibnamefont {Steinhauser}},\
  }\bibfield  {title} {\bibinfo {title} {{On the Higgs boson pair production at
  the LHC}},\ }\href {https://doi.org/10.1016/j.nuclphysb.2013.06.024}
  {\bibfield  {journal} {\bibinfo  {journal} {Nucl. Phys. B}\ }\textbf
  {\bibinfo {volume} {875}},\ \bibinfo {pages} {1} (\bibinfo {year} {2013})},\
  \Eprint {https://arxiv.org/abs/1305.7340} {arXiv:1305.7340 [hep-ph]}
  \BibitemShut {NoStop}%
\bibitem [{\citenamefont {Maltoni}\ \emph {et~al.}(2014)\citenamefont
  {Maltoni}, \citenamefont {Vryonidou},\ and\ \citenamefont
  {Zaro}}]{Maltoni:2014eza}%
  \BibitemOpen
  \bibfield  {author} {\bibinfo {author} {\bibfnamefont {F.}~\bibnamefont
  {Maltoni}}, \bibinfo {author} {\bibfnamefont {E.}~\bibnamefont {Vryonidou}},\
  and\ \bibinfo {author} {\bibfnamefont {M.}~\bibnamefont {Zaro}},\ }\bibfield
  {title} {\bibinfo {title} {{Top-quark mass effects in double and triple Higgs
  production in gluon-gluon fusion at NLO}},\ }\href
  {https://doi.org/10.1007/JHEP11(2014)079} {\bibfield  {journal} {\bibinfo
  {journal} {JHEP}\ }\textbf {\bibinfo {volume} {11}},\ \bibinfo {pages}
  {079}},\ \Eprint {https://arxiv.org/abs/1408.6542} {arXiv:1408.6542 [hep-ph]}
  \BibitemShut {NoStop}%
\bibitem [{\citenamefont {Borowka}\ \emph
  {et~al.}(2016{\natexlab{a}})\citenamefont {Borowka}, \citenamefont {Greiner},
  \citenamefont {Heinrich}, \citenamefont {Jones}, \citenamefont {Kerner},
  \citenamefont {Schlenk},\ and\ \citenamefont {Zirke}}]{Borowka:2016ypz}%
  \BibitemOpen
  \bibfield  {author} {\bibinfo {author} {\bibfnamefont {S.}~\bibnamefont
  {Borowka}}, \bibinfo {author} {\bibfnamefont {N.}~\bibnamefont {Greiner}},
  \bibinfo {author} {\bibfnamefont {G.}~\bibnamefont {Heinrich}}, \bibinfo
  {author} {\bibfnamefont {S.~P.}\ \bibnamefont {Jones}}, \bibinfo {author}
  {\bibfnamefont {M.}~\bibnamefont {Kerner}}, \bibinfo {author} {\bibfnamefont
  {J.}~\bibnamefont {Schlenk}},\ and\ \bibinfo {author} {\bibfnamefont
  {T.}~\bibnamefont {Zirke}},\ }\bibfield  {title} {\bibinfo {title} {{Full top
  quark mass dependence in Higgs boson pair production at NLO}},\ }\href
  {https://doi.org/10.1007/JHEP10(2016)107} {\bibfield  {journal} {\bibinfo
  {journal} {JHEP}\ }\textbf {\bibinfo {volume} {10}},\ \bibinfo {pages}
  {107}},\ \Eprint {https://arxiv.org/abs/1608.04798} {arXiv:1608.04798
  [hep-ph]} \BibitemShut {NoStop}%
\bibitem [{\citenamefont {Borowka}\ \emph
  {et~al.}(2016{\natexlab{b}})\citenamefont {Borowka}, \citenamefont {Greiner},
  \citenamefont {Heinrich}, \citenamefont {Jones}, \citenamefont {Kerner},
  \citenamefont {Schlenk}, \citenamefont {Schubert},\ and\ \citenamefont
  {Zirke}}]{Borowka:2016ehy}%
  \BibitemOpen
  \bibfield  {author} {\bibinfo {author} {\bibfnamefont {S.}~\bibnamefont
  {Borowka}}, \bibinfo {author} {\bibfnamefont {N.}~\bibnamefont {Greiner}},
  \bibinfo {author} {\bibfnamefont {G.}~\bibnamefont {Heinrich}}, \bibinfo
  {author} {\bibfnamefont {S.~P.}\ \bibnamefont {Jones}}, \bibinfo {author}
  {\bibfnamefont {M.}~\bibnamefont {Kerner}}, \bibinfo {author} {\bibfnamefont
  {J.}~\bibnamefont {Schlenk}}, \bibinfo {author} {\bibfnamefont
  {U.}~\bibnamefont {Schubert}},\ and\ \bibinfo {author} {\bibfnamefont
  {T.}~\bibnamefont {Zirke}},\ }\bibfield  {title} {\bibinfo {title} {{Higgs
  Boson Pair Production in Gluon Fusion at Next-to-Leading Order with Full
  Top-Quark Mass Dependence}},\ }\href
  {https://doi.org/10.1103/PhysRevLett.117.079901} {\bibfield  {journal}
  {\bibinfo  {journal} {Phys. Rev. Lett.}\ }\textbf {\bibinfo {volume} {117}},\
  \bibinfo {pages} {012001} (\bibinfo {year} {2016}{\natexlab{b}})},\ \bibinfo
  {note} {[Erratum: Phys.Rev.Lett. 117, 079901 (2016)]},\ \Eprint
  {https://arxiv.org/abs/1604.06447} {arXiv:1604.06447 [hep-ph]} \BibitemShut
  {NoStop}%
\bibitem [{\citenamefont {Baglio}\ \emph {et~al.}(2019)\citenamefont {Baglio},
  \citenamefont {Campanario}, \citenamefont {Glaus}, \citenamefont
  {M\"uhlleitner}, \citenamefont {Spira},\ and\ \citenamefont
  {Streicher}}]{Baglio:2018lrj}%
  \BibitemOpen
  \bibfield  {author} {\bibinfo {author} {\bibfnamefont {J.}~\bibnamefont
  {Baglio}}, \bibinfo {author} {\bibfnamefont {F.}~\bibnamefont {Campanario}},
  \bibinfo {author} {\bibfnamefont {S.}~\bibnamefont {Glaus}}, \bibinfo
  {author} {\bibfnamefont {M.}~\bibnamefont {M\"uhlleitner}}, \bibinfo {author}
  {\bibfnamefont {M.}~\bibnamefont {Spira}},\ and\ \bibinfo {author}
  {\bibfnamefont {J.}~\bibnamefont {Streicher}},\ }\bibfield  {title} {\bibinfo
  {title} {{Gluon fusion into Higgs pairs at NLO QCD and the top mass
  scheme}},\ }\href {https://doi.org/10.1140/epjc/s10052-019-6973-3} {\bibfield
   {journal} {\bibinfo  {journal} {Eur. Phys. J. C}\ }\textbf {\bibinfo
  {volume} {79}},\ \bibinfo {pages} {459} (\bibinfo {year} {2019})},\ \Eprint
  {https://arxiv.org/abs/1811.05692} {arXiv:1811.05692 [hep-ph]} \BibitemShut
  {NoStop}%
\bibitem [{\citenamefont {Baglio}\ \emph {et~al.}(2020)\citenamefont {Baglio},
  \citenamefont {Campanario}, \citenamefont {Glaus}, \citenamefont
  {M\"uhlleitner}, \citenamefont {Ronca}, \citenamefont {Spira},\ and\
  \citenamefont {Streicher}}]{Baglio:2020ini}%
  \BibitemOpen
  \bibfield  {author} {\bibinfo {author} {\bibfnamefont {J.}~\bibnamefont
  {Baglio}}, \bibinfo {author} {\bibfnamefont {F.}~\bibnamefont {Campanario}},
  \bibinfo {author} {\bibfnamefont {S.}~\bibnamefont {Glaus}}, \bibinfo
  {author} {\bibfnamefont {M.}~\bibnamefont {M\"uhlleitner}}, \bibinfo {author}
  {\bibfnamefont {J.}~\bibnamefont {Ronca}}, \bibinfo {author} {\bibfnamefont
  {M.}~\bibnamefont {Spira}},\ and\ \bibinfo {author} {\bibfnamefont
  {J.}~\bibnamefont {Streicher}},\ }\bibfield  {title} {\bibinfo {title}
  {{Higgs-Pair Production via Gluon Fusion at Hadron Colliders: NLO QCD
  Corrections}},\ }\href {https://doi.org/10.1007/JHEP04(2020)181} {\bibfield
  {journal} {\bibinfo  {journal} {JHEP}\ }\textbf {\bibinfo {volume} {04}},\
  \bibinfo {pages} {181}},\ \Eprint {https://arxiv.org/abs/2003.03227}
  {arXiv:2003.03227 [hep-ph]} \BibitemShut {NoStop}%
\bibitem [{\citenamefont {de~Florian}\ and\ \citenamefont
  {Mazzitelli}(2015)}]{deFlorian:2015moa}%
  \BibitemOpen
  \bibfield  {author} {\bibinfo {author} {\bibfnamefont {D.}~\bibnamefont
  {de~Florian}}\ and\ \bibinfo {author} {\bibfnamefont {J.}~\bibnamefont
  {Mazzitelli}},\ }\bibfield  {title} {\bibinfo {title} {{Higgs pair production
  at next-to-next-to-leading logarithmic accuracy at the LHC}},\ }\href
  {https://doi.org/10.1007/JHEP09(2015)053} {\bibfield  {journal} {\bibinfo
  {journal} {JHEP}\ }\textbf {\bibinfo {volume} {09}},\ \bibinfo {pages}
  {053}},\ \Eprint {https://arxiv.org/abs/1505.07122} {arXiv:1505.07122
  [hep-ph]} \BibitemShut {NoStop}%
\bibitem [{\citenamefont {Grigo}\ \emph {et~al.}(2015)\citenamefont {Grigo},
  \citenamefont {Hoff},\ and\ \citenamefont {Steinhauser}}]{Grigo:2015dia}%
  \BibitemOpen
  \bibfield  {author} {\bibinfo {author} {\bibfnamefont {J.}~\bibnamefont
  {Grigo}}, \bibinfo {author} {\bibfnamefont {J.}~\bibnamefont {Hoff}},\ and\
  \bibinfo {author} {\bibfnamefont {M.}~\bibnamefont {Steinhauser}},\
  }\bibfield  {title} {\bibinfo {title} {{Higgs boson pair production: top
  quark mass effects at NLO and NNLO}},\ }\href
  {https://doi.org/10.1016/j.nuclphysb.2015.09.012} {\bibfield  {journal}
  {\bibinfo  {journal} {Nucl. Phys. B}\ }\textbf {\bibinfo {volume} {900}},\
  \bibinfo {pages} {412} (\bibinfo {year} {2015})},\ \Eprint
  {https://arxiv.org/abs/1508.00909} {arXiv:1508.00909 [hep-ph]} \BibitemShut
  {NoStop}%
\bibitem [{\citenamefont {de~Florian}\ \emph
  {et~al.}(2016{\natexlab{a}})\citenamefont {de~Florian}, \citenamefont
  {Grazzini}, \citenamefont {Hanga}, \citenamefont {Kallweit}, \citenamefont
  {Lindert}, \citenamefont {Maierh\"ofer}, \citenamefont {Mazzitelli},\ and\
  \citenamefont {Rathlev}}]{deFlorian:2016uhr}%
  \BibitemOpen
  \bibfield  {author} {\bibinfo {author} {\bibfnamefont {D.}~\bibnamefont
  {de~Florian}}, \bibinfo {author} {\bibfnamefont {M.}~\bibnamefont
  {Grazzini}}, \bibinfo {author} {\bibfnamefont {C.}~\bibnamefont {Hanga}},
  \bibinfo {author} {\bibfnamefont {S.}~\bibnamefont {Kallweit}}, \bibinfo
  {author} {\bibfnamefont {J.~M.}\ \bibnamefont {Lindert}}, \bibinfo {author}
  {\bibfnamefont {P.}~\bibnamefont {Maierh\"ofer}}, \bibinfo {author}
  {\bibfnamefont {J.}~\bibnamefont {Mazzitelli}},\ and\ \bibinfo {author}
  {\bibfnamefont {D.}~\bibnamefont {Rathlev}},\ }\bibfield  {title} {\bibinfo
  {title} {{Differential Higgs Boson Pair Production at Next-to-Next-to-Leading
  Order in QCD}},\ }\href {https://doi.org/10.1007/JHEP09(2016)151} {\bibfield
  {journal} {\bibinfo  {journal} {JHEP}\ }\textbf {\bibinfo {volume} {09}},\
  \bibinfo {pages} {151}},\ \Eprint {https://arxiv.org/abs/1606.09519}
  {arXiv:1606.09519 [hep-ph]} \BibitemShut {NoStop}%
\bibitem [{\citenamefont {Chen}\ \emph
  {et~al.}(2020{\natexlab{a}})\citenamefont {Chen}, \citenamefont {Li},
  \citenamefont {Shao},\ and\ \citenamefont {Wang}}]{Chen:2019lzz}%
  \BibitemOpen
  \bibfield  {author} {\bibinfo {author} {\bibfnamefont {L.-B.}\ \bibnamefont
  {Chen}}, \bibinfo {author} {\bibfnamefont {H.~T.}\ \bibnamefont {Li}},
  \bibinfo {author} {\bibfnamefont {H.-S.}\ \bibnamefont {Shao}},\ and\
  \bibinfo {author} {\bibfnamefont {J.}~\bibnamefont {Wang}},\ }\bibfield
  {title} {\bibinfo {title} {{Higgs boson pair production via gluon fusion at
  N$^3$LO in QCD}},\ }\href {https://doi.org/10.1016/j.physletb.2020.135292}
  {\bibfield  {journal} {\bibinfo  {journal} {Phys. Lett. B}\ }\textbf
  {\bibinfo {volume} {803}},\ \bibinfo {pages} {135292} (\bibinfo {year}
  {2020}{\natexlab{a}})},\ \Eprint {https://arxiv.org/abs/1909.06808}
  {arXiv:1909.06808 [hep-ph]} \BibitemShut {NoStop}%
\bibitem [{\citenamefont {Chen}\ \emph
  {et~al.}(2020{\natexlab{b}})\citenamefont {Chen}, \citenamefont {Li},
  \citenamefont {Shao},\ and\ \citenamefont {Wang}}]{Chen:2019fhs}%
  \BibitemOpen
  \bibfield  {author} {\bibinfo {author} {\bibfnamefont {L.-B.}\ \bibnamefont
  {Chen}}, \bibinfo {author} {\bibfnamefont {H.~T.}\ \bibnamefont {Li}},
  \bibinfo {author} {\bibfnamefont {H.-S.}\ \bibnamefont {Shao}},\ and\
  \bibinfo {author} {\bibfnamefont {J.}~\bibnamefont {Wang}},\ }\bibfield
  {title} {\bibinfo {title} {{The gluon-fusion production of Higgs boson pair:
  N$^3$LO QCD corrections and top-quark mass effects}},\ }\href
  {https://doi.org/10.1007/JHEP03(2020)072} {\bibfield  {journal} {\bibinfo
  {journal} {JHEP}\ }\textbf {\bibinfo {volume} {03}},\ \bibinfo {pages}
  {072}},\ \Eprint {https://arxiv.org/abs/1912.13001} {arXiv:1912.13001
  [hep-ph]} \BibitemShut {NoStop}%
\bibitem [{\citenamefont {Heinrich}\ \emph {et~al.}(2019)\citenamefont
  {Heinrich}, \citenamefont {Jones}, \citenamefont {Kerner}, \citenamefont
  {Luisoni},\ and\ \citenamefont {Scyboz}}]{Heinrich:2019bkc}%
  \BibitemOpen
  \bibfield  {author} {\bibinfo {author} {\bibfnamefont {G.}~\bibnamefont
  {Heinrich}}, \bibinfo {author} {\bibfnamefont {S.~P.}\ \bibnamefont {Jones}},
  \bibinfo {author} {\bibfnamefont {M.}~\bibnamefont {Kerner}}, \bibinfo
  {author} {\bibfnamefont {G.}~\bibnamefont {Luisoni}},\ and\ \bibinfo {author}
  {\bibfnamefont {L.}~\bibnamefont {Scyboz}},\ }\bibfield  {title} {\bibinfo
  {title} {{Probing the trilinear Higgs boson coupling in di-Higgs production
  at NLO QCD including parton shower effects}},\ }\href
  {https://doi.org/10.1007/JHEP06(2019)066} {\bibfield  {journal} {\bibinfo
  {journal} {JHEP}\ }\textbf {\bibinfo {volume} {06}},\ \bibinfo {pages}
  {066}},\ \Eprint {https://arxiv.org/abs/1903.08137} {arXiv:1903.08137
  [hep-ph]} \BibitemShut {NoStop}%
\bibitem [{\citenamefont {Baglio}\ \emph {et~al.}(2021)\citenamefont {Baglio},
  \citenamefont {Campanario}, \citenamefont {Glaus}, \citenamefont
  {M\"uhlleitner}, \citenamefont {Ronca},\ and\ \citenamefont
  {Spira}}]{Baglio:2020wgt}%
  \BibitemOpen
  \bibfield  {author} {\bibinfo {author} {\bibfnamefont {J.}~\bibnamefont
  {Baglio}}, \bibinfo {author} {\bibfnamefont {F.}~\bibnamefont {Campanario}},
  \bibinfo {author} {\bibfnamefont {S.}~\bibnamefont {Glaus}}, \bibinfo
  {author} {\bibfnamefont {M.}~\bibnamefont {M\"uhlleitner}}, \bibinfo {author}
  {\bibfnamefont {J.}~\bibnamefont {Ronca}},\ and\ \bibinfo {author}
  {\bibfnamefont {M.}~\bibnamefont {Spira}},\ }\bibfield  {title} {\bibinfo
  {title} {{$gg\to HH$ : Combined uncertainties}},\ }\href
  {https://doi.org/10.1103/PhysRevD.103.056002} {\bibfield  {journal} {\bibinfo
   {journal} {Phys. Rev. D}\ }\textbf {\bibinfo {volume} {103}},\ \bibinfo
  {pages} {056002} (\bibinfo {year} {2021})},\ \Eprint
  {https://arxiv.org/abs/2008.11626} {arXiv:2008.11626 [hep-ph]} \BibitemShut
  {NoStop}%
\bibitem [{\citenamefont {de~Florian}\ \emph
  {et~al.}(2016{\natexlab{b}})\citenamefont {de~Florian} \emph
  {et~al.}}]{LHCHiggsCrossSectionWorkingGroup:2016ypw}%
  \BibitemOpen
  \bibfield  {author} {\bibinfo {author} {\bibfnamefont {D.}~\bibnamefont
  {de~Florian}} \emph {et~al.} (\bibinfo {collaboration} {LHC Higgs Cross
  Section Working Group}),\ }\bibfield  {title} {\bibinfo {title} {{Handbook of
  LHC Higgs Cross Sections: 4. Deciphering the Nature of the Higgs Sector}}\
  }\textbf {\bibinfo {volume} {2/2017}},\ \href
  {https://doi.org/10.23731/CYRM-2017-002} {10.23731/CYRM-2017-002} (\bibinfo
  {year} {2016}{\natexlab{b}}),\ \Eprint {https://arxiv.org/abs/1610.07922}
  {arXiv:1610.07922 [hep-ph]} \BibitemShut {NoStop}%
\bibitem [{\citenamefont {Artoisenet}\ \emph {et~al.}(2013)\citenamefont
  {Artoisenet}, \citenamefont {Frederix}, \citenamefont {Mattelaer},\ and\
  \citenamefont {Rietkerk}}]{Artoisenet_2013}%
  \BibitemOpen
  \bibfield  {author} {\bibinfo {author} {\bibfnamefont {P.}~\bibnamefont
  {Artoisenet}}, \bibinfo {author} {\bibfnamefont {R.}~\bibnamefont
  {Frederix}}, \bibinfo {author} {\bibfnamefont {O.}~\bibnamefont
  {Mattelaer}},\ and\ \bibinfo {author} {\bibfnamefont {R.}~\bibnamefont
  {Rietkerk}},\ }\bibfield  {title} {\bibinfo {title} {Automatic spin-entangled
  decays of heavy resonances in monte carlo simulations},\ }\bibfield
  {journal} {\bibinfo  {journal} {Journal of High Energy Physics}\ }\textbf
  {\bibinfo {volume} {2013}},\ \href {https://doi.org/10.1007/jhep03(2013)015}
  {10.1007/jhep03(2013)015} (\bibinfo {year} {2013})\BibitemShut {NoStop}%
\bibitem [{\citenamefont {Bierlich}\ \emph {et~al.}(2022)\citenamefont
  {Bierlich} \emph {et~al.}}]{Bierlich:2022pfr}%
  \BibitemOpen
  \bibfield  {author} {\bibinfo {author} {\bibfnamefont {C.}~\bibnamefont
  {Bierlich}} \emph {et~al.},\ }\bibfield  {title} {\bibinfo {title} {{A
  comprehensive guide to the physics and usage of PYTHIA 8.3}},\ }\href
  {https://doi.org/10.21468/SciPostPhysCodeb.8} {\bibfield  {journal} {\bibinfo
   {journal} {SciPost Phys. Codeb.}\ }\textbf {\bibinfo {volume} {2022}},\
  \bibinfo {pages} {8} (\bibinfo {year} {2022})},\ \Eprint
  {https://arxiv.org/abs/2203.11601} {arXiv:2203.11601 [hep-ph]} \BibitemShut
  {NoStop}%
\bibitem [{\citenamefont {de~Favereau}\ \emph {et~al.}(2014)\citenamefont
  {de~Favereau}, \citenamefont {Delaere}, \citenamefont {Demin}, \citenamefont
  {Giammanco}, \citenamefont {Lemaître}, \citenamefont {Mertens},\ and\
  \citenamefont {Selvaggi}}]{de_Favereau_2014}%
  \BibitemOpen
  \bibfield  {author} {\bibinfo {author} {\bibfnamefont {J.}~\bibnamefont
  {de~Favereau}}, \bibinfo {author} {\bibfnamefont {C.}~\bibnamefont
  {Delaere}}, \bibinfo {author} {\bibfnamefont {P.}~\bibnamefont {Demin}},
  \bibinfo {author} {\bibfnamefont {A.}~\bibnamefont {Giammanco}}, \bibinfo
  {author} {\bibfnamefont {V.}~\bibnamefont {Lemaître}}, \bibinfo {author}
  {\bibfnamefont {A.}~\bibnamefont {Mertens}},\ and\ \bibinfo {author}
  {\bibfnamefont {M.}~\bibnamefont {Selvaggi}},\ }\bibfield  {title} {\bibinfo
  {title} {Delphes 3: a modular framework for fast simulation of a generic
  collider experiment},\ }\bibfield  {journal} {\bibinfo  {journal} {Journal of
  High Energy Physics}\ }\textbf {\bibinfo {volume} {2014}},\ \href
  {https://doi.org/10.1007/jhep02(2014)057} {10.1007/jhep02(2014)057} (\bibinfo
  {year} {2014})\BibitemShut {NoStop}%
\bibitem [{\citenamefont {Cacciari}\ \emph {et~al.}(2008)\citenamefont
  {Cacciari}, \citenamefont {Salam},\ and\ \citenamefont
  {Soyez}}]{Cacciari:2008gp}%
  \BibitemOpen
  \bibfield  {author} {\bibinfo {author} {\bibfnamefont {M.}~\bibnamefont
  {Cacciari}}, \bibinfo {author} {\bibfnamefont {G.~P.}\ \bibnamefont
  {Salam}},\ and\ \bibinfo {author} {\bibfnamefont {G.}~\bibnamefont {Soyez}},\
  }\bibfield  {title} {\bibinfo {title} {{The anti-$k_t$ jet clustering
  algorithm}},\ }\href {https://doi.org/10.1088/1126-6708/2008/04/063}
  {\bibfield  {journal} {\bibinfo  {journal} {JHEP}\ }\textbf {\bibinfo
  {volume} {04}},\ \bibinfo {pages} {063}},\ \Eprint
  {https://arxiv.org/abs/0802.1189} {arXiv:0802.1189 [hep-ph]} \BibitemShut
  {NoStop}%
\bibitem [{\citenamefont {Cacciari}\ \emph {et~al.}(2012)\citenamefont
  {Cacciari}, \citenamefont {Salam},\ and\ \citenamefont
  {Soyez}}]{Cacciari:2011ma}%
  \BibitemOpen
  \bibfield  {author} {\bibinfo {author} {\bibfnamefont {M.}~\bibnamefont
  {Cacciari}}, \bibinfo {author} {\bibfnamefont {G.~P.}\ \bibnamefont
  {Salam}},\ and\ \bibinfo {author} {\bibfnamefont {G.}~\bibnamefont {Soyez}},\
  }\bibfield  {title} {\bibinfo {title} {{FastJet User Manual}},\ }\href
  {https://doi.org/10.1140/epjc/s10052-012-1896-2} {\bibfield  {journal}
  {\bibinfo  {journal} {Eur. Phys. J. C}\ }\textbf {\bibinfo {volume} {72}},\
  \bibinfo {pages} {1896} (\bibinfo {year} {2012})},\ \Eprint
  {https://arxiv.org/abs/1111.6097} {arXiv:1111.6097 [hep-ph]} \BibitemShut
  {NoStop}%
\bibitem [{\citenamefont {Heinrich}\ and\ \citenamefont
  {Lang}(2024)}]{Heinrich:2023rsd}%
  \BibitemOpen
  \bibfield  {author} {\bibinfo {author} {\bibfnamefont {G.}~\bibnamefont
  {Heinrich}}\ and\ \bibinfo {author} {\bibfnamefont {J.}~\bibnamefont
  {Lang}},\ }\bibfield  {title} {\bibinfo {title} {{Combining chromomagnetic
  and four-fermion operators with leading SMEFT operators for gg
  \textrightarrow{} hh at NLO QCD}},\ }\href
  {https://doi.org/10.1007/JHEP05(2024)121} {\bibfield  {journal} {\bibinfo
  {journal} {JHEP}\ }\textbf {\bibinfo {volume} {05}},\ \bibinfo {pages}
  {121}},\ \Eprint {https://arxiv.org/abs/2311.15004} {arXiv:2311.15004
  [hep-ph]} \BibitemShut {NoStop}%
\bibitem [{\citenamefont {Heinrich}\ \emph {et~al.}(2020)\citenamefont
  {Heinrich}, \citenamefont {Jones}, \citenamefont {Kerner},\ and\
  \citenamefont {Scyboz}}]{Heinrich:2020ckp}%
  \BibitemOpen
  \bibfield  {author} {\bibinfo {author} {\bibfnamefont {G.}~\bibnamefont
  {Heinrich}}, \bibinfo {author} {\bibfnamefont {S.~P.}\ \bibnamefont {Jones}},
  \bibinfo {author} {\bibfnamefont {M.}~\bibnamefont {Kerner}},\ and\ \bibinfo
  {author} {\bibfnamefont {L.}~\bibnamefont {Scyboz}},\ }\bibfield  {title}
  {\bibinfo {title} {{A non-linear EFT description of $gg\to HH$ at NLO
  interfaced to POWHEG}},\ }\href {https://doi.org/10.1007/JHEP10(2020)021}
  {\bibfield  {journal} {\bibinfo  {journal} {JHEP}\ }\textbf {\bibinfo
  {volume} {10}},\ \bibinfo {pages} {021}},\ \Eprint
  {https://arxiv.org/abs/2006.16877} {arXiv:2006.16877 [hep-ph]} \BibitemShut
  {NoStop}%
\bibitem [{\citenamefont {Heinrich}\ \emph {et~al.}(2022)\citenamefont
  {Heinrich}, \citenamefont {Lang},\ and\ \citenamefont
  {Scyboz}}]{Heinrich:2022idm}%
  \BibitemOpen
  \bibfield  {author} {\bibinfo {author} {\bibfnamefont {G.}~\bibnamefont
  {Heinrich}}, \bibinfo {author} {\bibfnamefont {J.}~\bibnamefont {Lang}},\
  and\ \bibinfo {author} {\bibfnamefont {L.}~\bibnamefont {Scyboz}},\
  }\bibfield  {title} {\bibinfo {title} {{SMEFT predictions for gg
  \textrightarrow{} hh at full NLO QCD and truncation uncertainties}},\ }\href
  {https://doi.org/10.1007/JHEP08(2022)079} {\bibfield  {journal} {\bibinfo
  {journal} {JHEP}\ }\textbf {\bibinfo {volume} {08}},\ \bibinfo {pages}
  {079}},\ \bibinfo {note} {[Erratum: JHEP 10, 086 (2023)]},\ \Eprint
  {https://arxiv.org/abs/2204.13045} {arXiv:2204.13045 [hep-ph]} \BibitemShut
  {NoStop}%
\bibitem [{\citenamefont {de~Florian}\ \emph {et~al.}(2021)\citenamefont
  {de~Florian}, \citenamefont {Fabre}, \citenamefont {Heinrich}, \citenamefont
  {Mazzitelli},\ and\ \citenamefont {Scyboz}}]{deFlorian:2021azd}%
  \BibitemOpen
  \bibfield  {author} {\bibinfo {author} {\bibfnamefont {D.}~\bibnamefont
  {de~Florian}}, \bibinfo {author} {\bibfnamefont {I.}~\bibnamefont {Fabre}},
  \bibinfo {author} {\bibfnamefont {G.}~\bibnamefont {Heinrich}}, \bibinfo
  {author} {\bibfnamefont {J.}~\bibnamefont {Mazzitelli}},\ and\ \bibinfo
  {author} {\bibfnamefont {L.}~\bibnamefont {Scyboz}},\ }\bibfield  {title}
  {\bibinfo {title} {{Anomalous couplings in Higgs-boson pair production at
  approximate NNLO QCD}},\ }\href {https://doi.org/10.1007/JHEP09(2021)161}
  {\bibfield  {journal} {\bibinfo  {journal} {JHEP}\ }\textbf {\bibinfo
  {volume} {09}},\ \bibinfo {pages} {161}},\ \Eprint
  {https://arxiv.org/abs/2106.14050} {arXiv:2106.14050 [hep-ph]} \BibitemShut
  {NoStop}%
\bibitem [{\citenamefont {F\"ah}\ and\ \citenamefont
  {Greiner}(2017)}]{Fah:2017wlf}%
  \BibitemOpen
  \bibfield  {author} {\bibinfo {author} {\bibfnamefont {D.}~\bibnamefont
  {F\"ah}}\ and\ \bibinfo {author} {\bibfnamefont {N.}~\bibnamefont
  {Greiner}},\ }\bibfield  {title} {\bibinfo {title} {{Diphoton production in
  association with two bottom jets}},\ }\href
  {https://doi.org/10.1140/epjc/s10052-017-5296-5} {\bibfield  {journal}
  {\bibinfo  {journal} {Eur. Phys. J. C}\ }\textbf {\bibinfo {volume} {77}},\
  \bibinfo {pages} {750} (\bibinfo {year} {2017})},\ \Eprint
  {https://arxiv.org/abs/1706.08309} {arXiv:1706.08309 [hep-ph]} \BibitemShut
  {NoStop}%
\bibitem [{\citenamefont {Kim}\ \emph {et~al.}(2024)\citenamefont {Kim},
  \citenamefont {Lee}, \citenamefont {Jung}, \citenamefont {Kim}, \citenamefont
  {Kim},\ and\ \citenamefont {Song}}]{Kim:2024ppt}%
  \BibitemOpen
  \bibfield  {author} {\bibinfo {author} {\bibfnamefont {D.}~\bibnamefont
  {Kim}}, \bibinfo {author} {\bibfnamefont {S.}~\bibnamefont {Lee}}, \bibinfo
  {author} {\bibfnamefont {H.}~\bibnamefont {Jung}}, \bibinfo {author}
  {\bibfnamefont {D.}~\bibnamefont {Kim}}, \bibinfo {author} {\bibfnamefont
  {J.}~\bibnamefont {Kim}},\ and\ \bibinfo {author} {\bibfnamefont
  {J.}~\bibnamefont {Song}},\ }\bibfield  {title} {\bibinfo {title} {{A
  panoramic study of K-factors for 111 processes at the 14 TeV LHC}},\ }\href
  {https://doi.org/10.1007/s40042-024-01072-0} {\bibfield  {journal} {\bibinfo
  {journal} {J. Korean Phys. Soc.}\ }\textbf {\bibinfo {volume} {84}},\
  \bibinfo {pages} {914} (\bibinfo {year} {2024})},\ \Eprint
  {https://arxiv.org/abs/2402.16276} {arXiv:2402.16276 [hep-ph]} \BibitemShut
  {NoStop}%
\bibitem [{\citenamefont {Mangano}\ \emph {et~al.}(2016)\citenamefont {Mangano}
  \emph {et~al.}}]{Mangano:2016jyj}%
  \BibitemOpen
  \bibfield  {author} {\bibinfo {author} {\bibfnamefont {M.~L.}\ \bibnamefont
  {Mangano}} \emph {et~al.},\ }\bibfield  {title} {\bibinfo {title} {{Physics
  at a 100 TeV pp Collider: Standard Model Processes}}\ }\href
  {https://doi.org/10.23731/CYRM-2017-003.1} {10.23731/CYRM-2017-003.1}
  (\bibinfo {year} {2016}),\ \Eprint {https://arxiv.org/abs/1607.01831}
  {arXiv:1607.01831 [hep-ph]} \BibitemShut {NoStop}%
\bibitem [{\citenamefont {Sirunyan}\ \emph {et~al.}(2021)\citenamefont
  {Sirunyan} \emph {et~al.}}]{CMS:2020tkr}%
  \BibitemOpen
  \bibfield  {author} {\bibinfo {author} {\bibfnamefont {A.~M.}\ \bibnamefont
  {Sirunyan}} \emph {et~al.} (\bibinfo {collaboration} {CMS}),\ }\bibfield
  {title} {\bibinfo {title} {{Search for nonresonant Higgs boson pair
  production in final states with two bottom quarks and two photons in
  proton-proton collisions at $ \sqrt{s} $ = 13 TeV}},\ }\href
  {https://doi.org/10.1007/JHEP03(2021)257} {\bibfield  {journal} {\bibinfo
  {journal} {JHEP}\ }\textbf {\bibinfo {volume} {03}},\ \bibinfo {pages}
  {257}},\ \Eprint {https://arxiv.org/abs/2011.12373} {arXiv:2011.12373
  [hep-ex]} \BibitemShut {NoStop}%
\bibitem [{\citenamefont {Jia}(2024)}]{Jia:2024nbv}%
  \BibitemOpen
  \bibfield  {author} {\bibinfo {author} {\bibfnamefont {Z.}~\bibnamefont
  {Jia}} (\bibinfo {collaboration} {ATLAS}),\ }\bibfield  {title} {\bibinfo
  {title} {{Studies of new Higgs boson interactions through nonresonant $HH$
  production in the $b\bar{b}\gamma\gamma$ final state in $pp$ collisions at
  $\sqrt{s}$ = 13 TeV with the ATLAS detector}},\ }\href
  {https://doi.org/10.22323/1.449.0422} {\bibfield  {journal} {\bibinfo
  {journal} {PoS}\ }\textbf {\bibinfo {volume} {EPS-HEP2023}},\ \bibinfo
  {pages} {422} (\bibinfo {year} {2024})}\BibitemShut {NoStop}%
\bibitem [{\citenamefont {Chatterjee}\ \emph {et~al.}(2022)\citenamefont
  {Chatterjee}, \citenamefont {Rohshap}, \citenamefont {Sch\"ofbeck},\ and\
  \citenamefont {Schwarz}}]{Chatterjee:2022oco}%
  \BibitemOpen
  \bibfield  {author} {\bibinfo {author} {\bibfnamefont {S.}~\bibnamefont
  {Chatterjee}}, \bibinfo {author} {\bibfnamefont {S.}~\bibnamefont {Rohshap}},
  \bibinfo {author} {\bibfnamefont {R.}~\bibnamefont {Sch\"ofbeck}},\ and\
  \bibinfo {author} {\bibfnamefont {D.}~\bibnamefont {Schwarz}},\ }\bibfield
  {title} {\bibinfo {title} {{Learning the EFT likelihood with tree
  boosting}},\ }\href@noop {} {\  (\bibinfo {year} {2022})},\ \Eprint
  {https://arxiv.org/abs/2205.12976} {arXiv:2205.12976 [hep-ph]} \BibitemShut
  {NoStop}%
\bibitem [{\citenamefont {Carrazza}\ \emph {et~al.}(2021)\citenamefont
  {Carrazza}, \citenamefont {Cruz-Martinez}, \citenamefont {Rossi},\ and\
  \citenamefont {Zaro}}]{Carrazza:2021zug}%
  \BibitemOpen
  \bibfield  {author} {\bibinfo {author} {\bibfnamefont {S.}~\bibnamefont
  {Carrazza}}, \bibinfo {author} {\bibfnamefont {J.}~\bibnamefont
  {Cruz-Martinez}}, \bibinfo {author} {\bibfnamefont {M.}~\bibnamefont
  {Rossi}},\ and\ \bibinfo {author} {\bibfnamefont {M.}~\bibnamefont {Zaro}},\
  }\bibfield  {title} {\bibinfo {title} {{MadFlow: towards the automation of
  Monte Carlo simulation on GPU for particle physics processes}},\ }\href
  {https://doi.org/10.1051/epjconf/202125103022} {\bibfield  {journal}
  {\bibinfo  {journal} {EPJ Web Conf.}\ }\textbf {\bibinfo {volume} {251}},\
  \bibinfo {pages} {03022} (\bibinfo {year} {2021})},\ \Eprint
  {https://arxiv.org/abs/2105.10529} {arXiv:2105.10529 [physics.comp-ph]}
  \BibitemShut {NoStop}%
\bibitem [{\citenamefont {Heinrich}\ and\ \citenamefont
  {Kagan}(2023)}]{Heinrich:2022xfa}%
  \BibitemOpen
  \bibfield  {author} {\bibinfo {author} {\bibfnamefont {L.}~\bibnamefont
  {Heinrich}}\ and\ \bibinfo {author} {\bibfnamefont {M.}~\bibnamefont
  {Kagan}},\ }\bibfield  {title} {\bibinfo {title} {{Differentiable Matrix
  Elements with MadJax}},\ }\href
  {https://doi.org/10.1088/1742-6596/2438/1/012137} {\bibfield  {journal}
  {\bibinfo  {journal} {J. Phys. Conf. Ser.}\ }\textbf {\bibinfo {volume}
  {2438}},\ \bibinfo {pages} {012137} (\bibinfo {year} {2023})},\ \Eprint
  {https://arxiv.org/abs/2203.00057} {arXiv:2203.00057 [hep-ph]} \BibitemShut
  {NoStop}%
\bibitem [{\citenamefont {Nachman}\ and\ \citenamefont
  {Prestel}(2022)}]{Nachman:2022jbj}%
  \BibitemOpen
  \bibfield  {author} {\bibinfo {author} {\bibfnamefont {B.}~\bibnamefont
  {Nachman}}\ and\ \bibinfo {author} {\bibfnamefont {S.}~\bibnamefont
  {Prestel}},\ }\bibfield  {title} {\bibinfo {title} {{Morphing parton showers
  with event derivatives}},\ }\href@noop {} {\  (\bibinfo {year} {2022})},\
  \Eprint {https://arxiv.org/abs/2208.02274} {arXiv:2208.02274 [hep-ph]}
  \BibitemShut {NoStop}%
\bibitem [{\citenamefont {Neyman}\ and\ \citenamefont
  {Pearson}(1933)}]{neyman1933ix}%
  \BibitemOpen
  \bibfield  {author} {\bibinfo {author} {\bibfnamefont {J.}~\bibnamefont
  {Neyman}}\ and\ \bibinfo {author} {\bibfnamefont {E.~S.}\ \bibnamefont
  {Pearson}},\ }\bibfield  {title} {\bibinfo {title} {On the problem of the
  most efficient tests of statistical hypotheses},\ }\href@noop {} {\bibfield
  {journal} {\bibinfo  {journal} {Phil. Trans. R. Soc. Lond. A}\ }\textbf
  {\bibinfo {volume} {231}},\ \bibinfo {pages} {289} (\bibinfo {year}
  {1933})}\BibitemShut {NoStop}%
\bibitem [{\citenamefont {Hastie}\ \emph {et~al.}(2001)\citenamefont {Hastie},
  \citenamefont {Tibshirani},\ and\ \citenamefont
  {Friedman}}]{hastie01statisticallearning}%
  \BibitemOpen
  \bibfield  {author} {\bibinfo {author} {\bibfnamefont {T.}~\bibnamefont
  {Hastie}}, \bibinfo {author} {\bibfnamefont {R.}~\bibnamefont {Tibshirani}},\
  and\ \bibinfo {author} {\bibfnamefont {J.}~\bibnamefont {Friedman}},\
  }\href@noop {} {\emph {\bibinfo {title} {The Elements of Statistical
  Learning}}},\ Springer Series in Statistics\ (\bibinfo  {publisher} {Springer
  New York Inc.},\ \bibinfo {address} {New York, NY, USA},\ \bibinfo {year}
  {2001})\BibitemShut {NoStop}%
\bibitem [{\citenamefont {Sugiyama}\ \emph {et~al.}(2012)\citenamefont
  {Sugiyama}, \citenamefont {Suzuki},\ and\ \citenamefont
  {Kanamori}}]{sugiyama_suzuki_kanamori_2012}%
  \BibitemOpen
  \bibfield  {author} {\bibinfo {author} {\bibfnamefont {M.}~\bibnamefont
  {Sugiyama}}, \bibinfo {author} {\bibfnamefont {T.}~\bibnamefont {Suzuki}},\
  and\ \bibinfo {author} {\bibfnamefont {T.}~\bibnamefont {Kanamori}},\ }\href
  {https://doi.org/10.1017/CBO9781139035613} {\emph {\bibinfo {title} {Density
  Ratio Estimation in Machine Learning}}}\ (\bibinfo  {publisher} {Cambridge
  University Press},\ \bibinfo {year} {2012})\BibitemShut {NoStop}%
\bibitem [{\citenamefont {Cranmer}\ \emph {et~al.}(2015)\citenamefont
  {Cranmer}, \citenamefont {Pavez},\ and\ \citenamefont {Louppe}}]{cranmer}%
  \BibitemOpen
  \bibfield  {author} {\bibinfo {author} {\bibfnamefont {K.}~\bibnamefont
  {Cranmer}}, \bibinfo {author} {\bibfnamefont {J.}~\bibnamefont {Pavez}},\
  and\ \bibinfo {author} {\bibfnamefont {G.}~\bibnamefont {Louppe}},\ }\href
  {https://doi.org/10.48550/ARXIV.1506.02169} {\bibinfo {title} {{Approximating
  Likelihood Ratios with Calibrated Discriminative Classifiers}}} (\bibinfo
  {year} {2015})\BibitemShut {NoStop}%
\bibitem [{\citenamefont {Rizvi}\ \emph {et~al.}(2024)\citenamefont {Rizvi},
  \citenamefont {Pettee},\ and\ \citenamefont {Nachman}}]{Rizvi:2023mws}%
  \BibitemOpen
  \bibfield  {author} {\bibinfo {author} {\bibfnamefont {S.}~\bibnamefont
  {Rizvi}}, \bibinfo {author} {\bibfnamefont {M.}~\bibnamefont {Pettee}},\ and\
  \bibinfo {author} {\bibfnamefont {B.}~\bibnamefont {Nachman}},\ }\bibfield
  {title} {\bibinfo {title} {{Learning likelihood ratios with neural network
  classifiers}},\ }\href {https://doi.org/10.1007/JHEP02(2024)136} {\bibfield
  {journal} {\bibinfo  {journal} {JHEP}\ }\textbf {\bibinfo {volume} {02}},\
  \bibinfo {pages} {136}},\ \Eprint {https://arxiv.org/abs/2305.10500}
  {arXiv:2305.10500 [hep-ph]} \BibitemShut {NoStop}%
\bibitem [{\citenamefont {Baldi}\ \emph {et~al.}(2016)\citenamefont {Baldi},
  \citenamefont {Cranmer}, \citenamefont {Faucett}, \citenamefont {Sadowski},\
  and\ \citenamefont {Whiteson}}]{Baldi:2016fzo}%
  \BibitemOpen
  \bibfield  {author} {\bibinfo {author} {\bibfnamefont {P.}~\bibnamefont
  {Baldi}}, \bibinfo {author} {\bibfnamefont {K.}~\bibnamefont {Cranmer}},
  \bibinfo {author} {\bibfnamefont {T.}~\bibnamefont {Faucett}}, \bibinfo
  {author} {\bibfnamefont {P.}~\bibnamefont {Sadowski}},\ and\ \bibinfo
  {author} {\bibfnamefont {D.}~\bibnamefont {Whiteson}},\ }\bibfield  {title}
  {\bibinfo {title} {{Parameterized neural networks for high-energy physics}},\
  }\href {https://doi.org/10.1140/epjc/s10052-016-4099-4} {\bibfield  {journal}
  {\bibinfo  {journal} {Eur. Phys. J. C}\ }\textbf {\bibinfo {volume} {76}},\
  \bibinfo {pages} {235} (\bibinfo {year} {2016})},\ \Eprint
  {https://arxiv.org/abs/1601.07913} {arXiv:1601.07913 [hep-ex]} \BibitemShut
  {NoStop}%
\bibitem [{\citenamefont {Paszke}\ \emph {et~al.}(2019)\citenamefont {Paszke},
  \citenamefont {Gross}, \citenamefont {Massa}, \citenamefont {Lerer},
  \citenamefont {Bradbury}, \citenamefont {Chanan}, \citenamefont {Killeen},
  \citenamefont {Lin}, \citenamefont {Gimelshein}, \citenamefont {Antiga},
  \citenamefont {Desmaison}, \citenamefont {Kopf}, \citenamefont {Yang},
  \citenamefont {DeVito}, \citenamefont {Raison}, \citenamefont {Tejani},
  \citenamefont {Chilamkurthy}, \citenamefont {Steiner}, \citenamefont {Fang},
  \citenamefont {Bai},\ and\ \citenamefont {Chintala}}]{NEURIPS2019_9015}%
  \BibitemOpen
  \bibfield  {author} {\bibinfo {author} {\bibfnamefont {A.}~\bibnamefont
  {Paszke}}, \bibinfo {author} {\bibfnamefont {S.}~\bibnamefont {Gross}},
  \bibinfo {author} {\bibfnamefont {F.}~\bibnamefont {Massa}}, \bibinfo
  {author} {\bibfnamefont {A.}~\bibnamefont {Lerer}}, \bibinfo {author}
  {\bibfnamefont {J.}~\bibnamefont {Bradbury}}, \bibinfo {author}
  {\bibfnamefont {G.}~\bibnamefont {Chanan}}, \bibinfo {author} {\bibfnamefont
  {T.}~\bibnamefont {Killeen}}, \bibinfo {author} {\bibfnamefont
  {Z.}~\bibnamefont {Lin}}, \bibinfo {author} {\bibfnamefont {N.}~\bibnamefont
  {Gimelshein}}, \bibinfo {author} {\bibfnamefont {L.}~\bibnamefont {Antiga}},
  \bibinfo {author} {\bibfnamefont {A.}~\bibnamefont {Desmaison}}, \bibinfo
  {author} {\bibfnamefont {A.}~\bibnamefont {Kopf}}, \bibinfo {author}
  {\bibfnamefont {E.}~\bibnamefont {Yang}}, \bibinfo {author} {\bibfnamefont
  {Z.}~\bibnamefont {DeVito}}, \bibinfo {author} {\bibfnamefont
  {M.}~\bibnamefont {Raison}}, \bibinfo {author} {\bibfnamefont
  {A.}~\bibnamefont {Tejani}}, \bibinfo {author} {\bibfnamefont
  {S.}~\bibnamefont {Chilamkurthy}}, \bibinfo {author} {\bibfnamefont
  {B.}~\bibnamefont {Steiner}}, \bibinfo {author} {\bibfnamefont
  {L.}~\bibnamefont {Fang}}, \bibinfo {author} {\bibfnamefont {J.}~\bibnamefont
  {Bai}},\ and\ \bibinfo {author} {\bibfnamefont {S.}~\bibnamefont
  {Chintala}},\ }\bibfield  {title} {\bibinfo {title} {Pytorch: An imperative
  style, high-performance deep learning library},\ }in\ \href
  {http://papers.neurips.cc/paper/9015-pytorch-an-imperative-style-high-performance-deep-learning-library.pdf}
  {\emph {\bibinfo {booktitle} {Advances in Neural Information Processing
  Systems 32}}}\ (\bibinfo  {publisher} {Curran Associates, Inc.},\ \bibinfo
  {year} {2019})\ pp.\ \bibinfo {pages} {8024--8035}\BibitemShut {NoStop}%
\bibitem [{\citenamefont {Kingma}\ and\ \citenamefont
  {Ba}(2017)}]{kingma2017adam}%
  \BibitemOpen
  \bibfield  {author} {\bibinfo {author} {\bibfnamefont {D.~P.}\ \bibnamefont
  {Kingma}}\ and\ \bibinfo {author} {\bibfnamefont {J.}~\bibnamefont {Ba}},\
  }\href@noop {} {\bibinfo {title} {Adam: A method for stochastic
  optimization}} (\bibinfo {year} {2017}),\ \Eprint
  {https://arxiv.org/abs/1412.6980} {arXiv:1412.6980 [cs.LG]} \BibitemShut
  {NoStop}%
\end{thebibliography}%

\end{document}